\documentclass[12pt,english]{article}
\usepackage{times}
\usepackage[T1]{fontenc}
\usepackage{geometry}
\usepackage{rotating}
\usepackage{color}
\usepackage{graphicx}
\usepackage{setspace}
\usepackage{amssymb}

\makeatletter

\providecommand{\tabularnewline}{\\}

\usepackage{bm}

\usepackage{babel}
\makeatother
\begin{document}

\section*{\textcolor{black}{A Self-Replicating Single-Shape Tiling Technique
for the Design of Highly Modular Planar Phased Arrays - The Case of}
\textcolor{black}{\emph{L}}\textcolor{black}{-Shaped Rep-Tiles}}

\noindent ~

\noindent \vfill

\noindent N. Anselmi,$^{(1)}$ \emph{Senior Member, IEEE}, L. Tosi,$^{(1)}$,
P. Rocca,$^{(1)(2)}$ \emph{Senior Member, IEEE}, G. Toso,$^{(3)}$
\emph{Senior Member, IEEE}, and A. Massa,$^{(1)(4)(5)}$ \emph{Fellow,
IEEE}

\noindent \vfill

\noindent ~

\noindent ~

\noindent \textcolor{black}{\scriptsize $^{(1)}$} \textcolor{black}{\emph{\scriptsize ELEDIA
Research Center}} \textcolor{black}{\scriptsize (}\textcolor{black}{\emph{\scriptsize ELEDIA}}\textcolor{black}{\scriptsize @}\textcolor{black}{\emph{\scriptsize UniTN}}
\textcolor{black}{\scriptsize - University of Trento)}{\scriptsize \par}

\noindent \textcolor{black}{\scriptsize DICAM - Department of Civil,
Environmental, and Mechanical Engineering}{\scriptsize \par}

\noindent \textcolor{black}{\scriptsize Via Mesiano 77, 38123 Trento
- Italy}{\scriptsize \par}

\noindent \textit{\textcolor{black}{\emph{\scriptsize E-mail:}}} \textcolor{black}{\scriptsize \{}\textcolor{black}{\emph{\scriptsize nicola.anselmi}}\textcolor{black}{\scriptsize ,}
\textcolor{black}{\emph{\scriptsize luca.tosi}}\textcolor{black}{\scriptsize ,}
\textcolor{black}{\emph{\scriptsize paolo.rocca}}\textcolor{black}{\scriptsize ,}
\textcolor{black}{\emph{\scriptsize andrea.massa}}\textcolor{black}{\scriptsize \}@}\textcolor{black}{\emph{\scriptsize unitn.it}}{\scriptsize \par}

\noindent \textcolor{black}{\scriptsize Website:} \textcolor{black}{\emph{\scriptsize www.eledia.org/eledia-unitn}}{\scriptsize \par}

\noindent \textcolor{black}{\scriptsize ~}{\scriptsize \par}

\noindent \textcolor{black}{\scriptsize $^{(2)}$} \textcolor{black}{\emph{\scriptsize ELEDIA
Research Center}} \textcolor{black}{\scriptsize (}\textcolor{black}{\emph{\scriptsize ELEDIA@XIDIAN}}
\textcolor{black}{\scriptsize - Xidian University)}{\scriptsize \par}

\noindent \textcolor{black}{\scriptsize P.O. Box 191, No.2 South Tabai
Road, 710071 Xi'an, Shaanxi Province - China}{\scriptsize \par}

\noindent \textcolor{black}{\scriptsize E-mail:} \textcolor{black}{\emph{\scriptsize paolo.rocca@xidian.edu.cn}}{\scriptsize \par}

\noindent \textcolor{black}{\scriptsize Website:} \textcolor{black}{\emph{\scriptsize www.eledia.org/eledia-xidian}}{\scriptsize \par}

\noindent \textcolor{black}{\scriptsize ~}{\scriptsize \par}

\noindent \textcolor{black}{\scriptsize $^{(3)}$} \textcolor{black}{\emph{\scriptsize European
Space Agency}} \textcolor{black}{\scriptsize (ESA - ESTEC)}{\scriptsize \par}

\noindent \textcolor{black}{\scriptsize Antenna and Submillimeter
Wave Section, Radio Frequency Payloads \& Technology Division, 2200
AG Noordwijk - The Netherlands}{\scriptsize \par}

\noindent \textit{\textcolor{black}{\emph{\scriptsize E-mail:}}} \textcolor{black}{\emph{\scriptsize giovanni.toso@esa.int}}{\scriptsize \par}

\noindent \textcolor{black}{\scriptsize Website:} \textcolor{black}{\emph{\scriptsize www.esa.int}}{\scriptsize \par}

\noindent \textcolor{black}{\scriptsize ~}{\scriptsize \par}

\noindent \textcolor{black}{\scriptsize $^{(4)}$} \textcolor{black}{\emph{\scriptsize ELEDIA
Research Center}} \textcolor{black}{\scriptsize (}\textcolor{black}{\emph{\scriptsize ELEDIA}}\textcolor{black}{\scriptsize @}\textcolor{black}{\emph{\scriptsize UESTC}}
\textcolor{black}{\scriptsize - UESTC)}{\scriptsize \par}

\noindent \textcolor{black}{\scriptsize School of Electronic Engineering,
Chengdu 611731 - China}{\scriptsize \par}

\noindent \textit{\textcolor{black}{\emph{\scriptsize E-mail:}}} \textcolor{black}{\emph{\scriptsize andrea.massa@uestc.edu.cn}}{\scriptsize \par}

\noindent \textcolor{black}{\scriptsize Website:} \textcolor{black}{\emph{\scriptsize www.eledia.org/eledia}}\textcolor{black}{\scriptsize -}\textcolor{black}{\emph{\scriptsize uestc}}{\scriptsize \par}

\noindent \textcolor{black}{\scriptsize ~}{\scriptsize \par}

\noindent \textcolor{black}{\scriptsize $^{(5)}$} \textcolor{black}{\emph{\scriptsize ELEDIA
Research Center}} \textcolor{black}{\scriptsize (}\textcolor{black}{\emph{\scriptsize ELEDIA@TSINGHUA}}
\textcolor{black}{\scriptsize - Tsinghua University)}{\scriptsize \par}

\noindent \textcolor{black}{\scriptsize 30 Shuangqing Rd, 100084 Haidian,
Beijing - China}{\scriptsize \par}

\noindent \textcolor{black}{\scriptsize E-mail:} \textcolor{black}{\emph{\scriptsize andrea.massa@tsinghua.edu.cn}}{\scriptsize \par}

\noindent \textcolor{black}{\scriptsize Website:} \textcolor{black}{\emph{\scriptsize www.eledia.org/eledia-tsinghua}}{\scriptsize \par}

\vfill

\textbf{\emph{This work has been submitted to the IEEE for possible
publication. Copyright may be transferred without notice, after which
this version may no longer be accessible.}}

\noindent \vfill
\newpage

\section*{\textcolor{black}{A Self-Replicating Single-Shape Tiling Technique
for the Design of Highly Modular Planar Phased Arrays - The Case of}
\textcolor{black}{\emph{L}}\textcolor{black}{-Shaped Rep-Tiles}}

\noindent \textcolor{red}{~}

\noindent \textcolor{red}{~}

\noindent \textcolor{red}{~}

\noindent \textcolor{black}{N. Anselmi, L. Tosi, P. Rocca, G. Toso,
and A. Massa}

\noindent \textcolor{black}{\vfill}

\begin{abstract}
\noindent \textcolor{black}{The design of irregular planar phased
arrays (}\textcolor{black}{\emph{PA}}\textcolor{black}{s) characterized
by a highly-modular architecture is addressed. By exploiting the property
of self-replicating tile shapes, also known as} \textcolor{black}{\emph{rep-tiles}}\textcolor{black}{,
the arising array layouts consist of tiles having different sizes,
but equal shape, all being generated by assembling a finite number
of smaller and congruent copies of a single elementary building-block.
Towards this end, a deterministic optimization strategy is used so
that the arising rep-tile arrangement of the planar} \textcolor{black}{\emph{PA}}
\textcolor{black}{is an optimal trade-off between complexity, costs,
and fitting of user-defined requirements on the radiated power pattern,
while guaranteeing the complete overlay of the array aperture. As
a representative instance, such a synthesis method is applied to tile
rectangular apertures with} \textcolor{black}{\emph{L}}\textcolor{black}{-shaped
tromino tiles. A set of representative results, concerned with ideal
and real antenna models, as well, is reported for validation purposes,
but also to point out the possibility/effectiveness of the proposed
approach, unlike state-of-the-art tiling techniques, to reliably handle
large-size array apertures.}
\end{abstract}
\noindent \textcolor{black}{\vfill}

\noindent \textbf{\textcolor{black}{Key words}}\textcolor{black}{:
Phased Array Antenna, Planar Array, Irregular Tiling, Rep-tiles.}
\newpage

\section{\textcolor{black}{Introduction}}

\noindent \textcolor{black}{The interest towards innovative phased
array antennas (}\textcolor{black}{\emph{PA}}\textcolor{black}{s)
and related technologies has constantly grown over the last years
owing to many possible applications as, for instance, millimeter wave
communications \cite{Hong.2021}\cite{Oliveri.2019} and sensing \cite{Yu.2020},
automotive \cite{Waldshmidt.2021} and drone \cite{Peng.2021} radars,
directional heating systems \cite{Yang.2020}, and microwave imaging
\cite{Li.2021}\cite{Gu.2015} just to mention a few. The capability
of quickly reconfiguring the radiation pattern and its features by
controlling the amplitude and/or the phase coefficients of the excitations
of the array elements makes} \textcolor{black}{\emph{PA}}\textcolor{black}{s
a very attractive and promising technology. However, regularly-spaced
digital} \textcolor{black}{\emph{PA}}\textcolor{black}{s, where the
radiating elements are arranged on a uniform lattice and each element
is equipped with an independent transmit/receive module (}\textcolor{black}{\emph{TRM}}\textcolor{black}{)
and a digital channel, are too expensive and complex for standard
applications \cite{Herd.2016}\cite{Abdellatif.2021}.}

\noindent \textcolor{black}{Unconventional architectures \cite{Rocca.2016},
such as thinned \cite{Oliveri.2020}\cite{Poli.2013}, sparse \cite{Carlin.2015}\cite{Viswanathan.2021},
and clustered \cite{Manica.2008}-\cite{Rapakula.2020} arrays, have
been recently considered to yield a suitable compromise between performance
and costs. In such a framework, tiled} \textcolor{black}{\emph{PA}}\textcolor{black}{s
(i.e., clustered} \textcolor{black}{\emph{PA}}\textcolor{black}{s
composed by one or few elementary building-blocks) are gaining more
and more attention thanks to their modularity that makes the whole
antenna system, including the feeding network and the base-band layer,
easier to fabricate and to maintain as well as controllable with a
reduced number of} \textcolor{black}{\emph{TRM}}\textcolor{black}{s
or digital channels by using a single output/input for all the radiating
elements clustered in a tile \cite{Rocca.2015}-\cite{Rapakula.2020}. }

\noindent \textcolor{black}{Nevertheless, two main issues/needs arise
when dealing with clustered} \textcolor{black}{\emph{PA}} \textcolor{black}{structures.
First, the clusters have to be large enough to allow one a non-negligible
cost saving (i.e., reduction of} \textcolor{black}{\emph{TRM}}\textcolor{black}{s
or digital channels), while the whole clustered} \textcolor{black}{\emph{PA}}
\textcolor{black}{must still afford a desired radiation pattern.}
\textcolor{black}{\emph{}}\textcolor{black}{In addition, the tile
shapes must permit the design of irregular clustering configurations,
which are required to avoid periodic quantization errors that generally
imply the presence of high secondary lobes and/or grating lobes in
the radiated pattern \cite{Mailloux.2009}. To properly tackle these
drawbacks, tiling methods using either single \cite{Rocca.2015}-\cite{Dong.2020}
or multiple \cite{Oliveri.2016}-\cite{Rapakula.2020} tile shapes
have been proposed. The} \textcolor{black}{\emph{PA}}\textcolor{black}{s
design has been generally cast as the optimization of user-defined
cost-function terms related to the desired pattern performance indexes
(e.g., the sidelobe level, the beamwidth, and the directivity) or
quantifying the mismatch with a reference mask by also guaranteeing
a full coverage of the whole antenna aperture. For instance, optimization
techniques based on the height-function coding strategy \cite{Anselmi.2017}\cite{Ma.2019}\cite{Rocca.2020.a},
evolutionary algorithms \cite{Anselmi.2017}\cite{Ma.2019}\cite{Rocca.2020.a}\cite{Oliveri.2017},
compressive-sensing \cite{Oliveri.2016}\cite{Bekers.2019}, convex-relaxation
programming \cite{Dong.2020}, random approaches \cite{Rapakula.2020}
as well as exhaustive search \cite{Xiong.2013} have been successfully
applied to tile a finite area (i.e., the array aperture) with pre-selected
tile shapes. However, such design methods have been usually developed
for specific tile geometries or they turn out to be computationally
expensive when the cardinality of the tile alphabet is huge and/or
large apertures are at hand.}

\noindent \textcolor{black}{This paper proposes a novel and alternative
methodology for the design of tiled} \textcolor{black}{\emph{PA}}\textcolor{black}{s
featuring high modularity instead of exploiting multiple different-shape
tiles. It is based on the use and related properties of the self-replicating
tiles, also called} \textcolor{black}{\emph{rep-tiles}}\textcolor{black}{,
to perform the irregular clustering of the array aperture. More specifically,
the aperture is tiled with tiles having different sizes, but equal
shape, all being generated by assembling a finite number of smaller
and congruent copies of a single elementary building-block. Towards
this end, a deterministic optimization strategy is used so that the
arising rep-tile arrangement of the planar} \textcolor{black}{\emph{PA}}
\textcolor{black}{is an optimal trade-off between complexity, costs,
and fitting of user-defined requirements on the radiated power pattern,
while guaranteeing the complete overlay of the array aperture. As
a representative instance, such a synthesis method is applied to tile
rectangular apertures with} \textcolor{black}{\emph{L}}\textcolor{black}{-shaped
tromino tiles.}

\noindent \textcolor{black}{To the best of the authors knowledge,
the main novelties of the proposed technique over the state-of-the-art
array tiling comprise (}\textcolor{black}{\emph{i}}\textcolor{black}{)
the exploitation of the concept of self-replicating tiles for the
design of planar} \textcolor{black}{\emph{PA}}\textcolor{black}{s;
(}\textcolor{black}{\emph{ii}}\textcolor{black}{) the development
of a novel deterministic tiling method to exploit the rep-tiling property
for synthesizing clustered arrays with an optimal trade-off between
architecture complexity and the fulfilment of user-defined radiation
targets; (}\textcolor{black}{\emph{iii}}\textcolor{black}{) the theoretical
analysis of the properties of} \textcolor{black}{\emph{L}}\textcolor{black}{-shape
tromino rep-tiles and their engineering exploitation for the application
of the proposed tiling method.}

\noindent \textcolor{black}{The rest of the paper is organized as
follows. The mathematical formulation of the design of highly-modular
planar} \textcolor{black}{\emph{PA}}\textcolor{black}{s with self-replicating
tiles is reported in Sect. \ref{sec:Mathematical-Formulation}, while
the proposed deterministic synthesis method is described in Sect.
\ref{sec:RTA-Design-Method} and customized to} \textcolor{black}{\emph{L}}\textcolor{black}{-shape
tromino rep-tiles, as well. Section 4 is devoted to the numerical
validation against both ideal and real antenna models including mutual-coupling
effects, as well. Eventually, some conclusions and final remarks are
drawn (Sect. 5).}

\section{\noindent \textcolor{black}{Mathematical Formulation\label{sec:Mathematical-Formulation}}}

\noindent \textcolor{black}{Let us consider a planar} \textcolor{black}{\emph{PA}}
\textcolor{black}{covering a rectangular aperture $A$ and composed
by $M\times N$ radiating elements arranged on a regular lattice with
inter-element spacing $d_{x}$ and $d_{y}$ along the $x-$ and $y-$axes,
respectively (Fig. 1). The power pattern radiated in far field by
such a} \textcolor{black}{\emph{PA}} \textcolor{black}{is given by
$\mathcal{P}\left(u,v\right)=\left|\mathbf{e}\left(u,v\right)\times\mathcal{A}\left(u,v\right)\right|^{2}$
where $\mathbf{e}\left(u,v\right)$ is the embedded/active element
pattern \cite{Mailloux.2005}, which is assumed, without loss of generality,
but only for the sake of notation simplicity, identical for all antennas,
while $\mathcal{A}\left(u,v\right)$ is the array factor that depends
of the array architecture at hand as detailed in the following, $u$
($u\triangleq\sin\theta\cos\phi$) and $v$ ($v\triangleq\sin\theta\sin\phi$)
being the direction cosines ($0$ {[}deg{]} $\le$ $\theta$ $\le$
$90$ {[}deg{]}, $0$ {[}deg{]} $\le$ $\phi$ $\le$ $360$ {[}deg{]}).}

\noindent \textcolor{black}{The surface $A$ is partitioned into $Q$
($Q<M\times N$) tiles, $\bm{\varsigma}=\left\{ \varsigma_{q};\, q=1,...,Q\right\} $,
each belonging to the} \textcolor{black}{\emph{rep-tile}} \textcolor{black}{\cite{Golomb.1964}\cite{Gardner.2001}
alphabet $\bm{\Sigma}$ (i.e., $\varsigma_{q}\in\bm{\Sigma}$). More
specifically, the alphabet $\bm{\Sigma}$ is composed by $B\times R$
elements (i.e., $\bm{\Sigma}=\left\{ \sigma^{\left(r,\, b\right)};\, r=1,...,R;\, b=1,...,B\right\} $),
$B$ and $R$ being the number of admissible rep-tile orientations/flips
and the maximum rep-tile order/level, respectively.}

\noindent \textcolor{black}{Let the smallest building-block of the
rep-tile alphabet $\bm{\Sigma}$, namely the generating rep-tile of
order $r=1$ with $b$-th ($1\le b\le B$) orientation, $\left.\sigma^{\left(r,\, b\right)}\right\rfloor _{r=1}$
, be a cluster of $\left.I^{\left(r\right)}\right\rfloor _{r=1}$
radiating elements. The rep-tile element of $\bm{\Sigma}$ of order
$\left(r+1\right)$-th ($r=1,...,R-1$) and orientation $b$-th ($1\le b\le B$)
is then yielded by the union of $S$ tiles of the previous $r$-th
($r=1,...,R-1$) order\begin{equation}
\sigma^{\left(r+1,\, b\right)}=\bigcup_{s=1}^{S}\sigma^{\left(r,\, b_{s}\right)}\label{eq:_rep-tile.generation}\end{equation}
where $S$ is a function of the rep-tile geometry at hand {[}e.g.,}
\textcolor{black}{\emph{L}}\textcolor{black}{-shape tromino $\to$
$S=4$ - Fig. 2(}\textcolor{black}{\emph{b}}\textcolor{black}{){]}
\cite{Gardner.2001}, $b_{s}$ ($1\le b_{s}\le B$) is the orientation
of the $s$-th ($s=1,...,S$) generating tile of $r$-th order ($r=1,...,R-1$).
Therefore, a larger tile is yielded by assembling a finite number
of $S$ smaller and congruent copies with its same shape, but rotated
in one of $B$ possible positions \cite{Robinson.1999} {[}Fig. 2(}\textcolor{black}{\emph{b}}\textcolor{black}{){]}.}

\noindent \textcolor{black}{Moreover, the number of array elements
belonging to a rep-tile of $r$-th ($r=2,...,R$) order is given by}

\noindent \textcolor{black}{\begin{equation}
I^{\left(r\right)}=S^{r-1}\times I^{\left(1\right)},\label{eq:_elements.number}\end{equation}
thus the aperture $A$ of the array turns out to be tiled by multiple
tiles with different sizes, but all having the same (even if scaled)
shape {[}e.g.,} \textcolor{black}{\emph{L}}\textcolor{black}{-shape
- Fig. 2(}\textcolor{black}{\emph{a}}\textcolor{black}{){]}.}

\noindent \textcolor{black}{The array elements grouped into the $q$-th
($q=1,...,Q$) tile, $\varsigma_{q}$, share a unique input/output
port, which is connected to the feeding network through a common}
\textcolor{black}{\emph{TRM}} \textcolor{black}{and/or a digital channel
that provides a controllable complex weight $w_{q}$ (Fig. 1) of amplitude
$\alpha_{q}$ and phase $\beta_{q}$ (i.e., $w_{q}=\alpha_{q}e^{j\beta_{q}}$).
Accordingly, the antenna array factor is given by\begin{equation}
\mathcal{A}\left(u,v\right)=\sum_{q=1}^{Q}w_{q}\mathcal{S}_{q}^{\left(r,\, b\right)}\left(u,v\right)e^{jk\left(x_{q}u+y_{q}v\right)}\label{eq:_array.factor}\end{equation}
where $\left(x_{q},\, y_{q}\right)$ is the center of $\varsigma_{q}$
within the aperture $A$ and $k$ is the wavenumber ($k\triangleq\frac{2\pi}{\lambda}$),
$\lambda$ being the wavelength at the working frequency $f$. Moreover,
$\mathcal{S}_{q}^{\left(r,\, b\right)}$ is the space factor of the
$q$-th ($q=1,...,Q$) tile whose expression is\begin{equation}
\mathcal{S}_{q}^{\left(r,\, b\right)}\left(u,v\right)=\sum_{i=1}^{I_{q}^{\left(r\right)}}e^{jk\left(x_{i}^{\left(r,\, b\right)}u+y_{i}^{\left(r,\, b\right)}v\right)}\label{eq:_subarray.space.factor}\end{equation}
where $\left(x_{i}^{\left(r,\, b\right)},\, y_{i}^{\left(r,\, b\right)}\right)$
is the center of the $i$-th ($i=1,...,I_{q}^{\left(r\right)}$) radiating
element of the alphabet entry, namely $\sigma_{q}^{\left(r,\, b\right)}=\Phi\left\{ \varsigma_{q}\right\} $,}%
\footnote{\noindent \textcolor{black}{The operator $\Phi$ is called {}``associative
operator'' since it defines the correspondence between the $q$-th
tile and the corresponding letter $\sigma^{\left(r,\, b\right)}$
of the alphabet $\bm{\Sigma}$ (i.e., $\varsigma_{q}\in\bm{\Sigma}$).}%
} \textcolor{black}{associated to $\varsigma_{q}$ ($\varsigma_{q}\in\bm{\Sigma}$).}

\noindent \textcolor{black}{In order to define the array synthesis
problem at hand, let us code the tiling configuration of the $M\times N$
elements into $Q$ rep-tiles, $\bm{\varsigma}=\left\{ \varsigma_{q};\, q=1,...,Q\right\} $,
with the clustering vector $\mathbf{c}$ of dimension $M\times N$
whose ($m$, $n$)-th ($m=1,...,M$; $n=1,...,N$) entry, $c_{mn}$,
is given by $c_{mn}=\delta_{c_{mn}q}q$ where $\delta_{c_{mn}q}=1$($0$)
if the ($m$, $n$)-th radiating element belongs(does not belong)
to the $q$-th rep-tile, $\varsigma_{q}$ (Fig. 3). Moreover, the
amplitude/phase coefficients of the $Q$-sized set of rep-tiles, $\bm{\varsigma}=\left\{ \varsigma_{q};\, q=1,...,Q\right\} $,
coded into the tiling vector $\mathbf{c}$, are chosen according to
the excitation matching strategy \cite{Manica.2008}\cite{Rocca.2019}
as follows}

\noindent \textcolor{black}{\begin{equation}
\alpha_{q}=\frac{1}{I_{q}^{\left(r\right)}}\sum_{m=1}^{M}\sum_{n=1}^{N}\alpha_{mn}^{ref}\delta_{c_{mn}q}\label{eq:_subarray.amplitude}\end{equation}
and\begin{equation}
\beta_{q}=\frac{1}{I_{q}^{\left(r\right)}}\sum_{m=1}^{M}\sum_{n=1}^{N}\beta_{mn}^{ref}\delta_{c_{mn}q}.\label{eq:_subarray.phase}\end{equation}
Such a choice allows one to yield the optimal setup of the weights
of the tiled array, $\mathbf{w}$ ($\mathbf{w}$ $=$ \{$w_{q}$;
$q=1,...,Q$\}), in the least-square sense since they have the minimum
distance with respect to the {}``}\textcolor{black}{\emph{reference}}\textcolor{black}{''
excitations, $\mathbf{w}^{ref}$ ($\mathbf{w}^{ref}$ $=$ \{$w_{mn}^{ref}$;
$m=1,...,M$, $n=1,...,N$\}), of a fully-populated array of size
$M\times N$ with a dedicated} \textcolor{black}{\emph{TRM}} \textcolor{black}{for
each ($m$, $n$)-th array element ($w_{mn}^{ref}=\alpha_{mn}^{ref}e^{j\beta_{mn}^{ref}}$)
to afford an optimal power pattern, $\mathcal{P}^{ref}\left(u,v\right)$,
or to optimally approximate a user-defined power pattern mask, $\Psi\left(u,v\right)$
in the visible range $\Omega=\left\{ \left(u,v\right):\, u^{2}+v^{2}\leq1\right\} $.}

\noindent \textcolor{black}{Accordingly, the} \textcolor{black}{\emph{Rep-Tile
Array}} \textcolor{black}{(}\textcolor{black}{\emph{RTA}}\textcolor{black}{)}
\textcolor{black}{\emph{Synthesis Problem}} \textcolor{black}{can
be formulated as follows}

\begin{quotation}
\noindent \textcolor{black}{Given a planar} \textcolor{black}{\emph{PA}}
\textcolor{black}{of $M\times N$ radiating elements arranged on a
rectangular lattice, determine the optimal clustering of the array
elements, $\mathbf{c}_{opt}$, corresponding to the set of rep-tiles,
$\bm{\varsigma}_{opt}=\left\{ \varsigma_{q}^{opt};\, q=1,...,Q^{opt}\right\} $,
belonging to the alphabet $\bm{\Sigma}$ (i.e., $\varsigma_{q}^{opt}\in\bm{\Sigma}$),
and the sub-array weights, $\mathbf{w}_{opt}$ ($\mathbf{w}_{opt}$$=$
\{$w_{q}^{opt}=\alpha_{q}^{opt}e^{j\beta_{q}^{opt}}$; $q=1,...,Q^{opt}$\}),
according to (\ref{eq:_subarray.amplitude})(\ref{eq:_subarray.phase})
so that $Q^{opt}$ is the minimum number of rep-tiles covering the
aperture $A$ and the user-defined power upper mask $\Psi\left(u,v\right)$
is matched by minimizing the mask matching index\begin{equation}
\Gamma\left(\mathbf{c}\right)=\frac{\int_{\Omega}\left[P\left(u,v;\,\mathbf{c}\right)-\Psi\left(u,v\right)\right]\times\mathcal{H}\left\{ P\left(u,v;\,\mathbf{c}\right)-\Psi\left(u,v\right)\right\} dudv}{\int_{\Omega}\Psi\left(u,v\right)dudv}\label{eq:_mask.matching}\end{equation}
(i.e., $\mathbf{c}_{opt}=\arg\left\{ \min_{\mathbf{c}}\left[\Gamma\left(\mathbf{c}\right)\right]\right\} $),
$\mathcal{H}\left\{ \cdot\right\} $ being the Heaviside function.}
\end{quotation}
\textcolor{black}{Such a design problem is addressed with an innovative
deterministic strategy detailed in Sect. \ref{sec:RTA-Design-Method}.}

\section{\noindent \textcolor{black}{\emph{RTA}} \textcolor{black}{Design
Method (}\textcolor{black}{\emph{RTAM}}\textcolor{black}{)\label{sec:RTA-Design-Method}}}

\textcolor{black}{Under the hypothesis that the array aperture $A$
is fully tileable with $R$-th level rep-tiles, the} \textcolor{black}{\emph{RTAM}}
\textcolor{black}{is implemented by means of the following procedural
steps}\textcolor{black}{\emph{:}}

\begin{itemize}
\item \textbf{\textcolor{black}{Step 1.}} \textcolor{black}{\emph{Initial}}
\textcolor{black}{$R$}\textcolor{black}{\emph{-th Level Tiling}}
\textcolor{black}{- Depending on the cardinality $T^{\left(R\right)}$
of the problem at hand, either apply the Algorithm-X \cite{Knuth.2000}\cite{Anselmi.2022}
to generate the whole set of $T^{\left(R\right)}$ complete clusterings
of $A$ with $R$}\textcolor{black}{\emph{-th}} \textcolor{black}{order
rep-tiles only, \{$\hat{\mathbf{c}}_{t}$; $t=1,...,T^{\left(R\right)}$\}
and select the optimal configuration $\hat{\mathbf{c}}_{opt}$ of
$\hat{Q}$ tiles ($\hat{Q}\triangleq\frac{M\times N}{I^{\left(R\right)}}$)
as\begin{equation}
\hat{\mathbf{c}}_{opt}=\arg\left\{ \min_{t=1,...,T^{\left(R\right)}}\left[\Gamma\left(\hat{\mathbf{c}}_{t}\right)\right]\right\} ,\label{eq:_Initial.tiling}\end{equation}
or increase the tile order (i.e., $R\leftarrow R+1$) to reduce the
cardinality of the solution space ($T^{\left(R+1\right)}<$$T^{\left(R\right)}$)
if the aperture can be still fully partitioned, or exploit a global
optimization method \cite{Rocca.2015} to sample the $T^{\left(R\right)}$-size
solution space for retrieving the global optimum of the cost function
$\Gamma\left(\hat{\mathbf{c}}_{t}\right)$, $\hat{\mathbf{c}}_{best}$,
and set $\hat{\mathbf{c}}_{opt}=\hat{\mathbf{c}}_{best}$. Set the
initial ($h=0$, $h$ being the iteration index) tiling configuration
to that yielded in the} \textcolor{black}{\emph{Step 1}} \textcolor{black}{($\left.\mathbf{c}^{\left(h\right)}\right\rfloor _{h=0}\leftarrow\hat{\mathbf{c}}_{opt}$)
and composed by $R$}\textcolor{black}{\emph{-th}} \textcolor{black}{order
rep-tiles only, $\left.\bm{\varsigma}^{\left(h\right)}\right\rfloor _{h=0}=\left\{ \left.\varsigma_{q}^{\left(h\right)}\right\rfloor _{h=0}=\sigma_{q}^{\left(R,\, b\right)};\, q=1,...,\left.Q^{\left(h\right)}\right\rfloor _{h=0}\right\} $,
being $\left.Q^{\left(h\right)}\right\rfloor _{h=0}=\hat{Q}$;}
\item \textbf{\textcolor{black}{Step 2.}} \textcolor{black}{\emph{Rep-Tiling
Optimization}} \textcolor{black}{- Increase the iteration index ($h\leftarrow h+1$),
set the rep-tile level to $r=R$, and apply the following loop}

\begin{itemize}
\item \textbf{\textcolor{black}{Step 2.1.}} \textcolor{black}{\emph{Rep-Tile
Selection}} \textcolor{black}{- For each $q$-th ($q=1,...,Q^{\left(h\right)}$)
tile of the $h$-th clustering, $\mathbf{c}^{\left(h\right)}$, compute
the} \textcolor{black}{\emph{substitution tiling}} \textcolor{black}{metric
(}\textcolor{black}{\emph{STM}}\textcolor{black}{) \begin{equation}
\xi_{q}^{\left(h\right)}=\sum_{m=1}^{M}\sum_{n=1}^{N}\left|\mathcal{R}\left\{ w_{mn}^{ref}-w_{q}^{\left(h\right)}\right\} +j\mathcal{I}\left\{ w_{mn}^{ref}-w_{q}^{\left(h\right)}\right\} \right|\delta_{c_{mn}q}\label{eq:_reclustering.priority}\end{equation}
to quantify the mismatch between the reference, $\mathbf{w}^{ref}$,
and the current, $\mathbf{w}^{\left(h\right)}$ ($\mathbf{w}^{\left(h\right)}$$=$
\{$w_{q}^{\left(h\right)}=\alpha_{q}^{\left(h\right)}e^{j\beta_{q}^{\left(h\right)}}$;
$q=1,...,Q^{\left(h\right)}$\}) weight sets, $\mathcal{R}\left\{ \cdot\right\} $
and $\mathcal{I}\left\{ \cdot\right\} $ being the real part and the
imaginary one, respectively. Among the $Q^{\left(h\right)}$ reptiles
of $\bm{\varsigma}^{\left(h\right)}$ select the $\widehat{q}$-th
one of $r$-th ($2\le r\le R$) order with maximum} \textcolor{black}{\emph{STM}}
\textcolor{black}{value, $\xi_{q}^{\left(h\right)}$,\begin{equation}
\widehat{q}=\arg\max_{q=1,...,Q^{\left(h\right)}}\left\{ \left.\xi_{q}^{\left(h\right)}\right|\,\mathbb{R}\left\{ \sigma_{q}^{\left(r,\, b\right)}\right\} \ge2\right\} ,\label{eq:_tiles_selection}\end{equation}
$\mathbb{R}\left\{ .\right\} $ being the operator that returns the
rep-tile order of its argument;}
\item \textbf{\textcolor{black}{Step 2.2.}} \textcolor{black}{\emph{Rep-Tile
Split}} \textcolor{black}{- Subdivide the $\widehat{q}$-th tile of
the current array clustering, $\mathbf{c}^{\left(h\right)}$, into
$S$ smaller tiles of $\left(r-1\right)$-th order ($\sigma_{\widehat{q}}^{\left(r,\, b\right)}=\bigcup_{s=1}^{S}\sigma_{s}^{\left(r-1,\, b_{s}\right)}$)
to generate a new tiling, $\mathbf{c}^{\left(h+1\right)}$, of the
aperture $A$ into $Q^{\left(h+1\right)}=Q^{\left(h\right)}+\left(S-1\right)$
tiles;}
\item \textbf{\textcolor{black}{Step 2.4.}} \textcolor{black}{\emph{Excitations
Update}} \textcolor{black}{- Compute the amplitude, $\bm{\alpha}^{\left(h+1\right)}$
$=$ \{$\alpha_{q}^{\left(h+1\right)}$; $q=1,...,Q^{\left(h+1\right)}$\},
and the phase, $\bm{\beta}^{\left(h+1\right)}$ $=$ \{$\beta_{q}^{\left(h+1\right)}$;
$q=1,...,Q^{\left(h+1\right)}$\}, coefficients of the $Q^{\left(h+1\right)}$
tiles of $\mathbf{c}^{\left(h+1\right)}$ according to (\ref{eq:_subarray.amplitude})
and (\ref{eq:_subarray.phase}), respectively;}
\item \textbf{\textcolor{black}{Step 2.5.}} \textcolor{black}{\emph{Convergence
Check}} \textcolor{black}{- If the power pattern radiated by the tiled
array, $\bm{\varsigma}^{\left(h+1\right)}$, perfectly fulfils the
radiation mask {[}i.e., $\Gamma\left(\mathbf{c}^{\left(h+1\right)}\right)=0${]}
or the current number of clusters, $Q^{\left(h+1\right)}$, exceeds
the user-definer maximum, $Q_{max}$ (i.e., $Q^{\left(h+1\right)}\geq Q_{max}$
subject to $Q_{max}\le\frac{M\times N}{I^{\left(1\right)}}$), stop
the procedure and set the convergence iteration $H$ to the current
one (i.e., $H=h+1$) as well as the optimal tiling to the current
rep-tiles arrangement, $\mathbf{c}_{opt}=\mathbf{c}^{\left(h+1\right)}$.
Otherwise, update the iteration index ($h\leftarrow h+1$) and go
to Step 2.1.}
\end{itemize}
\end{itemize}
It is worth noticing that, besides the convergence solution $\mathbf{c}_{opt}$,
the iterative loop of \emph{Step 2} gives at each $h$-th ($h=1,...,H$)
iteration a compromise solution between feeding network complexity
(i.e., $Q$ $\to$ number of \emph{TRM}s) and fulfilment of the pattern-mask
constraint {[}i.e., \textcolor{black}{$\Gamma\left(\mathbf{c}^{\left(h+1\right)}\right)${]}
so that finally a {}``Pareto'' front of $H$ layouts can be defined
and profitably exploited by the array designer.}

\subsection{\noindent \textcolor{black}{The Case of} \textcolor{black}{\emph{L}}\textcolor{black}{-Shaped
Rep-Tiles\label{sub:The-Case-of}}}

\noindent \textcolor{black}{The} \textcolor{black}{\emph{RTAM}} \textcolor{black}{is
customized hereinafter to} \textcolor{black}{\emph{L}}\textcolor{black}{-tromino
tiles, which are rep-tiles \cite{Golomb.1964}\cite{Gardner.2001}
with $\left.I^{\left(r\right)}\right\rfloor _{r=1}$ $=$ $3$ and
$S=4$ {[}Fig. 2(}\textcolor{black}{\emph{a}}\textcolor{black}{){]},
as a representative example of self-replicating tile geometries {[}Fig.
2(}\textcolor{black}{\emph{b}}\textcolor{black}{){]}.}

\subsubsection*{\noindent \textcolor{black}{Tileability Theorem}}

\noindent \textcolor{black}{The} \textcolor{black}{\emph{RTAM}} \textcolor{black}{requires
the fulfilment of the complete tileability of the array aperture $A$
with the largest (i.e., highest order $R$) tiles of the alphabet
$\bm{\Sigma}$ (i.e., \{$\sigma^{\left(R,\, b\right)}$; $b=1,...,B$\})
since the highest order tiles can be certainly subdivided into smaller
ones thanks to the self-replicating property of the rep-tiles. The
covering theorem for an exact (i.e., complete) rectangular tessellation
with $R$-th order} \textcolor{black}{\emph{L}}\textcolor{black}{-tromino
tiles \cite{Chu.1958} reads as follows}

\begin{quotation}
\noindent \textcolor{black}{\emph{L-tromino Covering Theorem}} \textcolor{black}{-
Let $A$ a rectangular aperture of $M\times N$ elements/pixels, $M$
and $N$ being integer numbers such that $2\times l^{\left(R\right)}\leq M\leq N$,
where $l^{\left(R\right)}=\sqrt{\frac{I^{\left(R\right)}}{3}}$ is
the side length of the $R$-th order} \textcolor{black}{\emph{L}}\textcolor{black}{-tromino
tile. Then, $A$ can be fully covered with} \textcolor{black}{\emph{L}}\textcolor{black}{-tromino
tiles of order $R$ if and only if one of the following conditions
holds true}
\begin{itemize}
\item \noindent \textcolor{black}{$\widehat{M}=3$ and $\widehat{N}$ is
even {[}e.g., Fig. 4(}\textcolor{black}{\emph{a}}\textcolor{black}{){]};}
\item \noindent \textcolor{black}{$\widehat{M}\neq3$ and $\left\lfloor \frac{M\times N}{l^{\left(R\right)}}\right\rfloor =0$,
$\left\lfloor \,.\,\right\rfloor $ being the modulo operation {[}e.g.,
Fig. 4(}\textcolor{black}{\emph{c}}\textcolor{black}{) and Fig. 4(}\textcolor{black}{\emph{e}}\textcolor{black}{){]}}
\end{itemize}
\noindent \textcolor{black}{where $\widehat{M}\triangleq\frac{M}{l^{\left(R\right)}}$
and $\widehat{N}=\frac{N}{l^{\left(R\right)}}$. Otherwise, $A$ is
not fully tileable with $R$-th order} \textcolor{black}{\emph{L}}\textcolor{black}{-tromino
tiles {[}e.g., Fig. 4(}\textcolor{black}{\emph{b}}\textcolor{black}{),
Fig. 4(}\textcolor{black}{\emph{d}}\textcolor{black}{), and Fig. 4(}\textcolor{black}{\emph{f}}\textcolor{black}{){]}.}

\end{quotation}

\subsubsection*{\noindent \textcolor{black}{Problem Cardinality}}

\noindent \textcolor{black}{According to \cite{Ueno.2008}, the number
of tilings, $T^{\left(R\right)}$, that exactly/fully cover an aperture
$A$ of $M\times N$ pixels/elements with $R$-th order} \textcolor{black}{\emph{L}}\textcolor{black}{-trominoes
is equal to\begin{equation}
T^{\left(R\right)}=\left.\widehat{g}_{po}\right]_{p=1}^{o=2^{\widehat{M}}}\label{eq:_cardinality}\end{equation}
where $\widehat{g}_{po}$ is the ($p$, $o$)-th ($p$, $o$ $=$
$1,...,2^{\widehat{M}}$) entry of the matrix $\left(\mathbf{G}_{\widehat{M}}\right)^{\widehat{N}-1}$,
which is the square and symmetric matrix of size $2^{\widehat{M}}\times2^{\widehat{M}}$
given in the} \textcolor{black}{\emph{Appendix}} \textcolor{black}{(\ref{sec:Appendix-A}).
In order to give an idea of the cardinality of the solution space
of the tiling problem at hand, Table I gives the $T$ values for some
rectangular apertures with sizes \{$\widehat{M}\times\widehat{N}$
; $2\le\widehat{M}\le9$, $2\le\widehat{N}\le9$\}.}

\noindent \textcolor{black}{It is worth pointing out that} \textcolor{black}{\emph{a-priori}}
\textcolor{black}{knowing the value of $T^{\left(R\right)}$ (\ref{eq:_cardinality})
or, at least, its order of magnitude, is a very important information
for the array designer since it enables the choice of an exhaustive
enumerative strategy, which guarantees the retrieval of the optimal
solution in the} \textcolor{black}{\emph{Step 1}} \textcolor{black}{of
the} \textcolor{black}{\emph{RTAM}} \textcolor{black}{in Sect. \ref{sec:RTA-Design-Method}
since evaluating all $T^{\left(R\right)}$ exact tiling configurations,
or it forces to reformulate the problem as an optimization one when
the cardinality of the solution space turns out prohibitive for an
exhaustive search.}

\subsubsection*{\noindent \textcolor{black}{\emph{RTAM}} \textcolor{black}{Customization}}

\noindent \textcolor{black}{For illustrative purposes, let us consider
the application of the} \textcolor{black}{\emph{RTAM}} \textcolor{black}{($H=3$
iterations) to a planar array of $M\times N=8\times16$ elements (Fig.
5) by using the alphabet $\bm{\Sigma}$ of} \textcolor{black}{\emph{L}}\textcolor{black}{-tromino
rep-tiles in Fig. 2(}\textcolor{black}{\emph{a}}\textcolor{black}{)
being $R=3$ and $B=4$.}

\noindent \textcolor{black}{Starting ($h=0$) from the initial clustering
of $R$-th order} \textcolor{black}{\emph{L}}\textcolor{black}{-tromino
tiles found at the} \textcolor{black}{\emph{Step 1}} \textcolor{black}{(Sect.
\ref{sec:RTA-Design-Method}), $\hat{\mathbf{c}}_{opt}$, in Fig.
5(}\textcolor{black}{\emph{a}}\textcolor{black}{) where the normalized
values of the} \textcolor{black}{\emph{STM}} \textcolor{black}{of
the $\left.Q^{\left(h\right)}\right\rfloor _{h=0}=4$ tiles (i.e.,
\{$\xi_{q}^{\left(h\right)}$; $q=1,...,Q^{\left(h\right)}$\}) are
indicated, as well, the $I^{\left(3\right)}$-size ($\left.I^{\left(R\right)}\right\rfloor _{R=3}=48$)
tile with maximum} \textcolor{black}{\emph{STM}} \textcolor{black}{(i.e.,
$\xi_{2}^{\left(1\right)}=1.0$) {[}Fig. 5(}\textcolor{black}{\emph{a}}\textcolor{black}{){]}
is chosen for division according to (\ref{eq:_tiles_selection}) of}
\textcolor{black}{\emph{Step 2.1}} \textcolor{black}{(i.e., $\widehat{q}=2$)
to yield the new tiled arrangement $\mathbf{c}^{\left(1\right)}$
shown in Fig. 5(}\textcolor{black}{\emph{b}}\textcolor{black}{) of
$Q^{\left(1\right)}=7$ rep-tiles, namely $Q_{R}^{\left(1\right)}=3$
rep-tiles of order $R$, while $S=4$ rep-tiles of order $R-1$ ($Q_{R-1}^{\left(1\right)}=S$),
being $Q^{\left(h\right)}=\sum_{r=1}^{R}Q_{r}^{\left(h\right)}$.
Afterwards, other two tiles including $I^{\left(2\right)}$ ($I^{\left(2\right)}=12$)
{[}Fig. 5(}\textcolor{black}{\emph{c}}\textcolor{black}{){]} and $I^{\left(3\right)}$
{[}Fig. 5(}\textcolor{black}{\emph{e}}\textcolor{black}{){]} array
elements are the candidates for division (\ref{eq:_tiles_selection})
at the iteration $h=2$ and $h=H$, respectively, $\mathbf{c}^{\left(2\right)}$
of $Q^{\left(2\right)}=10$ ($Q_{R}^{\left(2\right)}=Q_{R-1}^{\left(2\right)}=3$
and $Q_{R-2}^{\left(2\right)}=S$) rep-tiles {[}Fig. 5(}\textcolor{black}{\emph{d}}\textcolor{black}{){]}
and $\mathbf{c}^{\left(3\right)}$ of $Q^{\left(3\right)}=13$ ($Q_{R}^{\left(3\right)}=2$,
$Q_{R-1}^{\left(3\right)}=7$, and $Q_{R-2}^{\left(3\right)}=S$)
rep-tiles {[}Fig. 5(}\textcolor{black}{\emph{f}}\textcolor{black}{){]}
being the generated array clusterings.}

\section{\noindent \textcolor{black}{Numerical Results}}

\noindent \textcolor{black}{This section is devoted to the analysis
of the behavior and the assessment of the performance of the proposed}
\textcolor{black}{\emph{RTAM}}\textcolor{black}{.}

\noindent \textcolor{black}{The first example deals with an array
comprising $M\times N=8\times12$ isotropic {[}i.e., $\mathbf{e}\left(u,v\right)=1${]}
radiating elements with uniform half-wavelength ($d_{x}=d_{y}=\frac{\lambda}{2}$)
inter-element spacing. The reference excitations have been computed
with a Convex Programming (}\textcolor{black}{\emph{CP}}\textcolor{black}{)
based approach \cite{Isernia.2004} to fit the upper-bound power mask
$\Psi\left(u,v\right)$ in Fig. 6(}\textcolor{black}{\emph{a}}\textcolor{black}{),
which is centered in $\left(u_{0},v_{0}\right)=\left(0,0\right)$
and it is characterized by a rectangular mainlobe region of dimensions
$BW_{u}=0.50$ and $BW_{v}=0.76$ along the $v=0$ and $u=0$ principal
planes, respectively. Figure 6(}\textcolor{black}{\emph{b}}\textcolor{black}{)
shows the real distribution of the reference excitations, \{$w_{mn}^{ref}=\alpha_{mn}^{ref}$;
$m=1,...,M$, $n=1,...,N$\}, while the corresponding power pattern,
affording a sidelobe level of $SLL=-25$ {[}dB{]} and matching the
mask $\Psi\left(u,v\right)$, is plotted along the principal planes
{[}i.e., $u=0$ - Fig. 7(}\textcolor{black}{\emph{b}}\textcolor{black}{);
$v=0$ - Fig. 7(}\textcolor{black}{\emph{c}}\textcolor{black}{){]}.}

\noindent \textcolor{black}{By considering the $R=2$-th level alphabet
of} \textcolor{black}{\emph{L}}\textcolor{black}{-tromino rep-tiles
with $\left.I^{\left(R\right)}\right\rfloor _{R=2}=12$ elements,
the $R$-order tileability of the aperture $A$ has been successfully
checked against the theorem in Sect. \ref{sub:The-Case-of}. Indeed,
the second condition of the} \textcolor{black}{\emph{L}}\textcolor{black}{-tromino
covering theorem holds true since $l^{\left(R\right)}=2$. The} \textcolor{black}{\emph{RTAM}}
\textcolor{black}{has been then executed by setting the maximum partitioning
of the array to $Q_{max}=14$ tiles. At the} \textcolor{black}{\emph{Step
1}} \textcolor{black}{($h=0$), $T^{\left(R\right)}=18$ (Tab. I -
$\widehat{M}=4$, $\widehat{N}=6$) different tilings of the aperture
with $I^{\left(R\right)}$-sized} \textcolor{black}{\emph{L}}\textcolor{black}{-trominoes
have been generated with the Algorithm-X \cite{Knuth.2000}\cite{Anselmi.2022}
and the optimal $R$-th order tiling, $\hat{\mathbf{c}}_{opt}$, which
minimizes the mask matching (\ref{eq:_Initial.tiling}), is shown
in Fig. 8(}\textcolor{black}{\emph{a}}\textcolor{black}{) where the
color level representation of the sub-array excitations, $\widehat{\mathbf{w}}_{opt}$
(i.e., $\widehat{\mathbf{w}}_{opt}$ $=$ \{$\widehat{w}_{q}^{opt}=\widehat{\alpha}_{q}^{opt}e^{j\widehat{\beta}_{q}^{opt}}$;
$q=1,...,\hat{Q}$\}), is given, as well. Afterwards ($h=1)$, the}
\textcolor{black}{\emph{STM}} \textcolor{black}{is computed (\ref{eq:_reclustering.priority})
for each $q$-th ($q=1,...,\hat{Q}$; $\hat{Q}=8$) tile of $\hat{\mathbf{c}}_{opt}$,
$\left.\xi_{q}^{\left(1\right)}\right\rfloor _{h=1}$ {[}Fig. 8(}\textcolor{black}{\emph{b}}\textcolor{black}{){]},
and the tile with the maximum value (i.e., $\widehat{q}=7$) has been
selected (\ref{eq:_tiles_selection}) and divided into $S=4$ rep-tiles
of level $\left(R-1\right)$ with $I^{\left(R-1\right)}=3$ elements
to yield the new array tiling $\left.\mathbf{c}^{\left(h\right)}\right\rfloor _{h=1}$
in Fig. 8(}\textcolor{black}{\emph{c}}\textcolor{black}{) ($\left.Q^{\left(h\right)}\right\rfloor _{h=1}=11$).
Once the $Q^{\left(1\right)}$ tiles of $\mathbf{c}^{\left(1\right)}$
{[}Fig. 8(}\textcolor{black}{\emph{d}}\textcolor{black}{){]} have
been ranked (\ref{eq:_tiles_selection}) according to their} \textcolor{black}{\emph{STM}}
\textcolor{black}{values ($\widehat{q}=2$) {[}Fig. 8(}\textcolor{black}{\emph{d}}\textcolor{black}{){]},
a further iteration ($h=2$) of the} \textcolor{black}{\emph{RTAM}}
\textcolor{black}{has been completed by deriving the tiled arrangement,
$\left.\mathbf{c}^{\left(h\right)}\right\rfloor _{h=2}$, in Fig.
8(}\textcolor{black}{\emph{e}}\textcolor{black}{) ($\left.Q^{\left(h\right)}\right\rfloor _{h=2}=14$
being $\left.Q_{r}^{\left(h\right)}\right]_{r=R}^{h=2}=6$ and $\left.Q_{r}^{\left(h\right)}\right]_{r=R-1}^{h=2}=8$).
The iterative procedure has been stopped (}\textcolor{black}{\emph{Step
2.5}}\textcolor{black}{) after $H=2$ iterations since the convergence
condition on the maximum number of sub-arrays has been reached (i.e.,
$Q^{\left(h\right)}=Q_{max}$) and the final clustered architecture
has been set to the current one ($\mathbf{c}_{opt}=\mathbf{c}^{\left(2\right)}$).
This latter affords a power pattern {[}Figs. 7(}\textcolor{black}{\emph{b}}\textcolor{black}{)-7(}\textcolor{black}{\emph{c}}\textcolor{black}{){]}
with a mask matching index equal to $\Gamma_{opt}=7.31\times10^{-4}$
{[}$\Gamma_{opt}\triangleq\Gamma\left(\mathbf{c}_{opt}\right)${]}.}

\noindent \textcolor{black}{To assess the effectiveness and the reliability
of the} \textcolor{black}{\emph{RTAM}} \textcolor{black}{to find the
optimal tiling of the aperture $A$, that is (in other words), to
prove that the solution $\mathbf{c}_{opt}$ is the global optimum
subject to the constraint $Q\le Q_{max}$, the Algorithm-X has been
used \cite{Knuth.2000}\cite{Anselmi.2022} in an exhaustive way to
generate all clustered layouts with $I^{\left(R-1\right)}$- and $I^{\left(R\right)}$-sized
($R=2$)} \textcolor{black}{\emph{L}}\textcolor{black}{-tromino tiles,
each having $B=4$ possible rotations. Once the whole set of $T\simeq5.92\times10^{7}$
tilings has been generated, the mask matching index (\ref{eq:_mask.matching})
of the $T_{Q_{max}}=6248$ partitions with $Q\leq Q_{max}$ tiles
(i.e., $Q_{R}=6$ and $Q_{R-1}=8$ being $Q=\sum_{r=1}^{R}Q_{r}$)
has been computed to determine the optimal solution of the problem
at hand, $\widetilde{\mathbf{c}}_{opt}$. The mask matching value
of the $T_{Q_{max}}$ solutions, sorted from the worst to the best,
are reported in Fig. 7(}\textcolor{black}{\emph{a}}\textcolor{black}{)
and compared to that from the} \textcolor{black}{\emph{RTAM}} \textcolor{black}{at
the convergence, $\Gamma_{opt}$. As it can be also inferred by the
plots of the power patterns in Figs. 7(}\textcolor{black}{\emph{b}}\textcolor{black}{)-7(}\textcolor{black}{\emph{c}}\textcolor{black}{),
the optimal solution found by the Algorithm-X exhaustive search coincides
with the} \textcolor{black}{\emph{RTAM}} \textcolor{black}{one (i.e.,
$\mathbf{c}_{opt}\equiv\widetilde{\mathbf{c}}_{opt}$), although this
latter has been yielded saving the $98.75$\% of the} \textcolor{black}{\emph{CPU}}
\textcolor{black}{costs (i.e., $\Delta t_{RTAM}<1$ {[}min{]} vs.
$\Delta t_{X}\approx80$ {[}min{]}) on a $1.6$GHz PC with $8$GB
of RAM.}

\noindent \textcolor{black}{The second example is aimed at pointing
out the advantages of using the} \textcolor{black}{\emph{L}}\textcolor{black}{-tromino
rep-tiles instead of the square ones. Indeed, the square geometry
belongs to the rep-tiles family since it can be represented as the
union of its $S=4$ smaller copies with halved edge length. Such a
comparison has been performed on a benchmark array of $M\times N=24\times24$
isotropic elements spaced by $d_{x}=d_{y}=\frac{\lambda}{2}$, while
the reference excitations {[}Fig. 9(}\textcolor{black}{\emph{b}}\textcolor{black}{){]}
have been still synthesized with the} \textcolor{black}{\emph{CP}}
\textcolor{black}{subject to the pattern constraints {[}i.e., $SLL_{1}=-25$
{[}dB{]}, $SLL_{2}=-30$ {[}dB{]}, $BW_{u}=BW_{v}=0.274$, and $\left(u_{0},v_{0}=0,0\right)${]}
coded by the mask $\Psi\left(u,v\right)$ in Fig. 9(}\textcolor{black}{\emph{a}}\textcolor{black}{).
Moreover, the maximum tiles order has been set to $R=3$ so that}
\textcolor{black}{\emph{L}}\textcolor{black}{-trominoes with $I_{L}^{\left(R-2\right)}=3$,
$I_{L}^{\left(R-1\right)}=12$, and $I_{L}^{\left(R\right)}=48$ elements
and squares with $I_{S}^{\left(R-2\right)}=4$, $I_{S}^{\left(R-1\right)}=16$,
and $I_{S}^{\left(R\right)}=64$ elements have been used for tiling
the array aperture $A$. Furthermore, $Q_{max}$ has been set to $Q_{max}=120$
to have a} \textcolor{black}{\emph{TRM}} \textcolor{black}{reduction
of about $\Delta_{TRM}^{min}\approx79$ \% ($\Delta_{TRM}^{min}=\left.\Delta_{TRM}^{Q}\right\rfloor _{Q=Q_{max}}$
being $\Delta_{TRM}^{Q}\triangleq1-\frac{Q}{M\times N}$).}

\noindent \textcolor{black}{Inasmuch as the $R$-order tileability
condition holds true, the} \textcolor{black}{\emph{RTAM}} \textcolor{black}{has
been applied with both} \textcolor{black}{\emph{L}}\textcolor{black}{-shaped
and square tiles. Figure 10 shows the behavior of the mask matching
versus the number of sub-arrays, $Q$, of the $H$ ($H=36$) layouts
generated during the} \textcolor{black}{\emph{RTAM}} \textcolor{black}{loop.
One can notice that the} \textcolor{black}{\emph{L}}\textcolor{black}{-shaped
arrangement always outperforms the corresponding (i.e., $Q_{L}=Q_{S}$)
square one, more and more as the heavier the clustering ratio is (i.e.,
$\Delta_{TRM}^{Q}\to100$ \%). Table II summarizes the results of
the comparative study by reporting the mask matching index along with
the key pattern descriptors (i.e., the $SLL$, the directivity, $D$,
and the half-power beamwidth along the azimuth, $HPBW_{az}$, and
the elevation, $HPBW_{el}$, planes) for the set of selected layouts
with $Q=\left\{ 39,\,81,\,120\right\} $ rep-tiles. For illustrative
purposes, the tilings synthesized at the $h=23$-th iteration the}
\textcolor{black}{\emph{RTAM}} \textcolor{black}{with $Q^{\left(h\right)}=81$
tiles are shown in Figs. 11(}\textcolor{black}{\emph{a}}\textcolor{black}{)-11(}\textcolor{black}{\emph{b}}\textcolor{black}{),
while the corresponding power patterns along the $v=0$ and the $u=0$
cuts are plotted in Fig. 11(}\textcolor{black}{\emph{c}}\textcolor{black}{)
and Fig. 11(}\textcolor{black}{\emph{d}}\textcolor{black}{), respectively.}

\noindent \textcolor{black}{The third experiment refers to a wider
rectangular aperture $A$ of $M\times N=24\times36$ elements. Because
of the array size, the use of an exhaustive tiling approach becomes
almost computationally unfeasible. Otherwise, the} \textcolor{black}{\emph{RTAM}}
\textcolor{black}{has been efficiently applied starting from the}
\textcolor{black}{\emph{CP}}\textcolor{black}{-synthesized reference
array in Fig. 12(}\textcolor{black}{\emph{c}}\textcolor{black}{),
which affords the power pattern of Fig. 12(}\textcolor{black}{\emph{b}}\textcolor{black}{)
that fulfils the pattern mask $\Psi\left(u,v\right)$ featuring the
following descriptors: $SLL_{1}=-25$ {[}dB{]}, $SLL_{2}=-30$ {[}dB{]},
$BW_{u}=0.164$, and $BW_{v}=0.274$ {[}Fig. 12(}\textcolor{black}{\emph{a}}\textcolor{black}{){]}.
More in detail, the} \textcolor{black}{\emph{RTAM}} \textcolor{black}{has
been executed with a rep-tiles alphabet of order $R=3$ ($I^{\left(R\right)}=48$,
$l^{\left(R\right)}=4$, $\widehat{M}=6$, and $\widehat{N}=9$) so
that the second condition of the covering theorem is fulfilled, while
the choice of rep-tiles of higher order (e.g., $R=4$ $\to$ $I^{\left(R\right)}=192$,
$l^{\left(R\right)}=8$, $\widehat{M}=3$, and $\widehat{N}=5$) would
not have guaranteed the complete tileability of $A$. Moreover, the
minimum} \textcolor{black}{\emph{TRM}} \textcolor{black}{saving has
been fixed to $\Delta_{TRM}^{min}\approx69$ \% (i.e., $Q_{max}=270$). }

\noindent \textcolor{black}{The} \textcolor{black}{\emph{RTAM}} \textcolor{black}{has
been initialized (}\textcolor{black}{\emph{Step 1}}\textcolor{black}{)
with the $T^{\left(R\right)}=4312$ (i.e., $\widehat{M}=6$, $\widehat{N}=9$
- Tab. I) uniform tilings of $\widehat{Q}=18$ $I^{\left(R\right)}$-sized}
\textcolor{black}{\emph{L}} \textcolor{black}{-tromino rep-tiles generated
by the Algorithm-X. The one, $\widehat{\mathbf{c}}_{opt}$, with minimum
mismatch (i.e., $\widehat{\Gamma}_{opt}=\left.\Gamma^{\left(h\right)}\right\rfloor _{h=0}=7.33\times10^{-5}$
being $\widehat{\Gamma}_{opt}\triangleq\Gamma\left(\widehat{\mathbf{c}}_{opt}\right)$
- Tab. III) is shown in Fig. 14(}\textcolor{black}{\emph{a}}\textcolor{black}{),
the corresponding power pattern being in Fig. 14(}\textcolor{black}{\emph{e}}\textcolor{black}{).
In the} \textcolor{black}{\emph{Step 2}}\textcolor{black}{, the} \textcolor{black}{\emph{RTAM}}
\textcolor{black}{loop has been runned for $H=84$ iterations by generating
at each $h$-th ($h=1,...,H$) iteration a different $Q$-tiled arrangement
(Fig. 13) until the convergence solution with $\left.Q^{\left(H\right)}\right\rfloor _{H=84}=270$
clusters.}

\noindent \textcolor{black}{Figure 14 includes the layouts {[}Figs.
14(}\textcolor{black}{\emph{b}}\textcolor{black}{)-14(}\textcolor{black}{\emph{d}}\textcolor{black}{){]}
and the corresponding power patterns {[}Figs. 14(}\textcolor{black}{\emph{f}}\textcolor{black}{)-14(}\textcolor{black}{\emph{h}}\textcolor{black}{){]}
for three representative solutions featuring different trade-offs
between the number of clusters {[}i.e., $Q=48$ - Fig. 14(}\textcolor{black}{\emph{b}}\textcolor{black}{)
and Fig. 14(}\textcolor{black}{\emph{f}}\textcolor{black}{); $Q=150$
- Fig. 14(}\textcolor{black}{\emph{c}}\textcolor{black}{) and Fig.
14(}\textcolor{black}{\emph{g}}\textcolor{black}{), and $Q=270$ -
Fig. 14(}\textcolor{black}{\emph{d}}\textcolor{black}{) and Fig. 14(}\textcolor{black}{\emph{h}}\textcolor{black}{){]}
and the pattern mask fitting (Fig. 13). More in detail, the solution
with $Q=150$ sub-arrays allows one to reduce of about $\Delta_{TRM}^{Q}\approx82.6$
\% the number of the} \textcolor{black}{\emph{TRM}}\textcolor{black}{s
of the reference fully-populated layout by radiating a power pattern
that deviates of $\Gamma=5.63\times10^{-5}$ from the mask, the $SLL$
being only $0.22$ {[}dB{]} above} \textcolor{black}{\emph{}}\textcolor{black}{the
reference value ($\left.SLL\right\rfloor _{Q=150}=-24.78$ {[}dB{]}
vs. $\left.SLL\right\rfloor _{ref}=-25.00$ {[}dB{]} - Tab. III).
Differently, the tiled configuration with $Q=48$ clusters further
reduces the architectural complexity up to $\Delta_{TRM}^{Q}\approx94.4$
\%, but it gets worse in terms of mask matching ($\Gamma=7.33\times10^{-5}$
- Tab. III) and sidelobe level ($\left.SLL\right\rfloor _{Q=48}=-21.33$
- Tab. III). As expected, it turns out that the more the sub-arrays,
the smaller the value of the mask matching index is at the cost of
a higher complexity of the feeding network (e.g., $Q=270$, $\Gamma=7.17\times10^{-6}$,
$SLL=-24.70$ {[}dB{]} - Tab. III). However, it is worth noting from
Fig. 13 that the mask-mismatch variations become almost negligible,
from a practical viewpoint, when $Q>150$, since the $SLL$ improves
less than $0.1$ {[}dB{]} as also confirmed by the comparison of the
power patterns along the $v=0$ and $u=0$ planes in Fig. 15(}\textcolor{black}{\emph{a}}\textcolor{black}{)
and Fig. 15(}\textcolor{black}{\emph{b}}\textcolor{black}{), respectively.}

\noindent \textcolor{black}{In order to validate the} \textcolor{black}{\emph{RTAM}}
\textcolor{black}{with arrays having non-zero phase excitations, the
next experiment addresses the design of the same size array of the
previous test case ($M\times N=24\times36$), but fitting a mask $\Psi\left(u,v\right)$
steered towards the direction $\left(u_{0},v_{0}\right)=\left(0.0755,\,0.0436\right)$
(i.e., $\left(\theta_{0},\phi_{0}\right)=\left(5,30\right)$ {[}deg{]})
having $SLL_{1}=-20$ {[}dB{]}, $SLL_{2}=-25${[}dB{]}, $BW_{u}=0.125$,
and $BW_{v}=0.2$ {[}Fig. 16(}\textcolor{black}{\emph{a}}\textcolor{black}{){]}.
Figure 16(}\textcolor{black}{\emph{b}}\textcolor{black}{) shows the
reference pattern radiated by the} \textcolor{black}{\emph{CP}}\textcolor{black}{-optimized
amplitude and phase coefficients in Fig. 16(}\textcolor{black}{\emph{c}}\textcolor{black}{)
and Fig. 16(}\textcolor{black}{\emph{d}}\textcolor{black}{), respectively.}

\noindent \textcolor{black}{The behaviour of the mask matching index
versus the number of clusters, $Q$, of the $H=84$ tiled arrangements
iteratively generated by the} \textcolor{black}{\emph{RTAM}} \textcolor{black}{is
shown in Fig. 17, while the pattern features of the layouts having
$Q=\left\{ 18,\,48,\,150,\,270\right\} $ are given in Tab. IV, as
well. For comparative purposes and analogously to the previous (non-steered)
case, the configuration with $Q=150$ sub-arrays is analyzed. Figure
18 indicates the aperture tiling within the color maps of the corresponding
sub-array amplitude {[}Fig. 18(}\textcolor{black}{\emph{a}}\textcolor{black}{){]}
and phase {[}Fig. 18(}\textcolor{black}{\emph{b}}\textcolor{black}{){]}
coefficients. Unlike the clustered layouts synthesized for the broadside-steered
arrays {[}e.g., Figs. 11(}\textcolor{black}{\emph{a}}\textcolor{black}{)-11(}\textcolor{black}{\emph{b}}\textcolor{black}{)
and Fig. 14(}\textcolor{black}{\emph{c}}\textcolor{black}{){]}, where
wider tiles are placed at the center of the array, the central part
of the aperture is here covered by smaller tiles {[}Figs. 18(}\textcolor{black}{\emph{a}}\textcolor{black}{)-18(}\textcolor{black}{\emph{b}}\textcolor{black}{){]}
to better match the complex reference excitations. As for the radiated
power pattern, the plots along the main planes in Figs. 18(}\textcolor{black}{\emph{c}}\textcolor{black}{)-(}\textcolor{black}{\emph{d}}\textcolor{black}{)
show non-negligible violations mainly in the mask region with $SLL_{2}=-25$
{[}dB{]}, while the mask matching is satisfactory near the main beam
with a $SLL$ violation of only $0.7$ {[}dB{]} (Tab. IV).}

\noindent \textcolor{black}{Next, the reliability of the} \textcolor{black}{\emph{RTAM}}\textcolor{black}{-optimized
tiled array when scanning the beam around the pointing direction has
been investigated by setting the sub-array phases, \{$\beta_{q}$;
$q=1,...,Q$\}, according to (\ref{eq:_subarray.phase}) starting
from the reference values}

\noindent \textcolor{black}{\begin{equation}
\beta_{mn}^{ref}=-k\left(x_{mn}u_{s}+y_{mn}v_{s}\right)\label{eq:_reference.phase}\end{equation}
($m=1,...,M$; $n=1,...,N$), where $u_{s}$ $\triangleq$ $\sin\left(\theta_{0}+\theta_{s}\right)$
$\cos\left(\phi_{0}+\phi_{s}\right)$ and $v_{s}$ $\triangleq$ $\sin\left(\theta_{0}+\theta_{s}\right)$
$\sin\left(\phi_{0}+\phi_{s}\right)$.}

\noindent \textcolor{black}{Figure 19 shows the color map of the $SLL$
when scanning the beam around the pointing direction $\left(\theta_{0},\,\phi_{0}\right)=\left(5,30\right)$
{[}deg{]} within the cone $0\leq\phi_{s}<360$ {[}deg{]} and $\theta_{s}\leq30$
{[}deg{]}. It turns out that $SLL_{max}\leq-14.5$ {[}dB{]} and $SLL_{avg}=-18.27$
{[}dB{]} for the beam scan region \{$\theta_{s}\leq5$ {[}deg{]},
$0$ {[}deg{]} $\le$ $\phi$ $\le$ $360$ {[}deg{]}\}, $SLL_{max}$
and $SLL_{avg}$ being maximum and the average} \textcolor{black}{\emph{SLL}}\textcolor{black}{,
respectively. }

\noindent \textcolor{black}{A further benchmark is concerned with
the asymmetric pattern mask of Fig. 20(}\textcolor{black}{\emph{a}}\textcolor{black}{),
which is centered at broadside with beamwidths $BW_{u}=0.080$ and
$BW_{v}=0.125$ and it is characterized by three sidelobe regions
(i.e., $SLL_{1}=-20$ {[}dB{]}, $SLL_{2}=-25$ {[}dB{]}, and $SLL_{3}=-30$
{[}dB{]}). Such a mask is fulfilled by the power pattern in Fig. 20(}\textcolor{black}{\emph{b}}\textcolor{black}{)
radiated by a fully-populated array of $M\times N$ elements with
excitations in Figs. 20(}\textcolor{black}{\emph{c}}\textcolor{black}{)-20(}\textcolor{black}{\emph{d}}\textcolor{black}{).}

\noindent \textcolor{black}{As a representative example, Figure 21
pictorially describes the compromise/intermediate solution, synthesized
after $h=44$ iterations by the} \textcolor{black}{\emph{RTAM}}\textcolor{black}{,
having $\left.Q^{\left(h\right)}\right\rfloor _{h=44}=150$ sub-arrays.
As expected, the rep-tiles layout presents the smaller tiles in the
regions of the array aperture where there are the most relevant variations
in the spatial distribution of the phase and the amplitude of the
reference excitations {[}Figs. 21(}\textcolor{black}{\emph{a}}\textcolor{black}{)-21(}\textcolor{black}{\emph{b}}\textcolor{black}{){]}.
As a result, the corresponding pattern faithfully fits the mask constraint
along the $v=0$ cut {[}Fig. 21(}\textcolor{black}{\emph{c}}\textcolor{black}{){]},
while there are two sidelobes in the $SLL_{3}=-30$ {[}dB{]} sidelobe
suppression region of the $u=0$ cut that exceed the target bound
of about $2$ {[}dB{]} {[}Fig. 21(}\textcolor{black}{\emph{d}}\textcolor{black}{){]}.}

\noindent \textcolor{black}{The last experiment is aimed at giving
some insights on the robustness of the} \textcolor{black}{\emph{RTAM}}\textcolor{black}{-synthesized
layout when dealing with non-ideal radiating elements. Towards this
end, a realistic antenna element with embedded pattern $\mathbf{e}\left(u,v\right)\neq1$
{[}Fig. 22(}\textcolor{black}{\emph{a}}\textcolor{black}{){]} has
been chosen for the analysis of the previous $Q=150$ element} \textcolor{black}{\emph{RTAM}}
\textcolor{black}{tiled array. In particular, such a radiating element
is a rectangular slot-fed patch antenna of length $1.029$ {[}mm{]}
and width $1.650$ {[}mm{]}, placed over a substrate with $\varepsilon_{r}=2.2$
and tan$\delta=9\times10^{-4}$, that resonates at the working frequency
$f=78.5$ {[}GHz{]}. Its embedded element pattern {[}Fig. 22(}\textcolor{black}{\emph{a}}\textcolor{black}{){]}
has been computed with the finite-element full-wave solver} \textcolor{black}{\emph{Ansys
HFSS}} \textcolor{black}{\cite{ANSYS 2019} by modeling the radiator
within a $5\times5$ set of identical elements {[}Fig. 22(}\textcolor{black}{\emph{b}}\textcolor{black}{){]}
to take into account the mutual coupling effects as well as other
electromagnetic interactions.}

\noindent \textcolor{black}{The comparison between the real and the
ideal pattern along the principal planes is reported in Figs. 21(}\textcolor{black}{\emph{c}}\textcolor{black}{)-21(}\textcolor{black}{\emph{d}}\textcolor{black}{).
One can observe that there are negligible differences in the lobes
regions close to end-fire, while the patterns in the mainlobe are
indistinguishable as confirmed by the values of the pattern metrics
in Tab. V ($\Gamma_{ideal}=4.44\times10^{-6}$ vs. $\Gamma_{real}=3.77\times10^{-6}$
and $SLL_{ideal}=-19.68$ {[}dB{]} vs. $SLL_{real}=-20.72$ {[}dB{]}).}

\section{\noindent \textcolor{black}{Conclusions}}

\textcolor{black}{In this work, the design of highly-modular planar}
\textcolor{black}{\emph{PA}}\textcolor{black}{s has been addressed
by exploiting the self-replicating property of rep-tiles. An innovative
deterministic method has been proposed to synthesize irregular rep-tiles
layouts fitting user-defined power pattern requirements, which are
mathematically coded in suitable pattern-masks, by jointly minimizing
the number of clusters to fully cover the array aperture.}

\noindent \textcolor{black}{From the numerical assessment, the following
outcomes and conclusions can be drawn:}

\begin{itemize}
\item \noindent \textcolor{black}{the proposed} \textcolor{black}{\emph{RTAM}}
\textcolor{black}{is able of converging towards optimal compromise
solutions of the constrained synthesis problem at hand by also enabling
a non-negligible saving of the computational cost with respect to
state-of-the-art exhaustive approaches;}
\item \noindent \textcolor{black}{since the iterative} \textcolor{black}{\emph{RTAM}}
\textcolor{black}{progressively generates multiple clustered array
layouts, which are different trade-offs between the number of clusters
$Q$ and the fulfilment of the user-defined pattern requirements,
it implicitly provides to the user, besides the optimal-matching arrangement
at the convergence, a Pareto front of compromise solutions;}
\item \noindent \textcolor{black}{thanks to the deterministic recursive
subdivision of wider rep-tiles into smaller ones in correspondence
with larger variations of the amplitude/phase distribution of the
reference excitations, the} \textcolor{black}{\emph{RTAM}} \textcolor{black}{is
an effective and reliable tool for the design of arbitrary-large/dense
arrays;}
\item \noindent \textcolor{black}{\emph{L}}\textcolor{black}{-tromino rep-tiles
turn out to be more effective than the square ones since they allow
one to better fit the user-defined power pattern mask with less sub-arrays.}
\end{itemize}
\noindent \textcolor{black}{Future research activities, beyond the
scope of this work, will include the investigation of different rep-tile
families and aperture geometries among those of interest for modern}
\textcolor{black}{\emph{PA}} \textcolor{black}{applications.}

\section*{\textcolor{black}{Acknowledgements}}

\textcolor{black}{This work benefited from the networking activities
carried out within the Project CYBER-PHYSICAL ELECTROMAGNETIC VISION:
Context-Aware Electromagnetic Sensing and Smart Reaction (EMvisioning)
(Grant no. 2017HZJXSZ) funded by the Italian Ministry of Education,
University, and Research under the PRIN2017 Program (CUP: E64I19002530001).
Moreover, it benefited from the networking activities carried out
within the Project SPEED (Grant No. 61721001) funded by National Science
Foundation of China under the Chang-Jiang Visiting Professorship Program,
the Project 'Inversion Design Method of Structural Factors of Conformal
Load-bearing Antenna Structure based on Desired EM Performance Interval'
(Grant no. 2017HZJXSZ) funded by the National Natural Science Foundation
of China, and the Project 'Research on Uncertainty Factors and Propagation
Mechanism of Conformal Loab-bearing Antenna Structure' (Grant No.
2021JZD-003) funded by the Department of Science and Technology of
Shaanxi Province within the Program Natural Science Basic Research
Plan in Shaanxi Province. A. Massa wishes to thank E. Vico for her
never-ending inspiration, support, guidance, and help.}

\section*{\textcolor{black}{Appendix \label{sec:Appendix-A}}}

\noindent \textcolor{black}{The matrix $\mathbf{G}_{\widehat{M}}$
($\widehat{M}\geq1$) is recursively computed as follows\begin{equation}
\mathbf{G}_{\widehat{M}}=\left[\begin{array}{cc}
\mathbf{S}_{\widehat{M}-1} & \mathbf{H}_{\widehat{M}-1}\\
\mathbf{G}_{\widehat{M}-1} & \mathbf{S}_{\widehat{M}-1}\end{array}\right]\label{eq:_appendix.A1}\end{equation}
where\begin{equation}
\mathbf{S}_{\widehat{M}}=\left[\begin{array}{cc}
\mathbf{Z}_{\widehat{M}-1} & \mathbf{G}_{\widehat{M}-1}\\
\mathbf{Z}_{\widehat{M}-1} & \mathbf{Z}_{\widehat{M}-1}\end{array}\right]\label{eq:_appendix.A2}\end{equation}
and\begin{equation}
\mathbf{H}_{\widehat{M}}=\left[\begin{array}{cc}
\mathbf{G}_{\widehat{M}-1} & 2\times\mathbf{S}_{\widehat{M}-1}\\
\mathbf{Z}_{\widehat{M}-1} & \mathbf{G}_{\widehat{M}-1}\end{array}\right],\label{eq:_appendix.A3}\end{equation}
$\mathbf{Z}_{\widehat{M}}$ being the $2^{\widehat{M}}\times2^{\widehat{M}}$
null matrix and setting $\left.\mathbf{G}_{\widehat{M}}\right\rfloor _{\widehat{M}=0}=1$,
$\left.\mathbf{S}_{\widehat{M}}\right\rfloor _{\widehat{M}=0}=0$,
and $\left.\mathbf{H}_{\widehat{M}}\right\rfloor _{\widehat{M}=0}=0$.}

\section*{\textcolor{black}{FIGURE CAPTIONS}}

\begin{itemize}
\item \textbf{\textcolor{black}{Figure 1.}} \textcolor{black}{Sketch of
a tiled architecture.}
\item \textbf{\textcolor{black}{Figure 2.}} \textcolor{black}{\emph{Illustrative
Example}} \textcolor{black}{($R=3$, $B=4$, $S=4$, $\left.I^{\left(r\right)}\right\rfloor _{r=1}=3$
-} \textcolor{black}{\emph{L}}\textcolor{black}{-tromino rep-tiles)
- Sketch of (}\textcolor{black}{\emph{a}}\textcolor{black}{) the alphabet
$\bm{\Sigma}$ and (}\textcolor{black}{\emph{b}}\textcolor{black}{)
the composition rule.}
\item \textbf{\textcolor{black}{Figure 3.}} \textcolor{black}{\emph{Illustrative
Example}} \textcolor{black}{($M=3$, $N=4$, $B=4$, $\left.I^{\left(r\right)}\right\rfloor _{r=1}=3$
-} \textcolor{black}{\emph{L}}\textcolor{black}{-tromino rep-tiles)
- Sketch of the tiling vector $\mathbf{c}=\left\{ 1,1,2,2,1,3,4,2,3,3,4,4\right\} $. }
\item \textbf{\textcolor{black}{Figure 4}}\textcolor{black}{.} \textcolor{black}{\emph{Illustrative
Example}} \textcolor{black}{($R=3$, $B=4$, $\left.I^{\left(r\right)}\right\rfloor _{r=1}=3$
$\to$ $I^{\left(R\right)}=48$ $\to$ $l^{\left(R\right)}=4$ -}
\textcolor{black}{\emph{L}}\textcolor{black}{-tromino rep-tiles) -
Sketch of (}\textcolor{black}{\emph{a}}\textcolor{black}{)(}\textcolor{black}{\emph{c}}\textcolor{black}{)(}\textcolor{black}{\emph{e}}\textcolor{black}{)
tileable and (}\textcolor{black}{\emph{b}}\textcolor{black}{)(}\textcolor{black}{\emph{d}}\textcolor{black}{)(}\textcolor{black}{\emph{f}}\textcolor{black}{)
non-tileable apertures: (}\textcolor{black}{\emph{a}}\textcolor{black}{)
$M=12$ and $N=16$ $\to$ $\widehat{M}=3$ ($\widehat{M}\triangleq\frac{M}{l^{\left(R\right)}}$)
and $\widehat{N}=4$ ($\widehat{N}\triangleq\frac{N}{l^{\left(R\right)}}$)
is even, (}\textcolor{black}{\emph{b}}\textcolor{black}{) $M=12$
and $N=20$ $\to$ $\widehat{M}=3$ but $\widehat{N}=5$ is odd, (}\textcolor{black}{\emph{c}}\textcolor{black}{)
$M=24$ and $N=8$ $\to$ $\widehat{M}\neq3$ and $\left\lfloor \frac{M\times N}{l^{\left(R\right)}}\right\rfloor =0$,
(}\textcolor{black}{\emph{d}}\textcolor{black}{) $M=20$ and $N=8$
$\to$ $\widehat{M}\neq3$ but $\left\lfloor \frac{M\times N}{l^{\left(R\right)}}\right\rfloor =16$,
(}\textcolor{black}{\emph{e}}\textcolor{black}{) $M=36$ and $N=24$
$\to$ $\widehat{M}\neq3$ and $\left\lfloor \frac{M\times N}{l^{\left(R\right)}}\right\rfloor =0$,
and (}\textcolor{black}{\emph{f}}\textcolor{black}{) $M=32$ and $N=28$
$\to$ $\widehat{M}\neq3$ but $\left\lfloor \frac{M\times N}{l^{\left(R\right)}}\right\rfloor =32$.}
\item \textbf{\textcolor{black}{Figure 5.}} \textcolor{black}{\emph{Illustrative
Example}} \textcolor{black}{($M=12$, $N=16,$ $R=3$, $B=4$, $S=4$,
$\left.I^{\left(r\right)}\right\rfloor _{r=1}=3$, $H=3$ -} \textcolor{black}{\emph{L}}\textcolor{black}{-tromino
rep-tiles) - Sketch of (}\textcolor{black}{\emph{a}}\textcolor{black}{)(}\textcolor{black}{\emph{c}}\textcolor{black}{)(}\textcolor{black}{\emph{e}}\textcolor{black}{)
rep-tiled array layout and (}\textcolor{black}{\emph{b}}\textcolor{black}{)(}\textcolor{black}{\emph{d}}\textcolor{black}{)(}\textcolor{black}{\emph{f}}\textcolor{black}{)
rep-tile splitting at the} \textcolor{black}{\emph{RTAM}} \textcolor{black}{iteration
(a) $h=0$, (}\textcolor{black}{\emph{b}}\textcolor{black}{)(}\textcolor{black}{\emph{c}}\textcolor{black}{)
$h=1$, (}\textcolor{black}{\emph{d}}\textcolor{black}{)(}\textcolor{black}{\emph{e}}\textcolor{black}{)
$h=2$, and (}\textcolor{black}{\emph{f}}\textcolor{black}{) $h=H$.}
\item \textbf{\textcolor{black}{Figure 6.}} \textcolor{black}{\emph{Numerical
Validation}} \textcolor{black}{($M=8$, $N=12$, $\mathbf{e}\left(u,v\right)=1$,
$d_{x}=d_{y}=\frac{\lambda}{2}$, $\left(u_{0},v_{0}\right)=\left(0,0\right)$
$\to$ $\left(\theta_{0},\phi_{0}\right)=\left(0,0\right)$ {[}deg{]})
- Color maps of (}\textcolor{black}{\emph{a}}\textcolor{black}{) the
power pattern mask $\Psi\left(u,v\right)$ and (}\textcolor{black}{\emph{b}}\textcolor{black}{)
the amplitude distribution of the reference excitations, \{$w_{mn}^{ref}=\alpha_{mn}^{ref}$;
$m=1,...,M$, $n=1,...,N$\}.}
\item \textbf{\textcolor{black}{Figure 7.}} \textcolor{black}{\emph{Numerical
Validation}} \textcolor{black}{($M=8$, $N=12$, $\mathbf{e}\left(u,v\right)=1$,
$d_{x}=d_{y}=\frac{\lambda}{2}$, $\left(u_{0},v_{0}\right)=\left(0,0\right)$
$\to$ $\left(\theta_{0},\phi_{0}\right)=\left(0,0\right)$ {[}deg{]};
$Q_{max}=14$, $R=2$, $I^{\left(R\right)}=12$, $I^{\left(R-1\right)}=3$)
- Plots of (}\textcolor{black}{\emph{a}}\textcolor{black}{) the ranked
values of the mask matching index of the $T_{Q_{max}}=6248$ aperture
partitions with $Q\leq Q_{max}$ tiles from the Algorithm-X together
with the one, $\Gamma_{opt}$, of the} \textcolor{black}{\emph{RTAM}}
\textcolor{black}{optimal solution, $\mathbf{c}_{opt}$, and (}\textcolor{black}{\emph{b}}\textcolor{black}{)(}\textcolor{black}{\emph{c}}\textcolor{black}{)
the normalized power patterns along the (}\textcolor{black}{\emph{b}}\textcolor{black}{)
$v=0$ (i.e., $\phi=0$ {[}deg{]}) and (}\textcolor{black}{\emph{c}}\textcolor{black}{)
$u=0$ (i.e., $\phi=90$ {[}deg{]}) planes.}
\item \textbf{\textcolor{black}{Figure 8.}} \textcolor{black}{\emph{Numerical
Validation}} \textcolor{black}{($M=8$, $N=12$, $\mathbf{e}\left(u,v\right)=1$,
$d_{x}=d_{y}=\frac{\lambda}{2}$, $\left(u_{0},v_{0}\right)=\left(0,0\right)$
$\to$ $\left(\theta_{0},\phi_{0}\right)=\left(0,0\right)$ {[}deg{]};
$Q_{max}=14$, $R=2$, $I^{\left(R\right)}=12$, $I^{\left(R-1\right)}=3$)
- Color maps of (}\textcolor{black}{\emph{a}}\textcolor{black}{)(}\textcolor{black}{\emph{c}}\textcolor{black}{)(}\textcolor{black}{\emph{e}}\textcolor{black}{)
amplitude distribution of the clustered excitations, \{$w_{q}^{\left(h\right)}=\alpha_{q}^{\left(h\right)}e^{j\beta_{q}^{\left(h\right)}}$;
$q=1,...,Q^{\left(h\right)}$\}, and (}\textcolor{black}{\emph{b}}\textcolor{black}{)(}\textcolor{black}{\emph{d}}\textcolor{black}{)}
\textcolor{black}{\emph{STM}} \textcolor{black}{value within the array
aperture at the} \textcolor{black}{\emph{RTAM}} \textcolor{black}{iterations
(}\textcolor{black}{\emph{a}}\textcolor{black}{)(}\textcolor{black}{\emph{b}}\textcolor{black}{)
$h=0$ ($Q^{\left(h\right)}=8$), (}\textcolor{black}{\emph{c}}\textcolor{black}{)(}\textcolor{black}{\emph{d}}\textcolor{black}{)
$h=1$ ($Q^{\left(h\right)}=11$), and (}\textcolor{black}{\emph{e}}\textcolor{black}{)
$h=H$ ($H=2$) ($Q^{\left(h\right)}=Q_{max}$).}
\item \textbf{\textcolor{black}{Figure 9.}} \textcolor{black}{\emph{Numerical
Validation}} \textcolor{black}{($M=24$, $N=24$, $\mathbf{e}\left(u,v\right)=1$,
$d_{x}=d_{y}=\frac{\lambda}{2}$, $\left(u_{0},v_{0}\right)=\left(0,0\right)$
$\to$ $\left(\theta_{0},\phi_{0}\right)=\left(0,0\right)$ {[}deg{]})
- Color maps of (}\textcolor{black}{\emph{a}}\textcolor{black}{) the
power pattern mask $\Psi\left(u,v\right)$ and (}\textcolor{black}{\emph{b}}\textcolor{black}{)
the amplitude distribution of the reference excitations, \{$w_{mn}^{ref}=\alpha_{mn}^{ref}$;
$m=1,...,M$, $n=1,...,N$\}.}
\item \textbf{\textcolor{black}{Figure 10.}} \textcolor{black}{\emph{Numerical
Validation}} \textcolor{black}{($M=24$, $N=24$, $\mathbf{e}\left(u,v\right)=1$,
$d_{x}=d_{y}=\frac{\lambda}{2}$, $\left(u_{0},v_{0}\right)=\left(0,0\right)$
$\to$ $\left(\theta_{0},\phi_{0}\right)=\left(0,0\right)$ {[}deg{]};
$Q_{max}=120$, $R=3$, $I_{L}^{\left(R\right)}=48$, $I_{L}^{\left(R-1\right)}=12$,
$I_{L}^{\left(R-2\right)}=3$, $I_{S}^{\left(R\right)}=64$, $I_{S}^{\left(R-1\right)}=16$,
$I_{S}^{\left(R-2\right)}=4$) - Plot of the mask matching value,
$\Gamma^{\left(h\right)}$, of the tiled array configuration, $\mathbf{c}^{\left(h\right)}$
{[}i.e., $\Gamma^{\left(h\right)}\triangleq\Gamma\left(\mathbf{c}^{\left(h\right)}\right)${]},
synthesized at the $h$-th ($h=0,...,H$; $H=36$)} \textcolor{black}{\emph{RTAM}}
\textcolor{black}{loop versus its number of sub-arrays, $Q^{\left(h\right)}$.}
\item \textbf{\textcolor{black}{Figure 11.}} \textcolor{black}{\emph{Numerical
Validation}} \textcolor{black}{($M=24$, $N=24$, $\mathbf{e}\left(u,v\right)=1$,
$d_{x}=d_{y}=\frac{\lambda}{2}$, $\left(u_{0},v_{0}\right)=\left(0,0\right)$
$\to$ $\left(\theta_{0},\phi_{0}\right)=\left(0,0\right)$ {[}deg{]};
$Q_{max}=120$, $R=3$, $I_{L}^{\left(R\right)}=48$, $I_{L}^{\left(R-1\right)}=12$,
$I_{L}^{\left(R-2\right)}=3$, $I_{S}^{\left(R\right)}=64$, $I_{S}^{\left(R-1\right)}=16$,
$I_{S}^{\left(R-2\right)}=4$, $h=60$, $\left.Q^{\left(h\right)}\right\rfloor _{h=60}=81$)
- Plots of} \textbf{\textcolor{black}{}}\textcolor{black}{(}\textcolor{black}{\emph{a}}\textcolor{black}{)(}\textcolor{black}{\emph{b}}\textcolor{black}{)
amplitude distributions of the excitations, \{$w_{q}^{\left(h\right)}=\alpha_{q}^{\left(h\right)}e^{j\beta_{q}^{\left(h\right)}}$;
$q=1,...,Q^{\left(h\right)}$\}, of the array tiled with (}\textcolor{black}{\emph{a}}\textcolor{black}{)}
\textcolor{black}{\emph{L}}\textcolor{black}{-trominoes or (}\textcolor{black}{\emph{b}}\textcolor{black}{)
squares and (}\textcolor{black}{\emph{c}}\textcolor{black}{)(}\textcolor{black}{\emph{d}}\textcolor{black}{)
the normalized power patterns along the (}\textcolor{black}{\emph{c}}\textcolor{black}{)
$v=0$ (i.e., $\phi=0$ {[}deg{]}) and (}\textcolor{black}{\emph{d}}\textcolor{black}{)
$u=0$ (i.e., $\phi=90$ {[}deg{]}) planes.}
\item \textbf{\textcolor{black}{Figure 12.}} \textcolor{black}{\emph{Numerical
Validation}} \textcolor{black}{($M=24$, $N=36$, $\mathbf{e}\left(u,v\right)=1$,
$d_{x}=d_{y}=\frac{\lambda}{2}$, $\left(u_{0},v_{0}\right)=\left(0,0\right)$
$\to$ $\left(\theta_{0},\phi_{0}\right)=\left(0,0\right)$ {[}deg{]})
- Plots of (}\textcolor{black}{\emph{a}}\textcolor{black}{) the power
pattern mask $\Psi\left(u,v\right)$ and (}\textcolor{black}{\emph{b}}\textcolor{black}{)
the normalized power pattern radiated by (}\textcolor{black}{\emph{c}}\textcolor{black}{)
the amplitude distribution of the reference excitations, \{$w_{mn}^{ref}=\alpha_{mn}^{ref}$;
$m=1,...,M$, $n=1,...,N$\}.}
\item \textbf{\textcolor{black}{Figure 13.}} \textcolor{black}{\emph{Numerical
Validation}} \textcolor{black}{($M=24$, $N=36$, $\mathbf{e}\left(u,v\right)=1$,
$d_{x}=d_{y}=\frac{\lambda}{2}$, $\left(u_{0},v_{0}\right)=\left(0,0\right)$
$\to$ $\left(\theta_{0},\phi_{0}\right)=\left(0,0\right)$ {[}deg{]};
$Q_{max}=270$, $R=3$, $I^{\left(R\right)}=48$, $I^{\left(R-1\right)}=12$,
$I^{\left(R-2\right)}=3$) - Plot of the mask matching value, $\Gamma^{\left(h\right)}$,
of the tiled array configuration, $\mathbf{c}^{\left(h\right)}$ {[}i.e.,
$\Gamma^{\left(h\right)}\triangleq\Gamma\left(\mathbf{c}^{\left(h\right)}\right)${]},
synthesized at the $h$-th ($h=0,...,H$; $H=84$)} \textcolor{black}{\emph{RTAM}}
\textcolor{black}{loop versus versus its number of sub-arrays, $Q^{\left(h\right)}$.}
\item \textbf{\textcolor{black}{Figure 14.}} \textcolor{black}{\emph{Numerical
Validation}} \textcolor{black}{($M=24$, $N=36$, $\mathbf{e}\left(u,v\right)=1$,
$d_{x}=d_{y}=\frac{\lambda}{2}$, $\left(u_{0},v_{0}\right)=\left(0,0\right)$
$\to$ $\left(\theta_{0},\phi_{0}\right)=\left(0,0\right)$ {[}deg{]};
$Q_{max}=270$, $R=3$, $I^{\left(R-2\right)}=48$, $I^{\left(R-1\right)}=12$,
$I^{\left(R\right)}=3$) - Plots of (}\textcolor{black}{\emph{a}}\textcolor{black}{)(}\textcolor{black}{\emph{b}}\textcolor{black}{)(}\textcolor{black}{\emph{c}}\textcolor{black}{)(}\textcolor{black}{\emph{d}}\textcolor{black}{)
the amplitude distribution of the clustered excitations, \{$w_{q}^{\left(h\right)}=\alpha_{q}^{\left(h\right)}$;
$q=1,...,Q^{\left(h\right)}$\}, and (}\textcolor{black}{\emph{e}}\textcolor{black}{)(}\textcolor{black}{\emph{f}}\textcolor{black}{)(}\textcolor{black}{\emph{g}}\textcolor{black}{)(}\textcolor{black}{\emph{h}}\textcolor{black}{)
the corresponding normalized power pattern at the} \textcolor{black}{\emph{RTAM}}
\textcolor{black}{iterations (}\textcolor{black}{\emph{a}}\textcolor{black}{)(}\textcolor{black}{\emph{e}}\textcolor{black}{)
$h=0$ ($Q^{\left(h\right)}=18$), (}\textcolor{black}{\emph{b}}\textcolor{black}{)(}\textcolor{black}{\emph{f}}\textcolor{black}{)
$h=10$ ($Q^{\left(h\right)}=48$), (}\textcolor{black}{\emph{c}}\textcolor{black}{)(}\textcolor{black}{\emph{g}}\textcolor{black}{)
$h=44$ ($Q^{\left(h\right)}=150$), and (}\textcolor{black}{\emph{d}}\textcolor{black}{)(}\textcolor{black}{\emph{h}}\textcolor{black}{)
$h=H$ ($H=84$) ($Q^{\left(h\right)}=Q_{max}$).}
\item \textbf{\textcolor{black}{Figure 15.}} \textcolor{black}{\emph{Numerical
Validation}} \textcolor{black}{($M=24$, $N=36$, $\mathbf{e}\left(u,v\right)=1$,
$d_{x}=d_{y}=\frac{\lambda}{2}$, $\left(u_{0},v_{0}\right)=\left(0,0\right)$
$\to$ $\left(\theta_{0},\phi_{0}\right)=\left(0,0\right)$ {[}deg{]};
$Q_{max}=270$, $R=3$, $I^{\left(R\right)}=48$, $I^{\left(R-1\right)}=12$,
$I^{\left(R-2\right)}=3$) - Plots of} \textbf{\textcolor{black}{}}\textcolor{black}{the
normalized power patterns along (}\textcolor{black}{\emph{a}}\textcolor{black}{)
the $v=0$ (i.e., $\phi=0$ {[}deg{]}) and (}\textcolor{black}{\emph{b}}\textcolor{black}{)
the $u=0$ (i.e., $\phi=90$ {[}deg{]}) planes.}
\item \textbf{\textcolor{black}{Figure 16.}} \textcolor{black}{\emph{Numerical
Validation}} \textcolor{black}{($M=24$, $N=36$, $\mathbf{e}\left(u,v\right)=1$,
$d_{x}=d_{y}=\frac{\lambda}{2}$, $\left(u_{0},v_{0}\right)=\left(0.0755,\,0.0436\right)$
$\to$ $\left(\theta_{0},\phi_{0}\right)=\left(5,\,30\right)$ {[}deg{]};
$Q_{max}=270$, $R=3$, $I^{\left(R\right)}=48$, $I^{\left(R-1\right)}=12$,
$I^{\left(R-2\right)}=3$) - Plots of (}\textcolor{black}{\emph{a}}\textcolor{black}{)
the power pattern mask $\Psi\left(u,v\right)$ and (}\textcolor{black}{\emph{b}}\textcolor{black}{)
the normalized power pattern radiated by (}\textcolor{black}{\emph{c}}\textcolor{black}{)
the amplitude and (}\textcolor{black}{\emph{d}}\textcolor{black}{)
the phase distributions of the reference excitations, \{$w_{mn}^{ref}=\alpha_{mn}^{ref}e^{j\beta_{mn}^{ref}}$;
$m=1,...,M$, $n=1,...,N$\}.}
\item \textbf{\textcolor{black}{Figure 17.}} \textcolor{black}{\emph{Numerical
Validation}} \textcolor{black}{($M=24$, $N=36$, $\mathbf{e}\left(u,v\right)=1$,
$d_{x}=d_{y}=\frac{\lambda}{2}$, $\left(u_{0},v_{0}\right)=\left(0.0755,\,0.0436\right)$
$\to$ $\left(\theta_{0},\phi_{0}\right)=\left(5,\,30\right)$ {[}deg{]};
$Q_{max}=270$, $R=3$, $I^{\left(R-2\right)}=48$, $I^{\left(R-1\right)}=12$,
$I^{\left(R\right)}=3$) - Plot of the mask matching value, $\Gamma^{\left(h\right)}$,
of the tiled array configuration, $\mathbf{c}^{\left(h\right)}$ {[}i.e.,
$\Gamma^{\left(h\right)}\triangleq\Gamma\left(\mathbf{c}^{\left(h\right)}\right)${]},
synthesized at the $h$-th ($h=0,...,H$, $H=90$)} \textcolor{black}{\emph{RTAM}}
\textcolor{black}{loop versus its number of sub-arrays, $Q^{\left(h\right)}$.}
\item \textbf{\textcolor{black}{Figure 18.}} \textcolor{black}{\emph{Numerical
Validation}} \textcolor{black}{($M=24$, $N=36$, $\mathbf{e}\left(u,v\right)=1$,
$d_{x}=d_{y}=\frac{\lambda}{2}$, $\left(u_{0},v_{0}\right)=\left(0.0755,\,0.0436\right)$
$\to$ $\left(\theta_{0},\phi_{0}\right)=\left(5,\,30\right)$ {[}deg{]};
$Q_{max}=270$, $R=3$, $I^{\left(R-2\right)}=48$, $I^{\left(R-1\right)}=12$,
$I^{\left(R\right)}=3$; $h=44$, $\left.Q^{\left(h\right)}\right\rfloor _{h=44}=150$)
- Plots of} \textbf{\textcolor{black}{}}\textcolor{black}{(}\textcolor{black}{\emph{a}}\textcolor{black}{)
the amplitude and (}\textcolor{black}{\emph{b}}\textcolor{black}{)
the phase distributions of the} \textcolor{black}{\emph{RTAM}}\textcolor{black}{-synthesized
excitations, \{$w_{q}^{\left(h\right)}=\alpha_{q}^{\left(h\right)}e^{j\beta_{q}^{\left(h\right)}}$;
$q=1,...,Q^{\left(h\right)}$\}, and (}\textcolor{black}{\emph{c}}\textcolor{black}{)(}\textcolor{black}{\emph{d}}\textcolor{black}{)
the normalized power patterns along the (}\textcolor{black}{\emph{c}}\textcolor{black}{)
$v=0$ (i.e., $\phi=0$ {[}deg{]}) and (}\textcolor{black}{\emph{d}}\textcolor{black}{)
$u=0$ (i.e., $\phi=90$ {[}deg{]}) planes.}
\item \textbf{\textcolor{black}{Figure 19.}} \textcolor{black}{\emph{Numerical
Validation}} \textcolor{black}{($M=24$, $N=36$, $\mathbf{e}\left(u,v\right)=1$,
$d_{x}=d_{y}=\frac{\lambda}{2}$, $\left(u_{0},v_{0}\right)=\left(0.0755,\,0.0436\right)$
$\to$ $\left(\theta_{0},\phi_{0}\right)=\left(5,\,30\right)$ {[}deg{]};
$Q_{max}=270$, $R=3$, $I^{\left(R-2\right)}=48$, $I^{\left(R-1\right)}=12$,
$I^{\left(R\right)}=3$; $h=44$, $\left.Q^{\left(h\right)}\right\rfloor _{h=44}=150$)
- Color map of the $SLL$ variations when scanning the beam around
the pointing direction $\left(\theta_{0},\,\phi_{0}\right)=\left(5,30\right)$
{[}deg{]} within the cone $0\leq\phi_{s}<360$ {[}deg{]} and $\theta_{s}\leq30$
{[}deg{]}.}
\item \textbf{\textcolor{black}{Figure 20.}} \textcolor{black}{\emph{Numerical
Validation}} \textcolor{black}{($M=24$, $N=36$, $\mathbf{e}\left(u,v\right)=1$,
$d_{x}=d_{y}=\frac{\lambda}{2}$, $\left(u_{0},v_{0}\right)=\left(0,0\right)$
$\to$ $\left(\theta_{0},\phi_{0}\right)=\left(0,0\right)$ {[}deg{]})
- Plots of (}\textcolor{black}{\emph{a}}\textcolor{black}{) the power
pattern mask $\Psi\left(u,v\right)$ and (}\textcolor{black}{\emph{b}}\textcolor{black}{)
the normalized power pattern radiated by (}\textcolor{black}{\emph{c}}\textcolor{black}{)
the amplitude and (}\textcolor{black}{\emph{d}}\textcolor{black}{)
the phase distributions of the reference excitations, \{$w_{mn}^{ref}=\alpha_{mn}^{ref}e^{j\beta_{mn}^{ref}}$;
$m=1,...,M$, $n=1,...,N$\}.}
\item \textbf{\textcolor{black}{Figure 21.}} \textcolor{black}{\emph{Numerical
Validation}} \textcolor{black}{($M=24$, $N=36$, $\mathbf{e}\left(u,v\right)=1$,
$d_{x}=d_{y}=\frac{\lambda}{2}$, $\left(u_{0},v_{0}\right)=\left(0,0\right)$
$\to$ $\left(\theta_{0},\phi_{0}\right)=\left(0,0\right)$ {[}deg{]};
; $Q_{max}=270$, $R=3$, $I^{\left(R-2\right)}=48$, $I^{\left(R-1\right)}=12$,
$I^{\left(R\right)}=3$; $h=44$, $\left.Q^{\left(h\right)}\right\rfloor _{h=44}=150$)
- Plots of} \textbf{\textcolor{black}{}}\textcolor{black}{(}\textcolor{black}{\emph{a}}\textcolor{black}{)
the amplitude and (}\textcolor{black}{\emph{b}}\textcolor{black}{)
the phase distributions of the} \textcolor{black}{\emph{RTAM}}\textcolor{black}{-synthesized
excitations, \{$w_{q}^{\left(h\right)}=\alpha_{q}^{\left(h\right)}e^{j\beta_{q}^{\left(h\right)}}$;
$q=1,...,Q^{\left(h\right)}$\}, and (}\textcolor{black}{\emph{c}}\textcolor{black}{)(}\textcolor{black}{\emph{d}}\textcolor{black}{)
the normalized power patterns along the (}\textcolor{black}{\emph{c}}\textcolor{black}{)
$v=0$ (i.e., $\phi=0$ {[}deg{]}) and (}\textcolor{black}{\emph{d}}\textcolor{black}{)
$u=0$ (i.e., $\phi=90$ {[}deg{]}) planes.}
\item \textbf{\textcolor{black}{Figure 22.}} \textcolor{black}{\emph{Numerical
Validation}} \textcolor{black}{- Plot of the (}\textcolor{black}{\emph{a}}\textcolor{black}{)
embedded element pattern of the real elementary radiator simulated
within (}\textcolor{black}{\emph{b}}\textcolor{black}{) a $5\times5$
set of identical elements.}
\end{itemize}

\section*{\textcolor{black}{TABLE CAPTIONS}}

\begin{itemize}
\item \textbf{\textcolor{black}{Table I.}} \textcolor{black}{\emph{Illustrative
Example}} \textcolor{black}{- Number of} \textcolor{black}{\emph{L}}\textcolor{black}{-tromino
tilings of $R$-th order (i.e., $T$ value).}
\item \textbf{\textcolor{black}{Table II.}} \textcolor{black}{\emph{Numerical
Validation}} \textcolor{black}{($M=24$, $N=24$, $\mathbf{e}\left(u,v\right)=1$,
$d_{x}=d_{y}=\frac{\lambda}{2}$, $\left(u_{0},v_{0}\right)=\left(0,0\right)$
$\to$ $\left(\theta_{0},\phi_{0}\right)=\left(0,0\right)$ {[}deg{]};
$Q_{max}=120$, $R=3$, $I_{L}^{\left(R\right)}=48$, $I_{L}^{\left(R-1\right)}=12$,
$I_{L}^{\left(R-2\right)}=3$, $I_{S}^{\left(R\right)}=64$, $I_{S}^{\left(R-1\right)}=16$,
$I_{S}^{\left(R-2\right)}=4$) - Pattern features and mask matching
index.}
\item \textbf{\textcolor{black}{Table III.}} \textcolor{black}{\emph{Numerical
Validation}} \textcolor{black}{($M=24$, $N=36$, $\mathbf{e}\left(u,v\right)=1$,
$d_{x}=d_{y}=\frac{\lambda}{2}$, $\left(u_{0},v_{0}\right)=\left(0.0755,\,0.0436\right)$
$\to$ $\left(\theta_{0},\phi_{0}\right)=\left(5,\,30\right)$ {[}deg{]};
$Q_{max}=270$, $R=3$, $I^{\left(R\right)}=48$, $I^{\left(R-1\right)}=12$,
$I^{\left(R-2\right)}=3$) - Pattern features and mask matching index.}
\item \textbf{\textcolor{black}{Table IV.}} \textcolor{black}{\emph{Numerical
Validation}} \textcolor{black}{($M=24$, $N=36$, $\mathbf{e}\left(u,v\right)=1$,
$d_{x}=d_{y}=\frac{\lambda}{2}$, $\left(u_{0},v_{0}\right)=\left(0.0755,\,0.0436\right)$
$\to$ $\left(\theta_{0},\phi_{0}\right)=\left(5,\,30\right)$ {[}deg{]};
$Q_{max}=270$, $R=3$, $I^{\left(R\right)}=48$, $I^{\left(R-1\right)}=12$,
$I^{\left(R-2\right)}=3$) - Pattern features and mask matching index.}
\item \textbf{\textcolor{black}{Table V.}} \textcolor{black}{\emph{Numerical
Validation}} \textcolor{black}{($M=24$, $N=36$, $d_{x}=d_{y}=\frac{\lambda}{2}$,
$\left(u_{0},v_{0}\right)=\left(0,0\right)$ $\to$ $\left(\theta_{0},\phi_{0}\right)=\left(0,0\right)$
{[}deg{]}; ; $Q_{max}=270$, $R=3$, $I^{\left(R-2\right)}=48$, $I^{\left(R-1\right)}=12$,
$I^{\left(R\right)}=3$) - Pattern features and mask matching index.}
\end{itemize}
\newpage
\textcolor{black}{~\vfill}

\begin{center}\textcolor{black}{}\begin{tabular}{c}
\textcolor{black}{\includegraphics[%
  width=0.70\columnwidth]{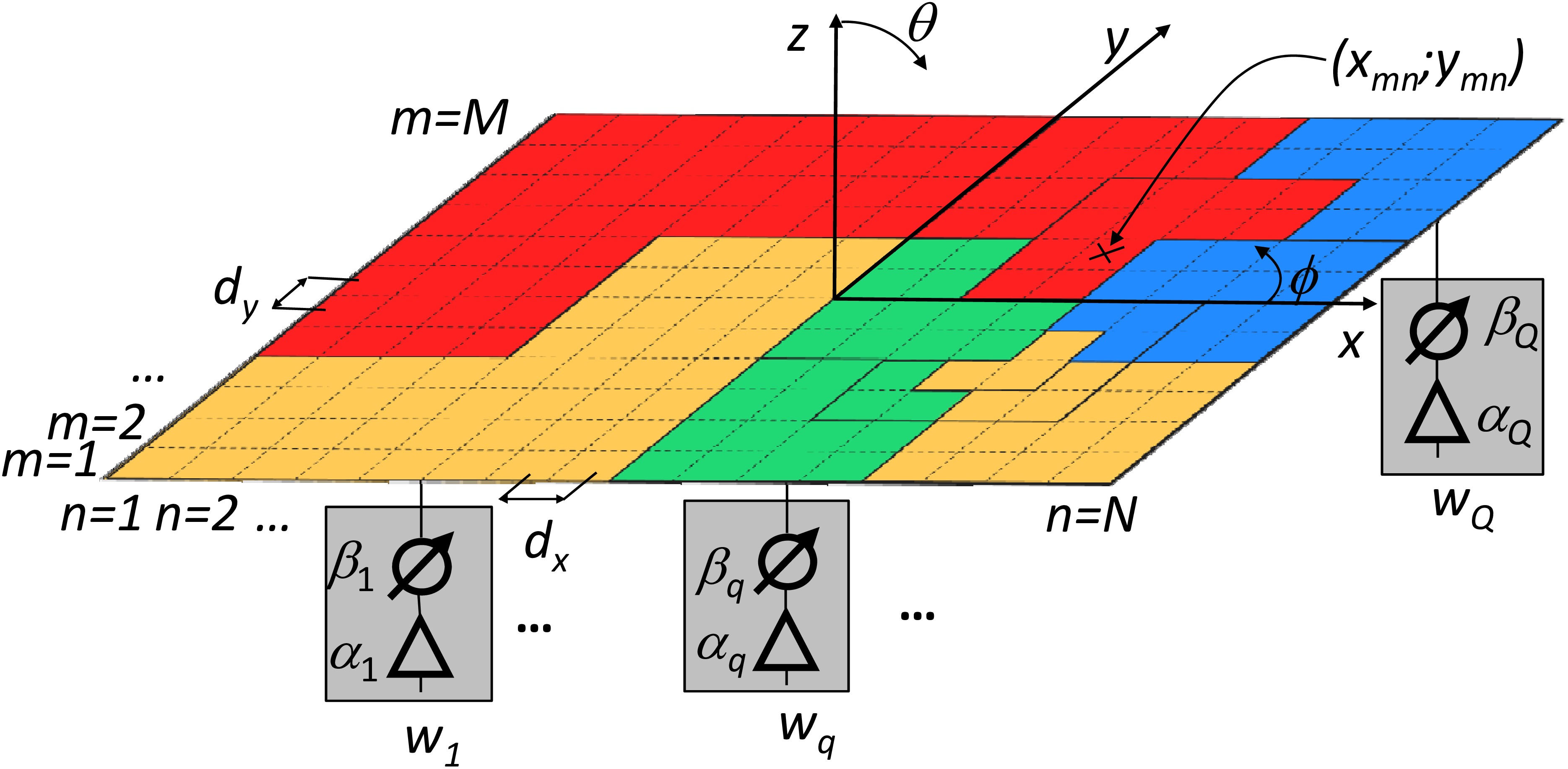}}\tabularnewline
\end{tabular}\end{center}

\begin{center}\textcolor{black}{~\vfill}\end{center}

\begin{center}\textbf{\textcolor{black}{Fig. 1 - N. Anselmi}} \textbf{\textcolor{black}{\emph{et
al.}}}\textbf{\textcolor{black}{,}} \textbf{\textcolor{black}{\emph{{}``}}}\textcolor{black}{A
Self-Replicating Single-Shape Tiling Technique ...''}\end{center}

\newpage
\begin{center}\textcolor{black}{}\begin{tabular}{c}
\textcolor{black}{\includegraphics[%
  width=0.90\columnwidth,
  keepaspectratio]{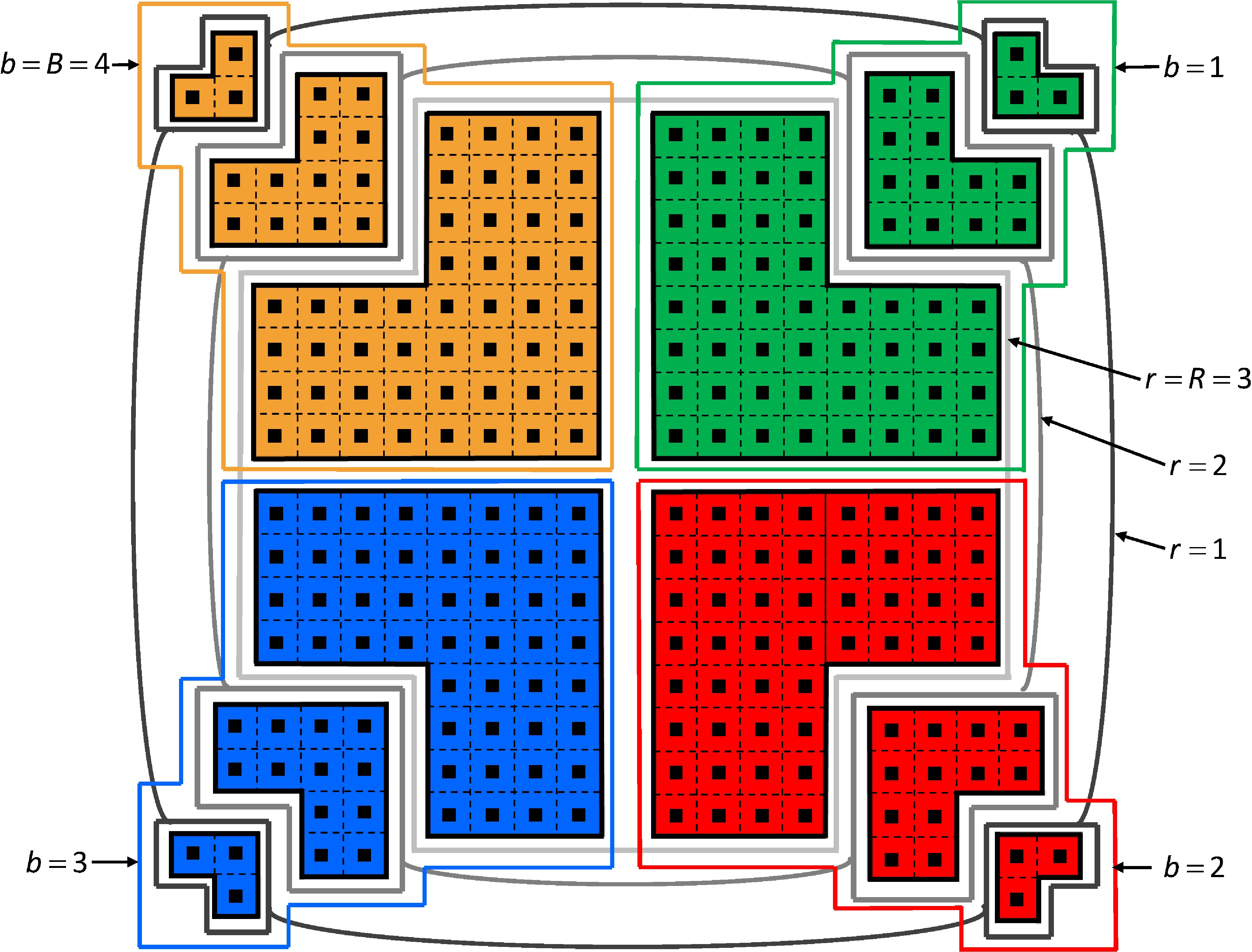}}\tabularnewline
\textcolor{black}{(}\textcolor{black}{\emph{a}}\textcolor{black}{)}\tabularnewline
\tabularnewline
\textcolor{black}{\includegraphics[%
  width=0.50\columnwidth,
  keepaspectratio]{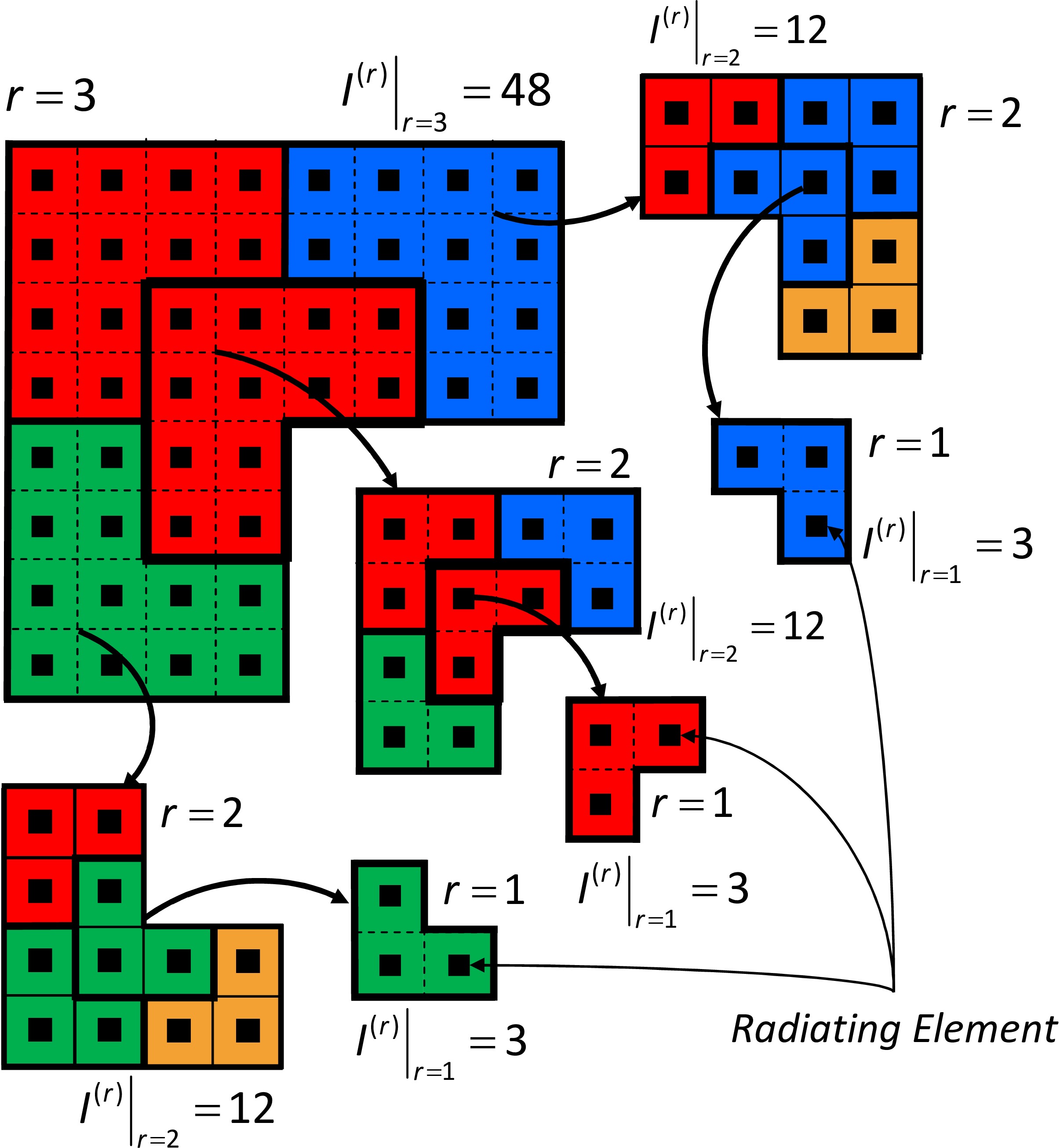}}\tabularnewline
\textcolor{black}{(}\textcolor{black}{\emph{b}}\textcolor{black}{)}\tabularnewline
\end{tabular}\end{center}

\begin{center}\textbf{\textcolor{black}{~}}\textcolor{black}{\vfill}\end{center}

\begin{center}\textbf{\textcolor{black}{Fig. 2 - N. Anselmi}} \textbf{\textcolor{black}{\emph{et
al.}}}\textbf{\textcolor{black}{,}} \textbf{\textcolor{black}{\emph{{}``}}}\textcolor{black}{A
Self-Replicating Single-Shape Tiling Technique ...''}\end{center}

\newpage
\textcolor{black}{~\vfill}

\begin{center}\textcolor{black}{}\begin{tabular}{c}
\textcolor{black}{\includegraphics[%
  width=0.70\columnwidth]{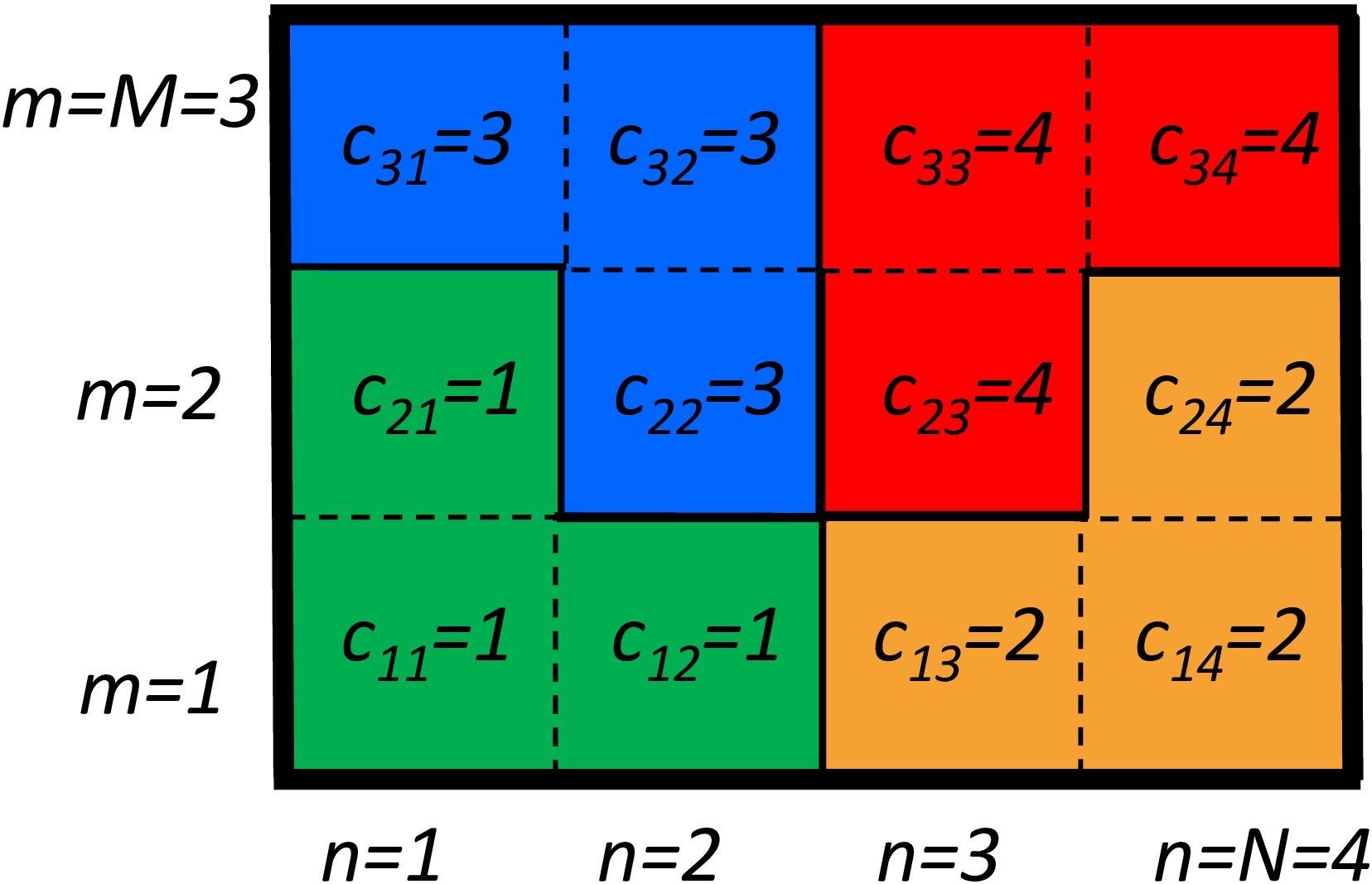}}\tabularnewline
\end{tabular}\end{center}

\begin{center}\textcolor{black}{~\vfill}\end{center}

\begin{center}\textbf{\textcolor{black}{Fig. 3 - N. Anselmi}} \textbf{\textcolor{black}{\emph{et
al.}}}\textbf{\textcolor{black}{,}} \textbf{\textcolor{black}{\emph{{}``}}}\textcolor{black}{A
Self-Replicating Single-Shape Tiling Technique ...''}\end{center}
\newpage

\begin{center}\textcolor{black}{}\begin{tabular}{cc}
\textcolor{black}{\includegraphics[%
  width=0.30\columnwidth]{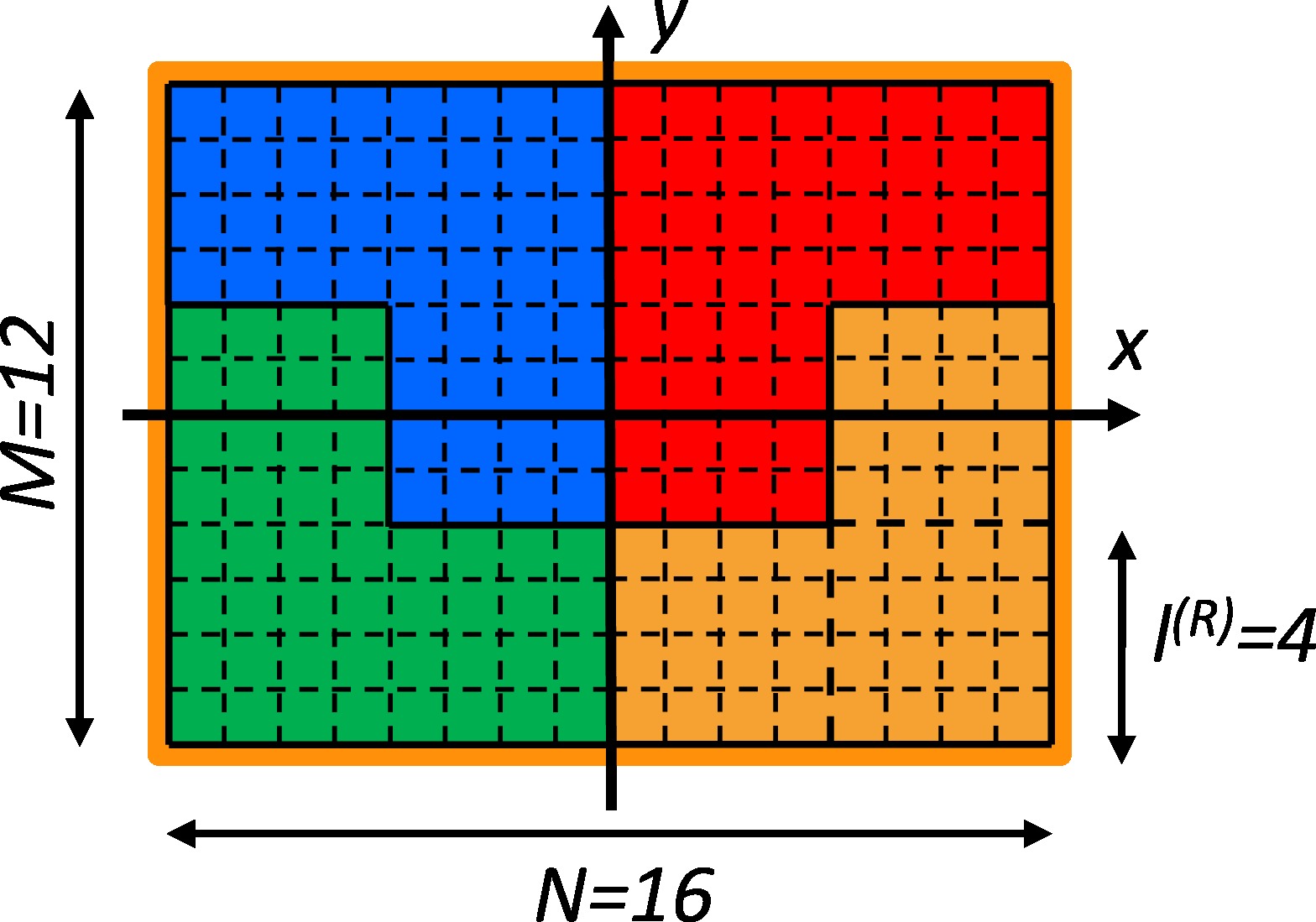}}&
\textcolor{black}{\includegraphics[%
  width=0.30\columnwidth]{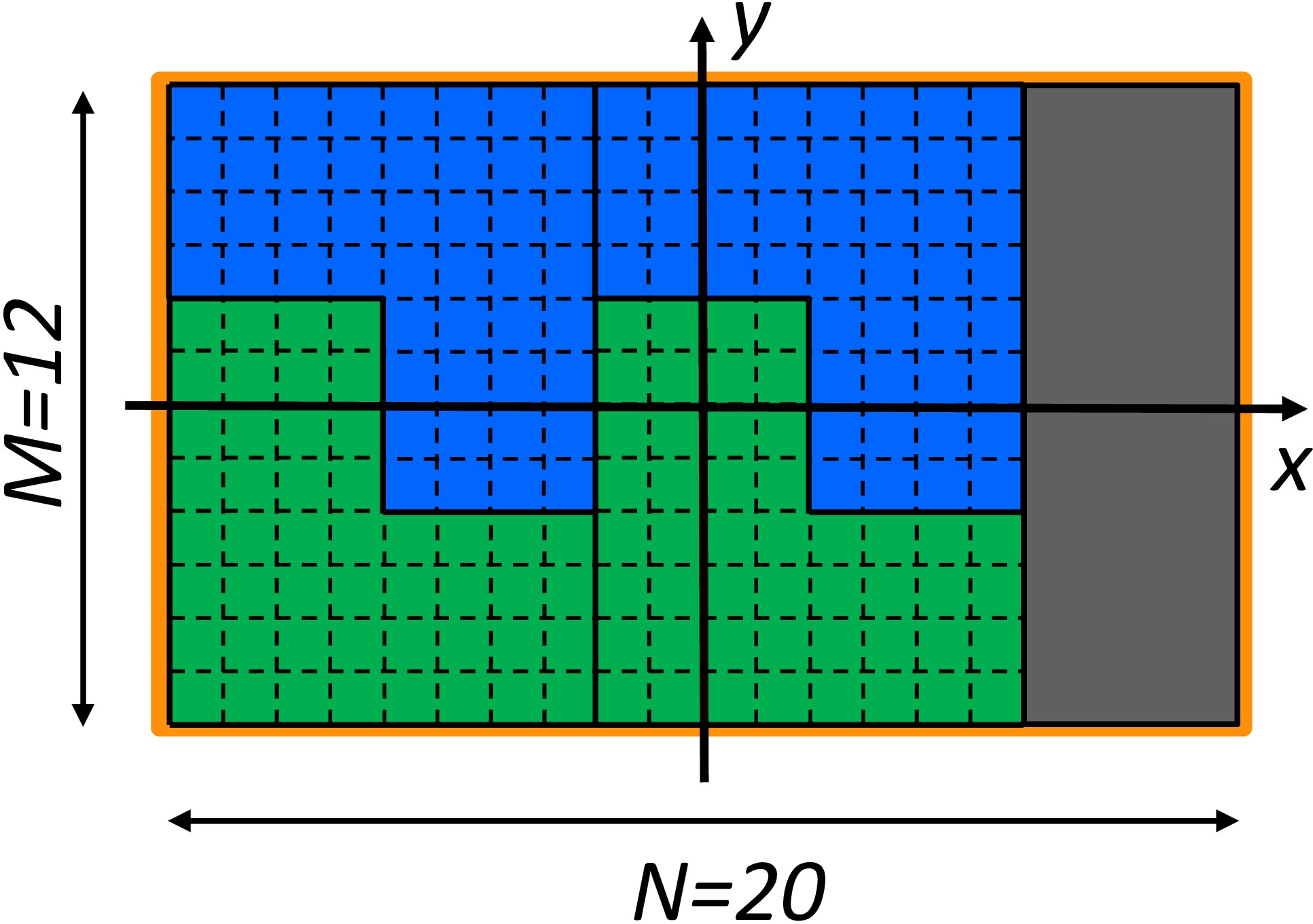}}\tabularnewline
\textcolor{black}{(}\textcolor{black}{\emph{a}}\textcolor{black}{)}&
\textcolor{black}{(}\textcolor{black}{\emph{b}}\textcolor{black}{)}\tabularnewline
&
\tabularnewline
\textcolor{black}{\includegraphics[%
  width=0.20\columnwidth]{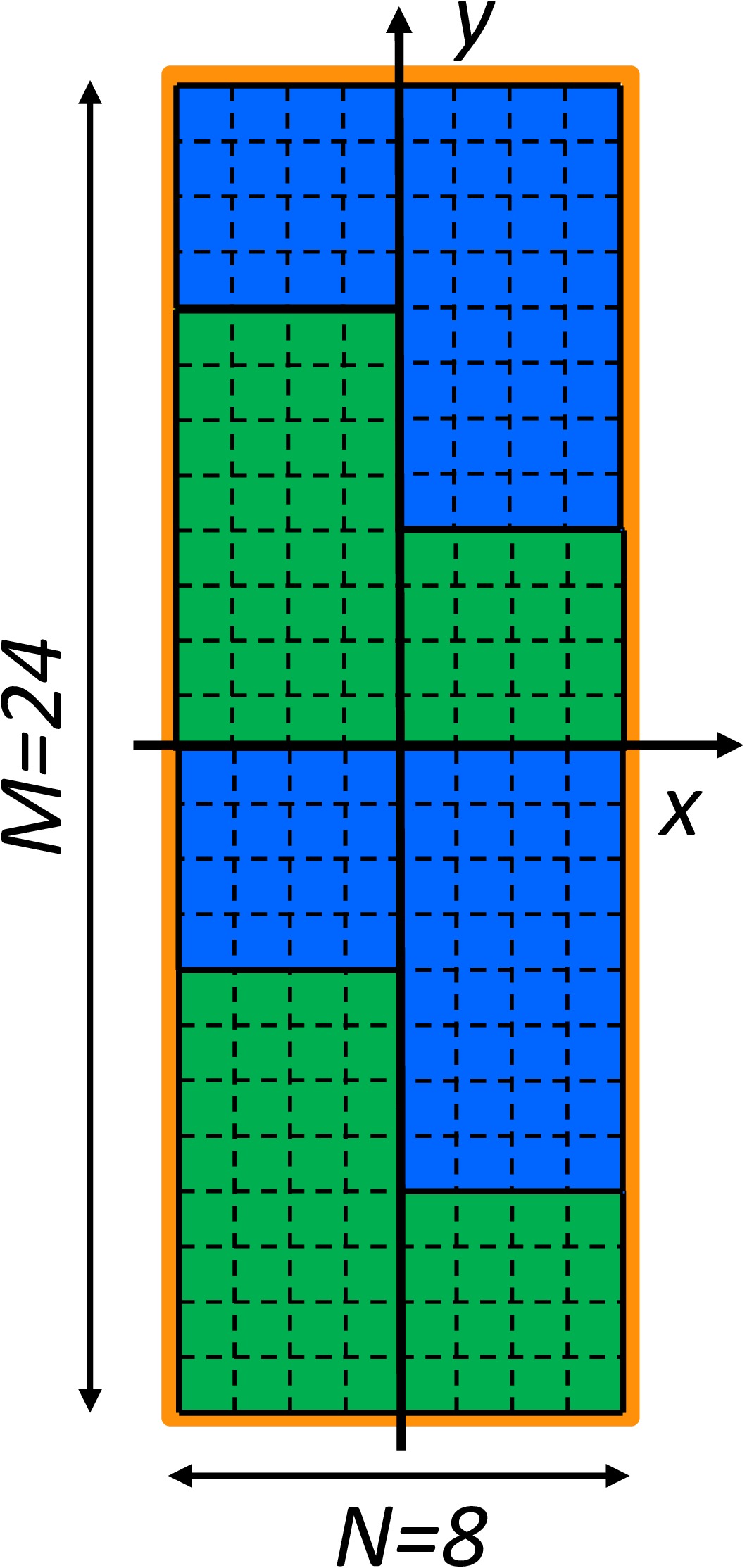}}&
\textcolor{black}{\includegraphics[%
  width=0.20\columnwidth]{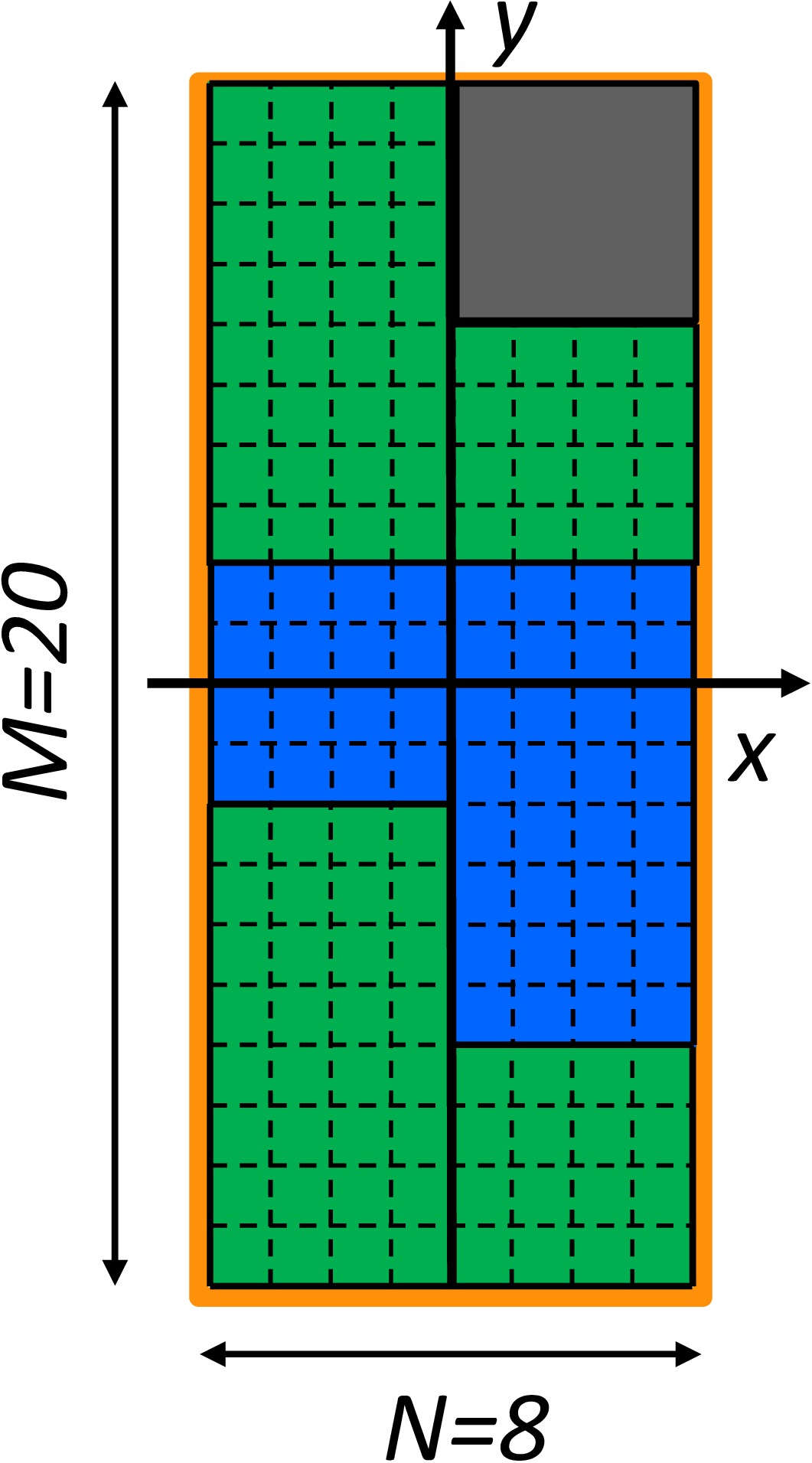}}\tabularnewline
\textcolor{black}{(}\textcolor{black}{\emph{c}}\textcolor{black}{)}&
\textcolor{black}{(}\textcolor{black}{\emph{d}}\textcolor{black}{)}\tabularnewline
&
\tabularnewline
\textcolor{black}{\includegraphics[%
  width=0.40\columnwidth]{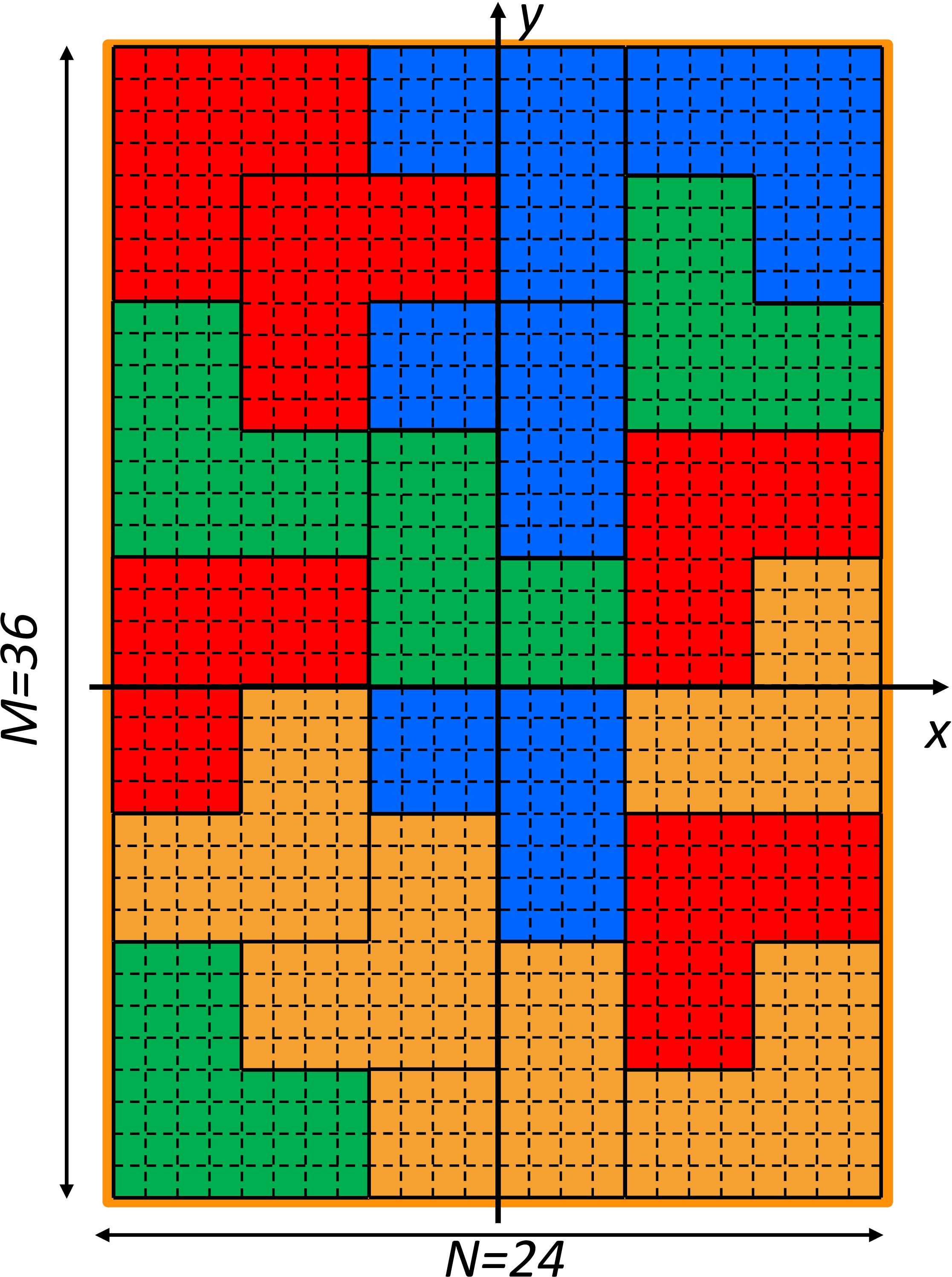}}&
\textcolor{black}{\includegraphics[%
  width=0.40\columnwidth]{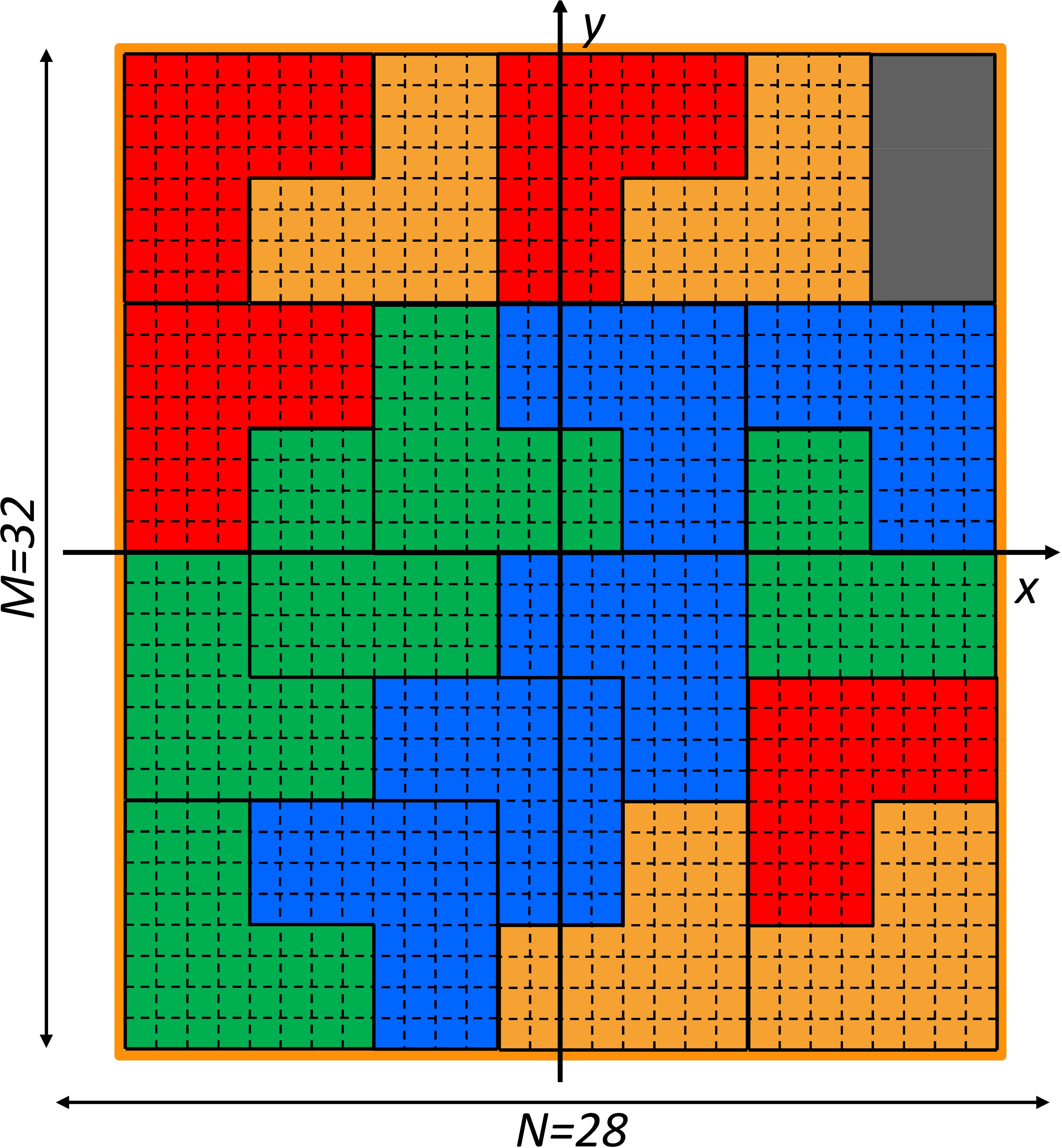}}\tabularnewline
\textcolor{black}{(}\textcolor{black}{\emph{e}}\textcolor{black}{)}&
\textcolor{black}{(}\textcolor{black}{\emph{f}}\textcolor{black}{)}\tabularnewline
\end{tabular}\end{center}

\begin{center}\textbf{\textcolor{black}{Fig. 4 - N. Anselmi}} \textbf{\textcolor{black}{\emph{et
al.}}}\textbf{\textcolor{black}{,}} \textbf{\textcolor{black}{\emph{{}``}}}\textcolor{black}{A
Self-Replicating Single-Shape Tiling Technique ...''}\end{center}

\newpage
\textcolor{black}{~\vfill}

\begin{center}\textcolor{black}{}\begin{tabular}{cc}
\textcolor{black}{\includegraphics[%
  width=0.40\columnwidth]{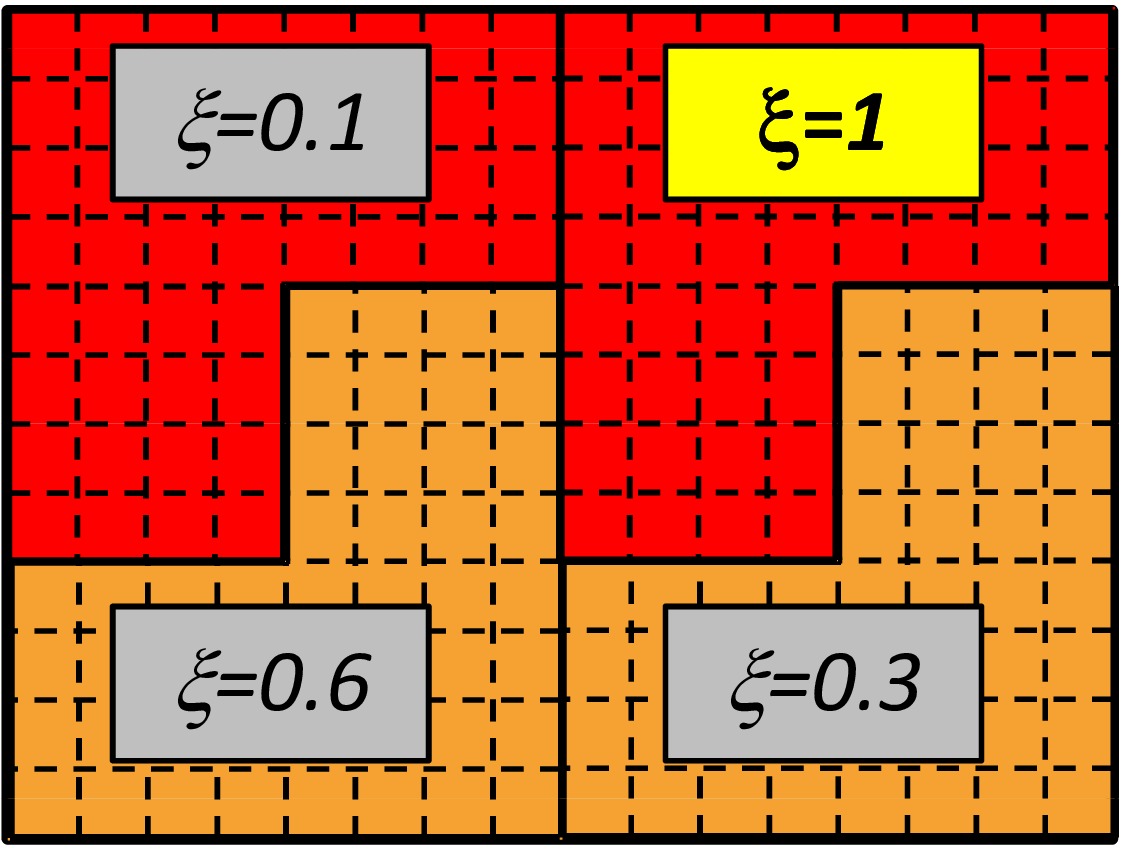}}&
\textcolor{black}{\includegraphics[%
  width=0.40\columnwidth]{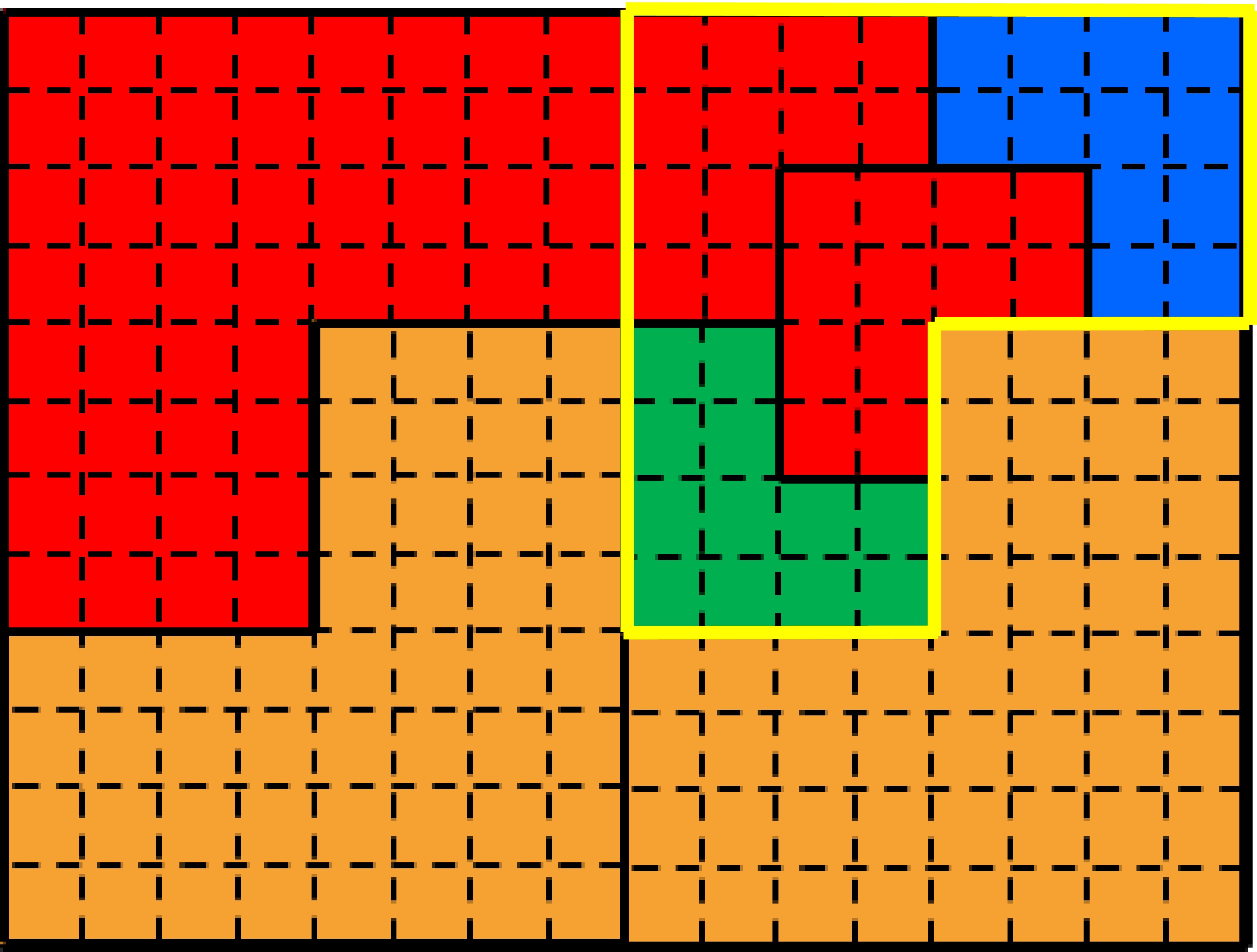}}\tabularnewline
\textcolor{black}{(}\textcolor{black}{\emph{a}}\textcolor{black}{)}&
\textcolor{black}{(}\textcolor{black}{\emph{b}}\textcolor{black}{)}\tabularnewline
&
\tabularnewline
\textcolor{black}{\includegraphics[%
  width=0.40\columnwidth]{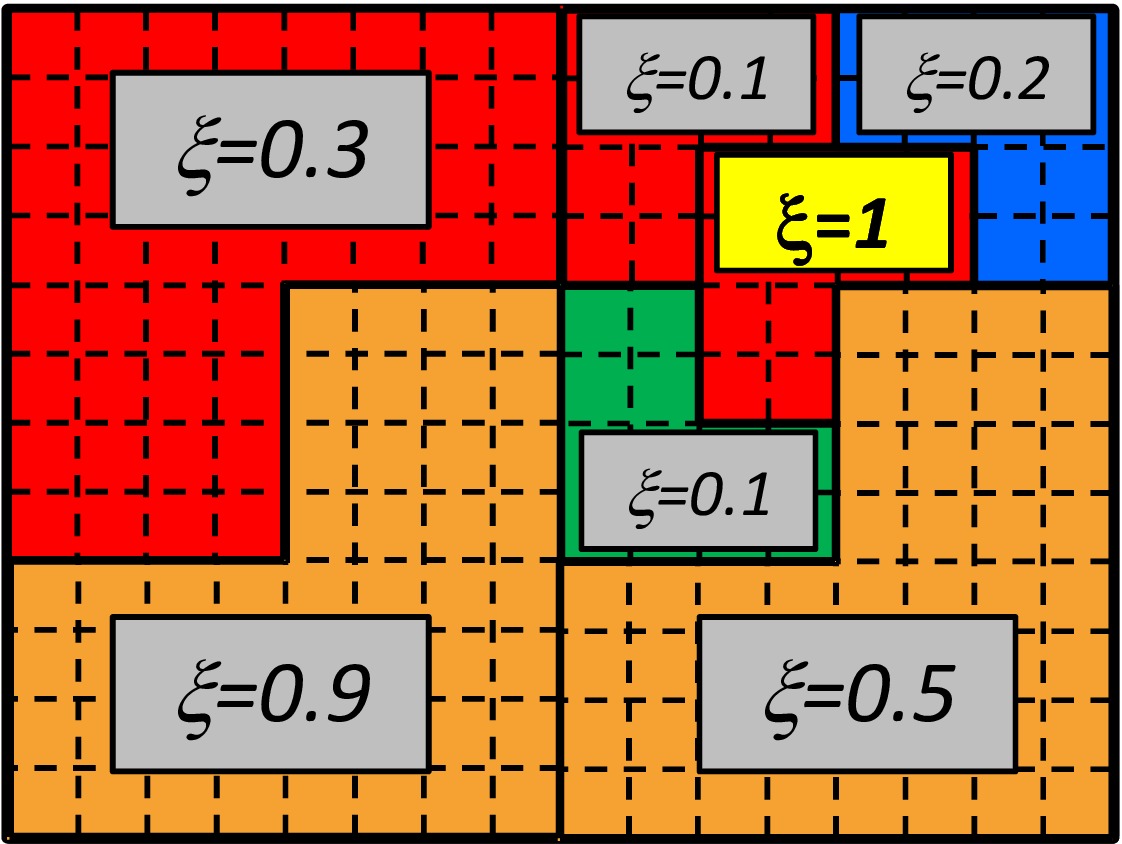}}&
\textcolor{black}{\includegraphics[%
  width=0.40\columnwidth]{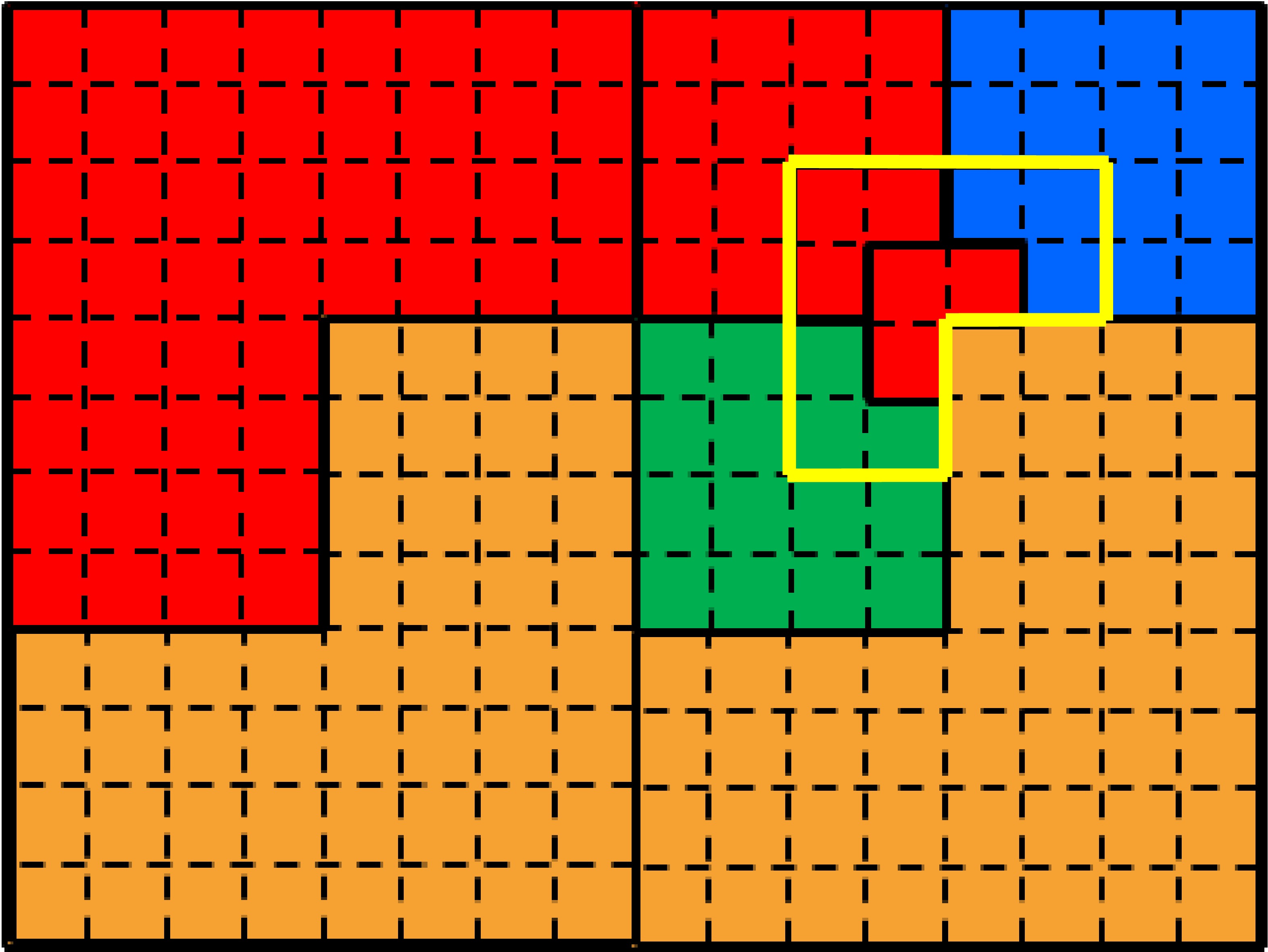}}\tabularnewline
\textcolor{black}{(}\textcolor{black}{\emph{c}}\textcolor{black}{)}&
\textcolor{black}{(}\textcolor{black}{\emph{d}}\textcolor{black}{)}\tabularnewline
&
\tabularnewline
\textcolor{black}{\includegraphics[%
  width=0.40\columnwidth]{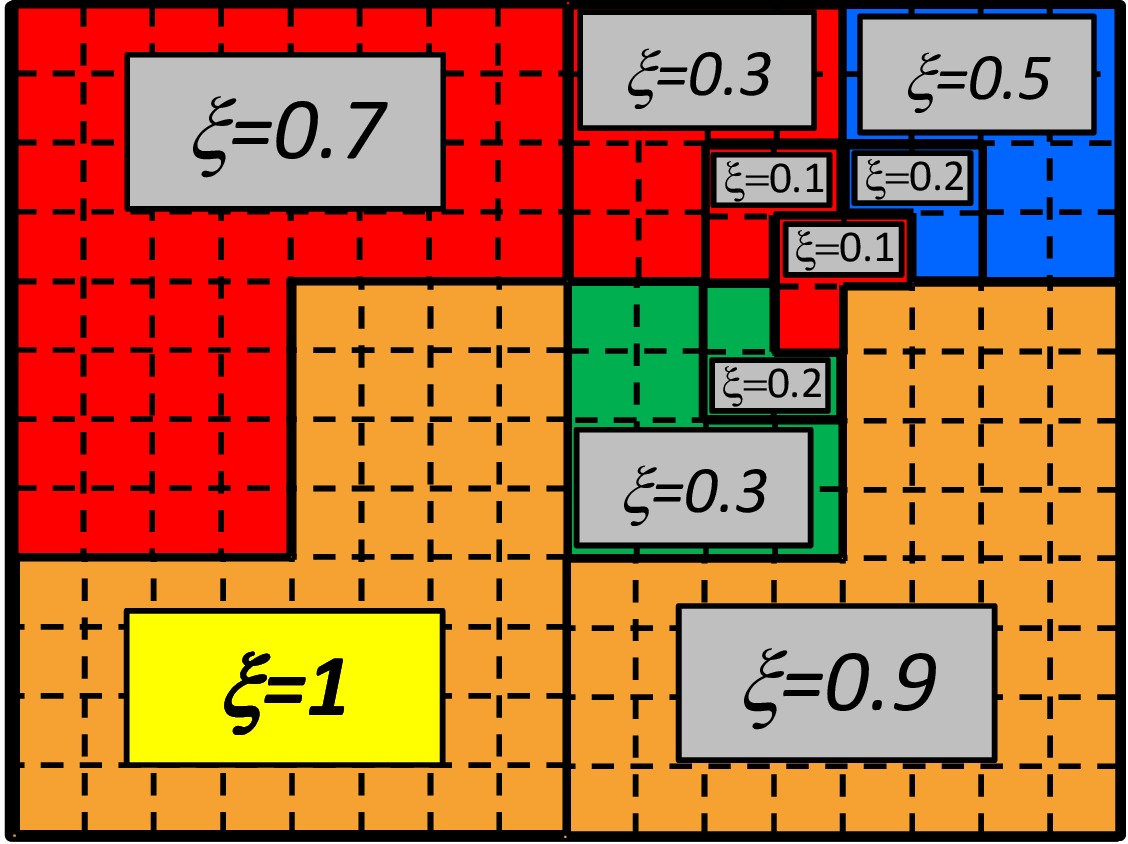}}&
\textcolor{black}{\includegraphics[%
  width=0.40\columnwidth]{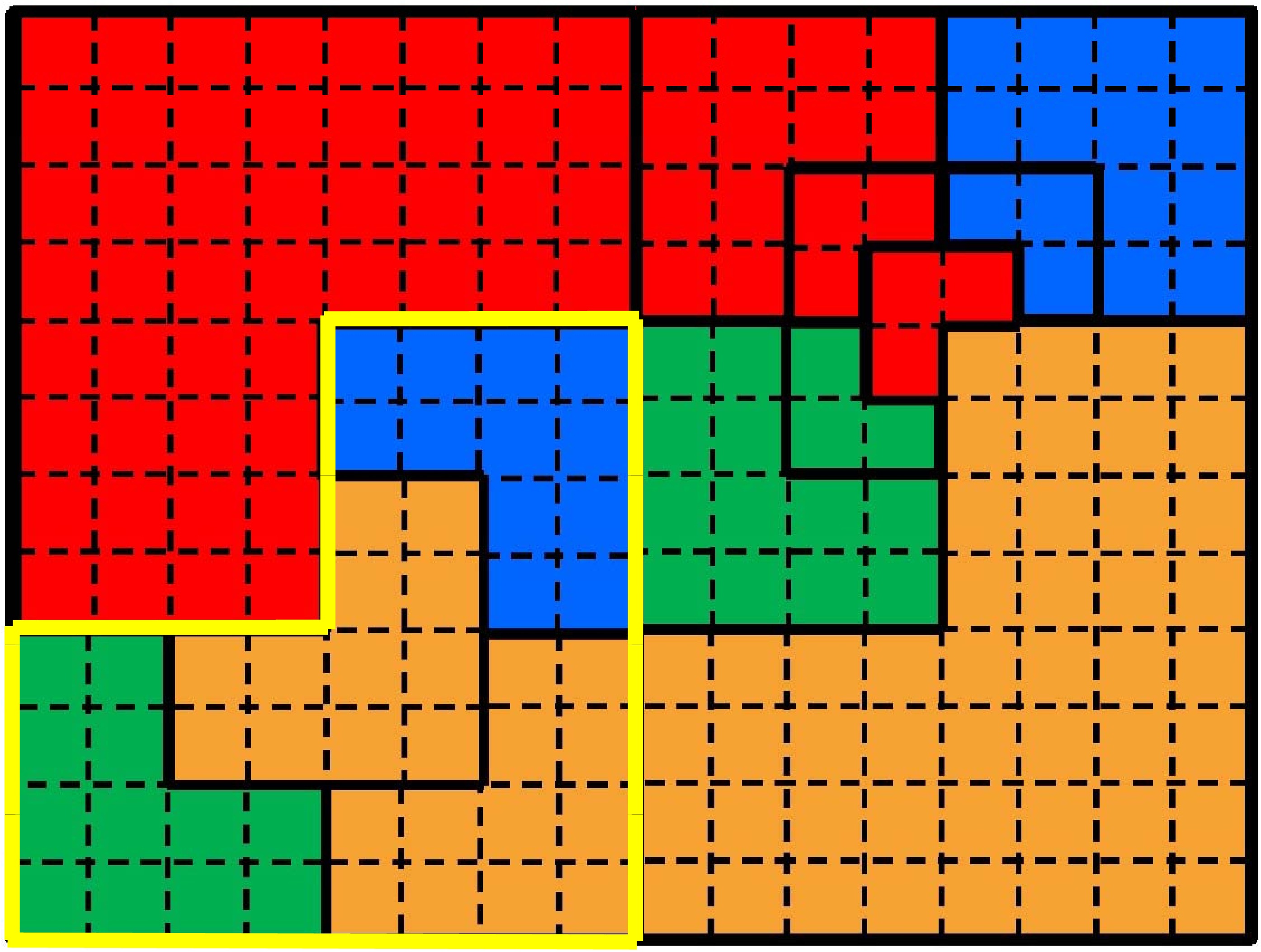}}\tabularnewline
\textcolor{black}{(}\textcolor{black}{\emph{e}}\textcolor{black}{)}&
\textcolor{black}{(}\textcolor{black}{\emph{f}}\textcolor{black}{)}\tabularnewline
\end{tabular}\end{center}

\begin{center}\textcolor{black}{~\vfill}\end{center}

\begin{center}\textbf{\textcolor{black}{Fig. 5 - N. Anselmi}} \textbf{\textcolor{black}{\emph{et
al.}}}\textbf{\textcolor{black}{,}} \textbf{\textcolor{black}{\emph{{}``}}}\textcolor{black}{A
Self-Replicating Single-Shape Tiling Technique ...''}\end{center}

\newpage
\begin{center}\textcolor{black}{~\vfill}\end{center}

\begin{center}\textcolor{black}{}\begin{tabular}{c}
\textcolor{black}{\includegraphics[%
  width=0.48\columnwidth]{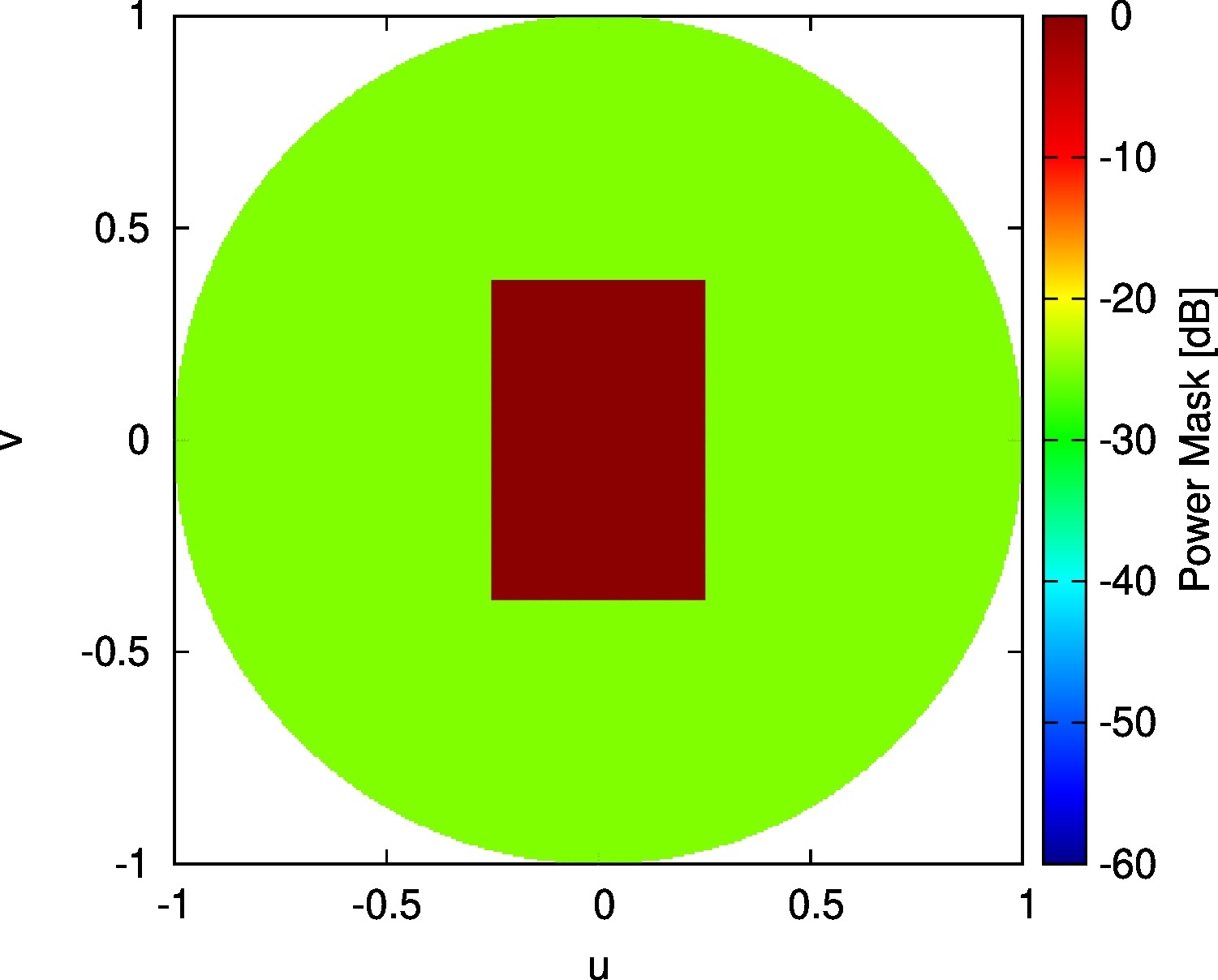}}\tabularnewline
\textcolor{black}{(}\textcolor{black}{\emph{a}}\textcolor{black}{)}\tabularnewline
\textcolor{black}{\includegraphics[%
  width=0.48\columnwidth]{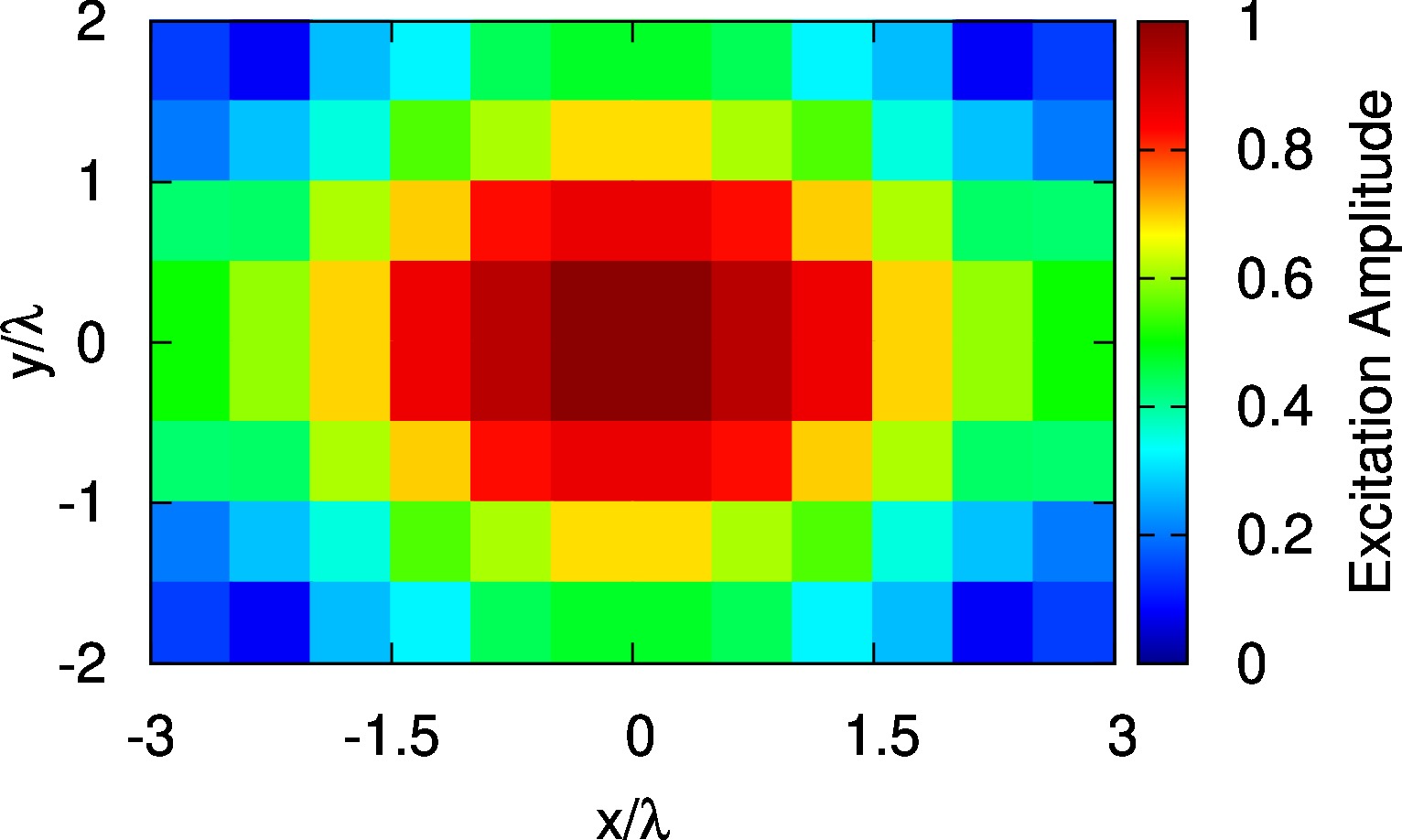}}\tabularnewline
\textcolor{black}{(}\textcolor{black}{\emph{b}}\textcolor{black}{)}\tabularnewline
\end{tabular}\end{center}

\begin{center}\textcolor{black}{~\vfill}\end{center}

\begin{center}\textbf{\textcolor{black}{Fig. 6 - N. Anselmi}} \textbf{\textcolor{black}{\emph{et
al.}}}\textbf{\textcolor{black}{,}} \textbf{\textcolor{black}{\emph{{}``}}}\textcolor{black}{A
Self-Replicating Single-Shape Tiling Technique ...''}\end{center}

\newpage
\begin{center}\textcolor{black}{~\vfill}\end{center}

\begin{center}\textcolor{black}{}\begin{tabular}{cc}
\multicolumn{2}{c}{\textcolor{black}{\includegraphics[%
  width=0.60\columnwidth]{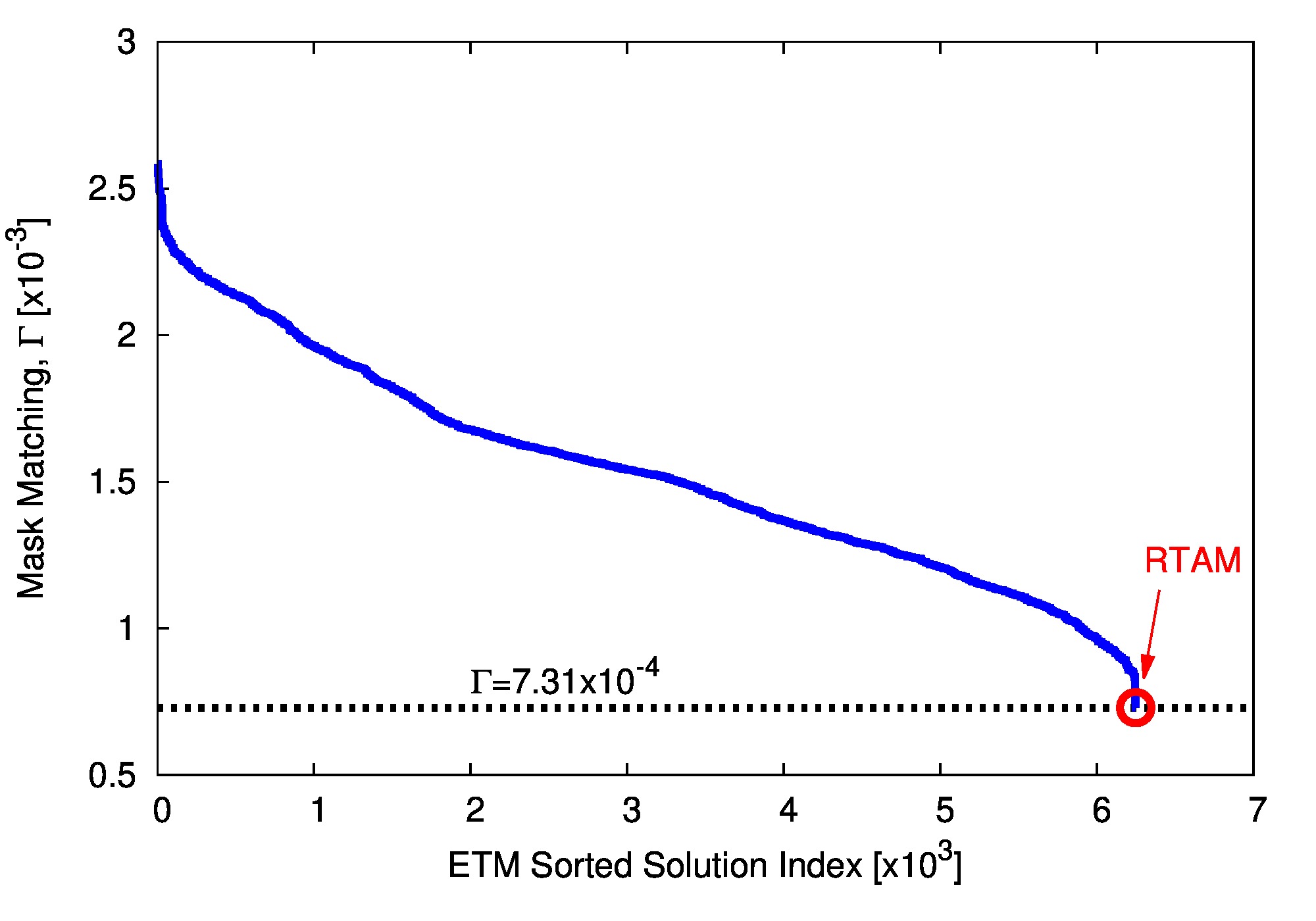}}}\tabularnewline
\multicolumn{2}{c}{\textcolor{black}{(}\textcolor{black}{\emph{a}}\textcolor{black}{)}}\tabularnewline
\textcolor{black}{\includegraphics[%
  width=0.48\columnwidth]{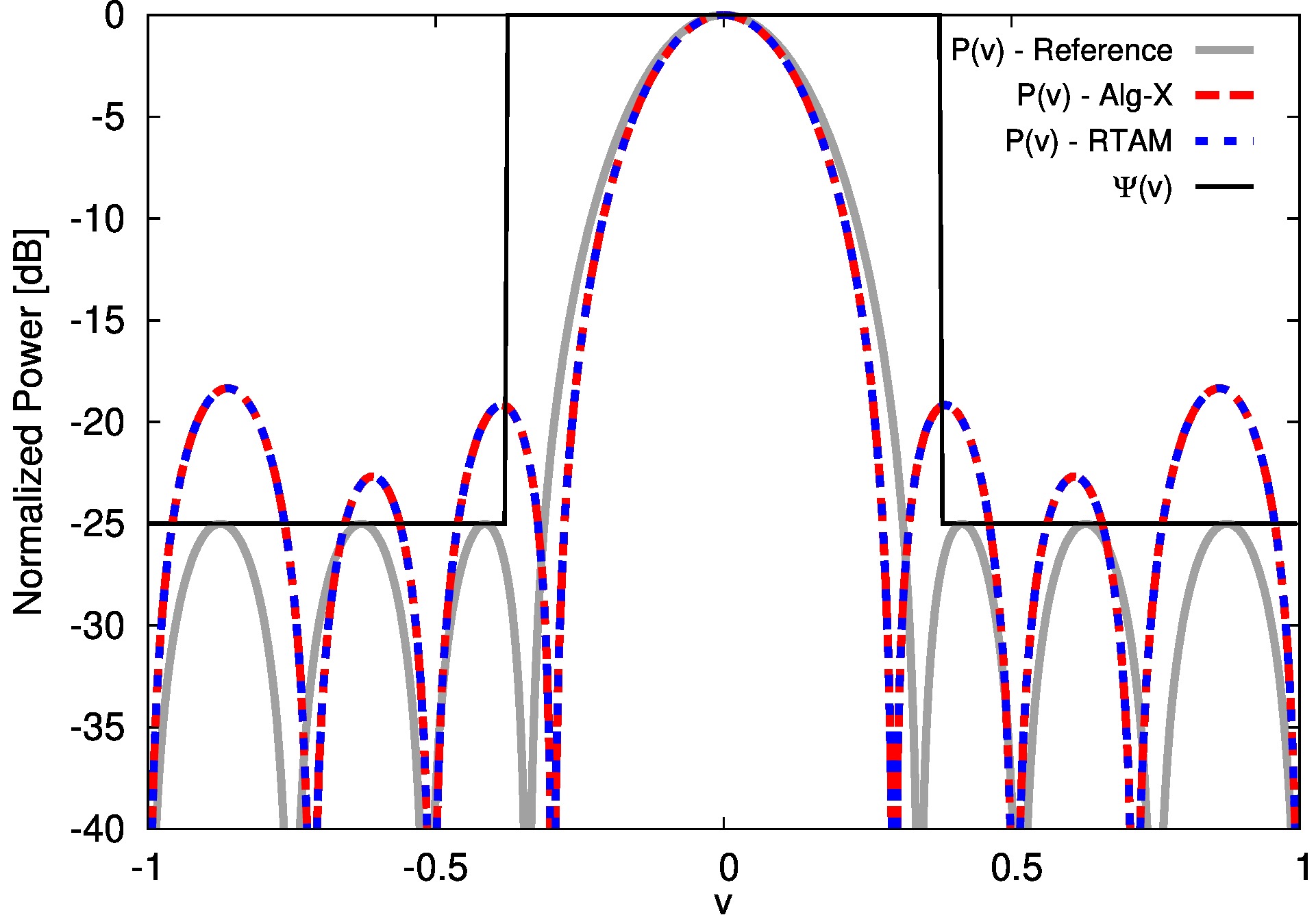}}&
\textcolor{black}{\includegraphics[%
  width=0.48\columnwidth]{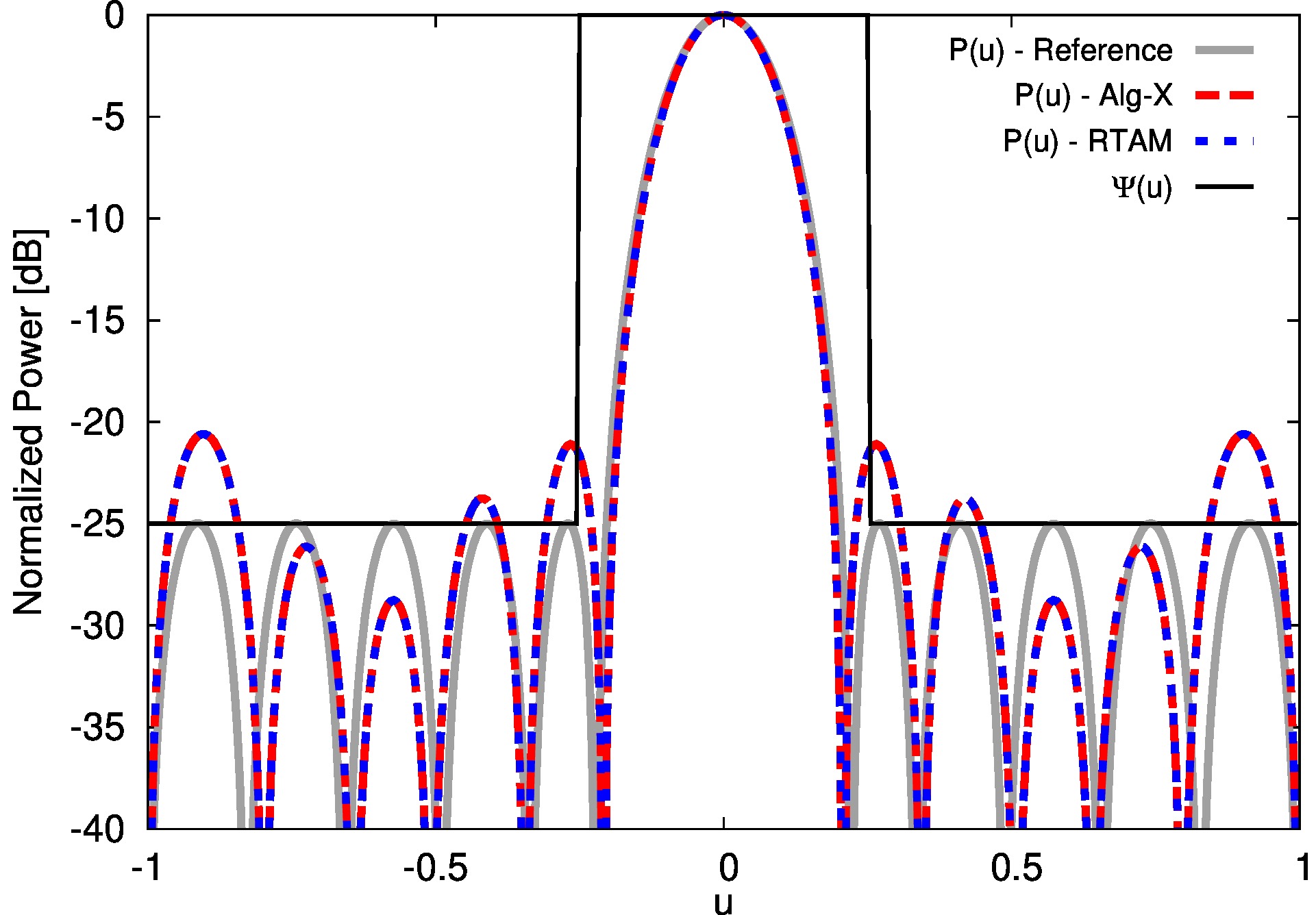}}\tabularnewline
\textcolor{black}{(}\textcolor{black}{\emph{b}}\textcolor{black}{)}&
\textcolor{black}{(}\textcolor{black}{\emph{c}}\textcolor{black}{)}\tabularnewline
\end{tabular}\end{center}

\begin{center}\textcolor{black}{~\vfill}\end{center}

\begin{center}\textbf{\textcolor{black}{Fig. 7 - N. Anselmi}} \textbf{\textcolor{black}{\emph{et
al.}}}\textbf{\textcolor{black}{,}} \textbf{\textcolor{black}{\emph{{}``}}}\textcolor{black}{A
Self-Replicating Single-Shape Tiling Technique ...''}\end{center}

\newpage
\textcolor{black}{~\vfill}

\begin{center}\textcolor{black}{}\begin{tabular}{cc}
\textcolor{black}{\includegraphics[%
  width=0.40\columnwidth]{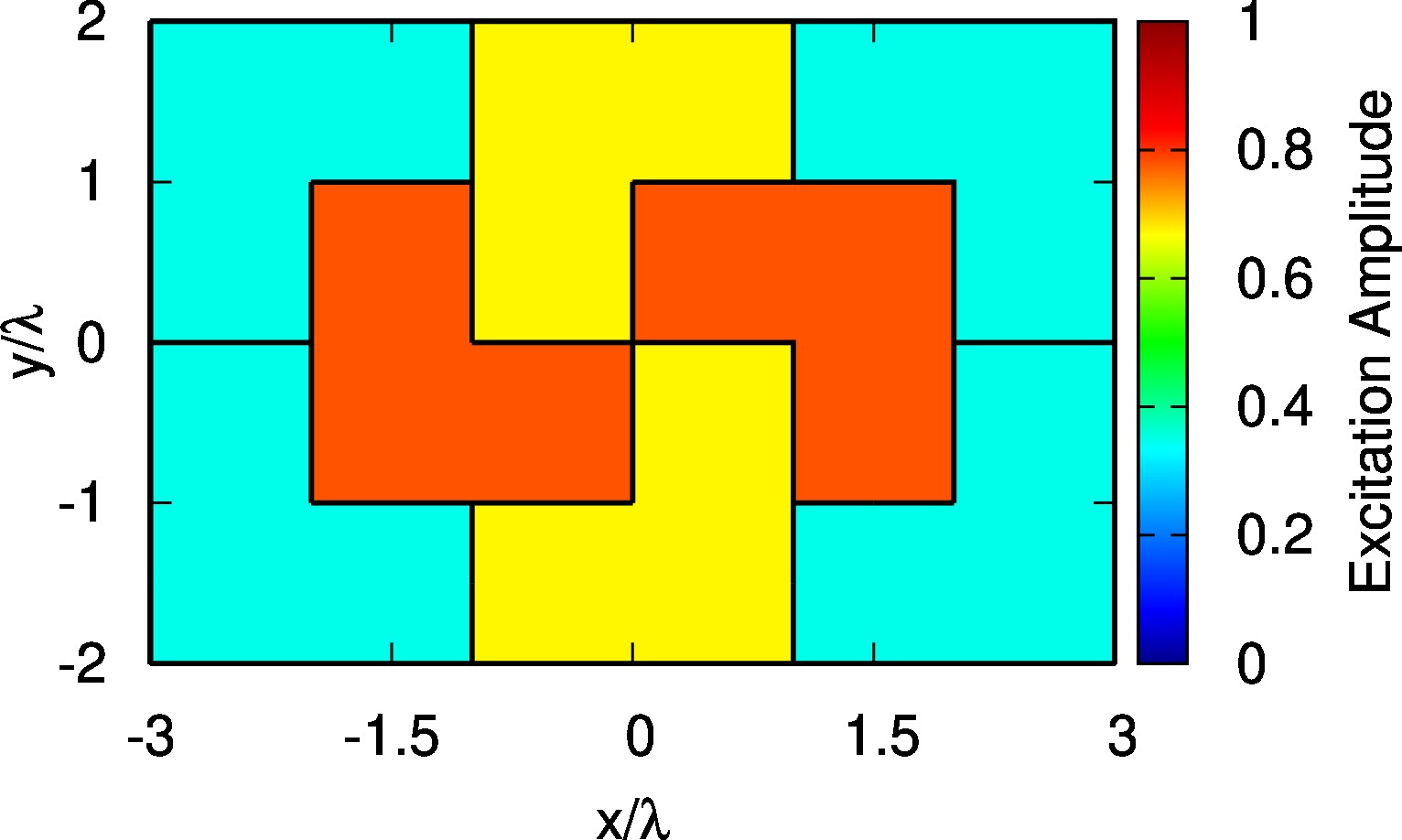}}&
\textcolor{black}{\includegraphics[%
  width=0.40\columnwidth]{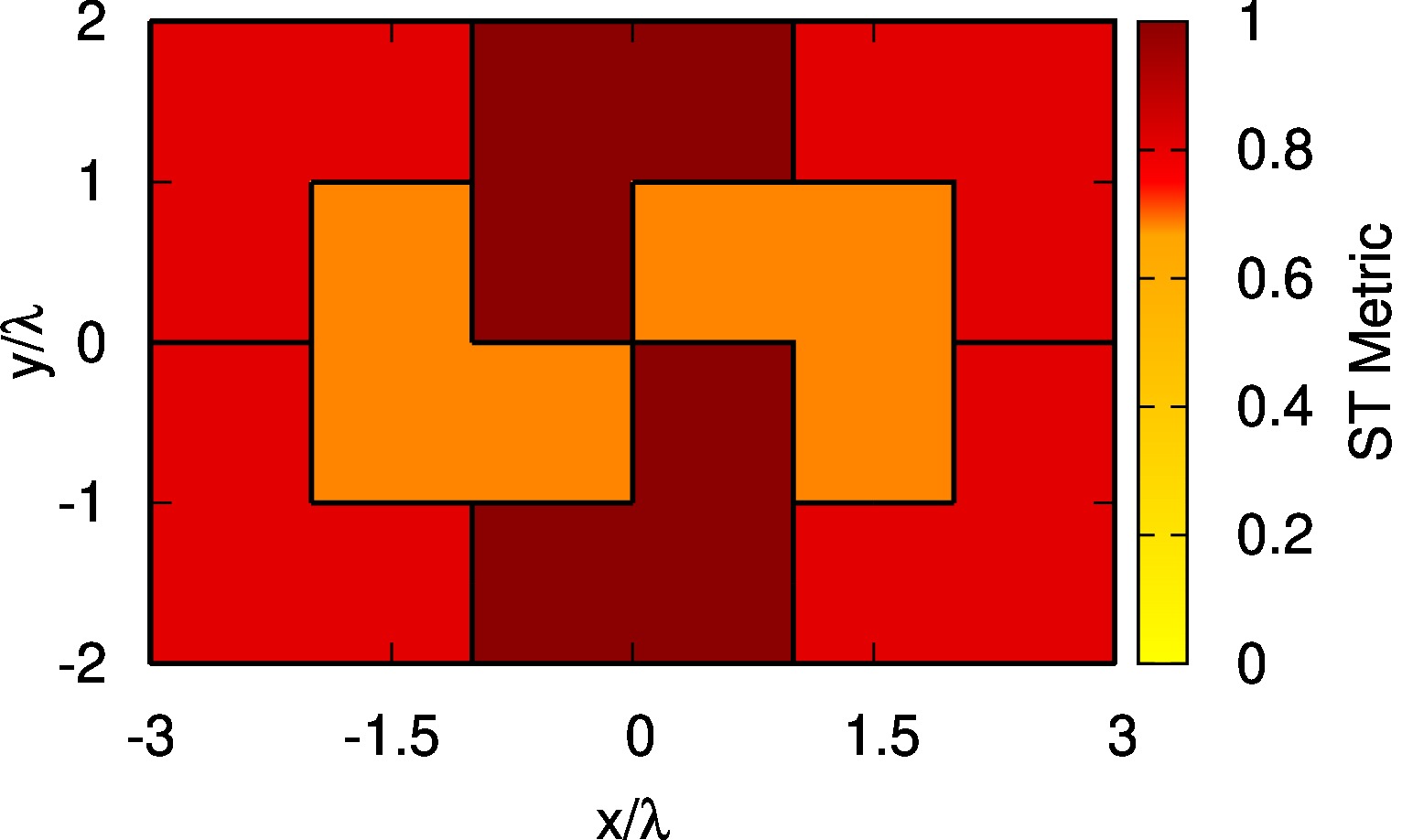}}\tabularnewline
\textcolor{black}{(}\textcolor{black}{\emph{a}}\textcolor{black}{)}&
\textcolor{black}{(}\textcolor{black}{\emph{b}}\textcolor{black}{)}\tabularnewline
\textcolor{black}{\includegraphics[%
  width=0.40\columnwidth]{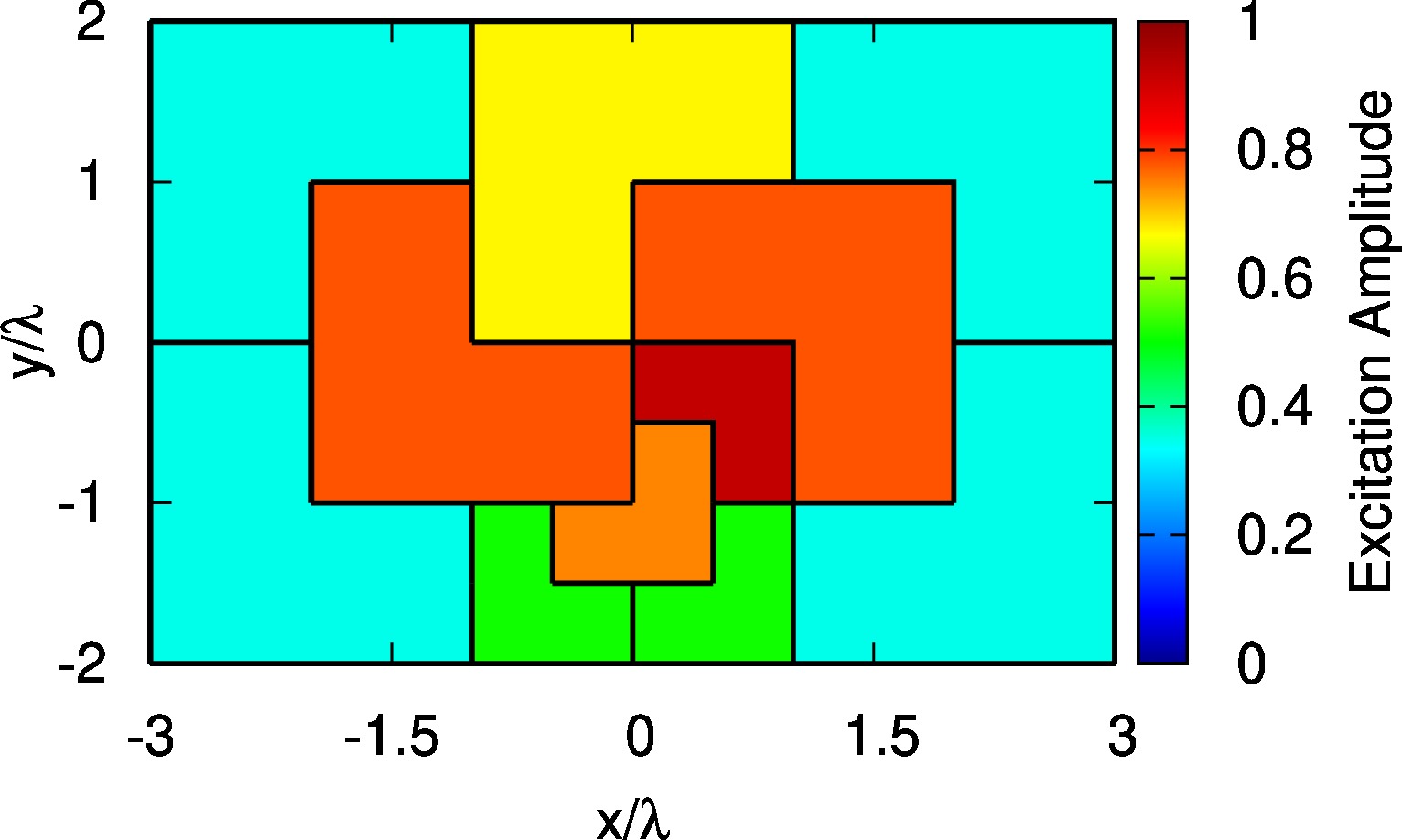}}&
\textcolor{black}{\includegraphics[%
  width=0.40\columnwidth]{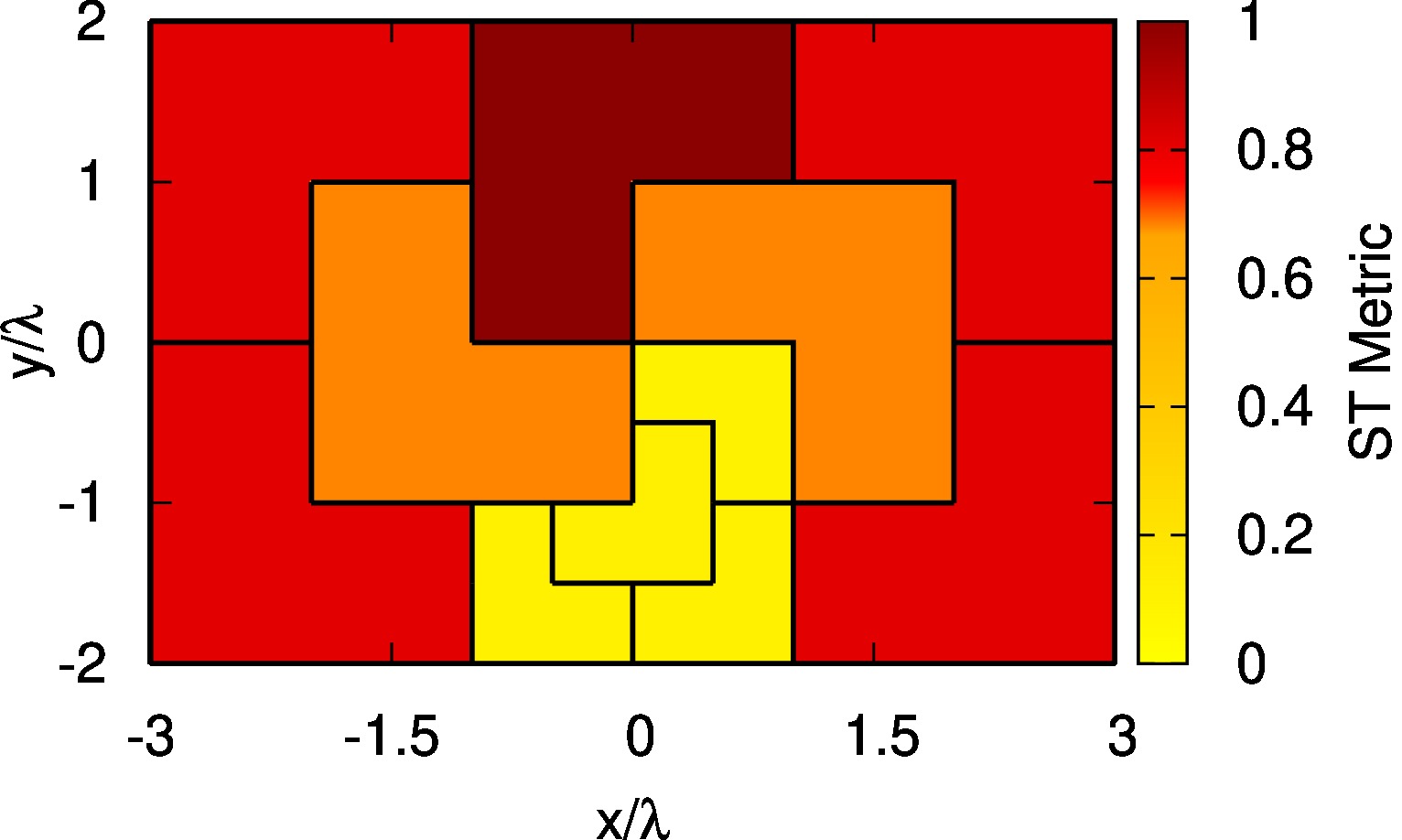}}\tabularnewline
\textcolor{black}{(}\textcolor{black}{\emph{c}}\textcolor{black}{)}&
\textcolor{black}{(}\textcolor{black}{\emph{d}}\textcolor{black}{)}\tabularnewline
\multicolumn{2}{c}{\textcolor{black}{\includegraphics[%
  width=0.40\columnwidth]{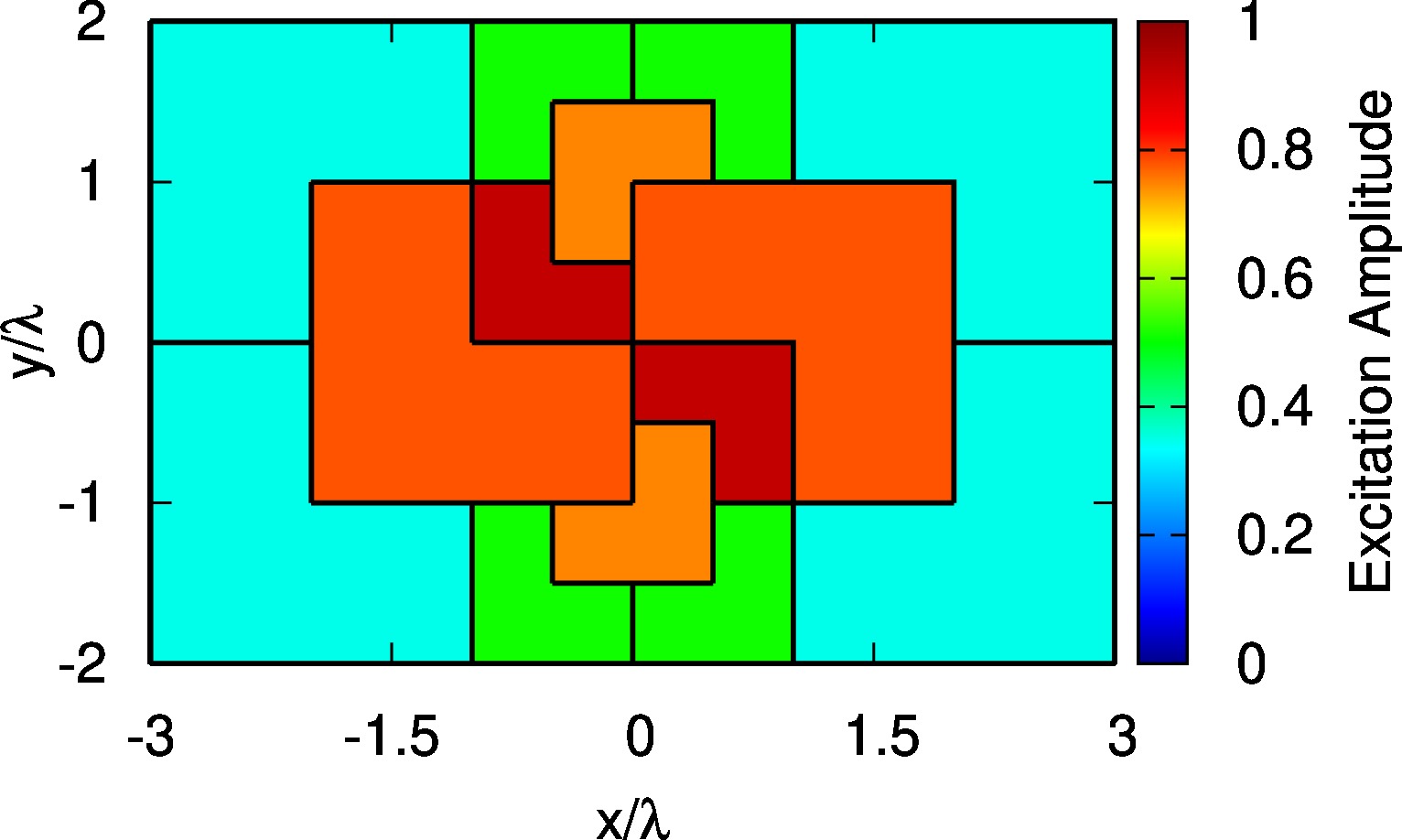}}}\tabularnewline
\multicolumn{2}{c}{\textcolor{black}{(}\textcolor{black}{\emph{e}}\textcolor{black}{)}}\tabularnewline
\end{tabular}\end{center}

\begin{center}\textcolor{black}{~\vfill}\end{center}

\begin{center}\textbf{\textcolor{black}{Fig. 8 - N. Anselmi}} \textbf{\textcolor{black}{\emph{et
al.}}}\textbf{\textcolor{black}{,}} \textbf{\textcolor{black}{\emph{{}``}}}\textcolor{black}{A
Self-Replicating Single-Shape Tiling Technique ...''}\end{center}

\newpage
\begin{center}\textcolor{black}{~\vfill}\end{center}

\begin{center}\textcolor{black}{}\begin{tabular}{c}
\textcolor{black}{\includegraphics[%
  width=0.48\columnwidth]{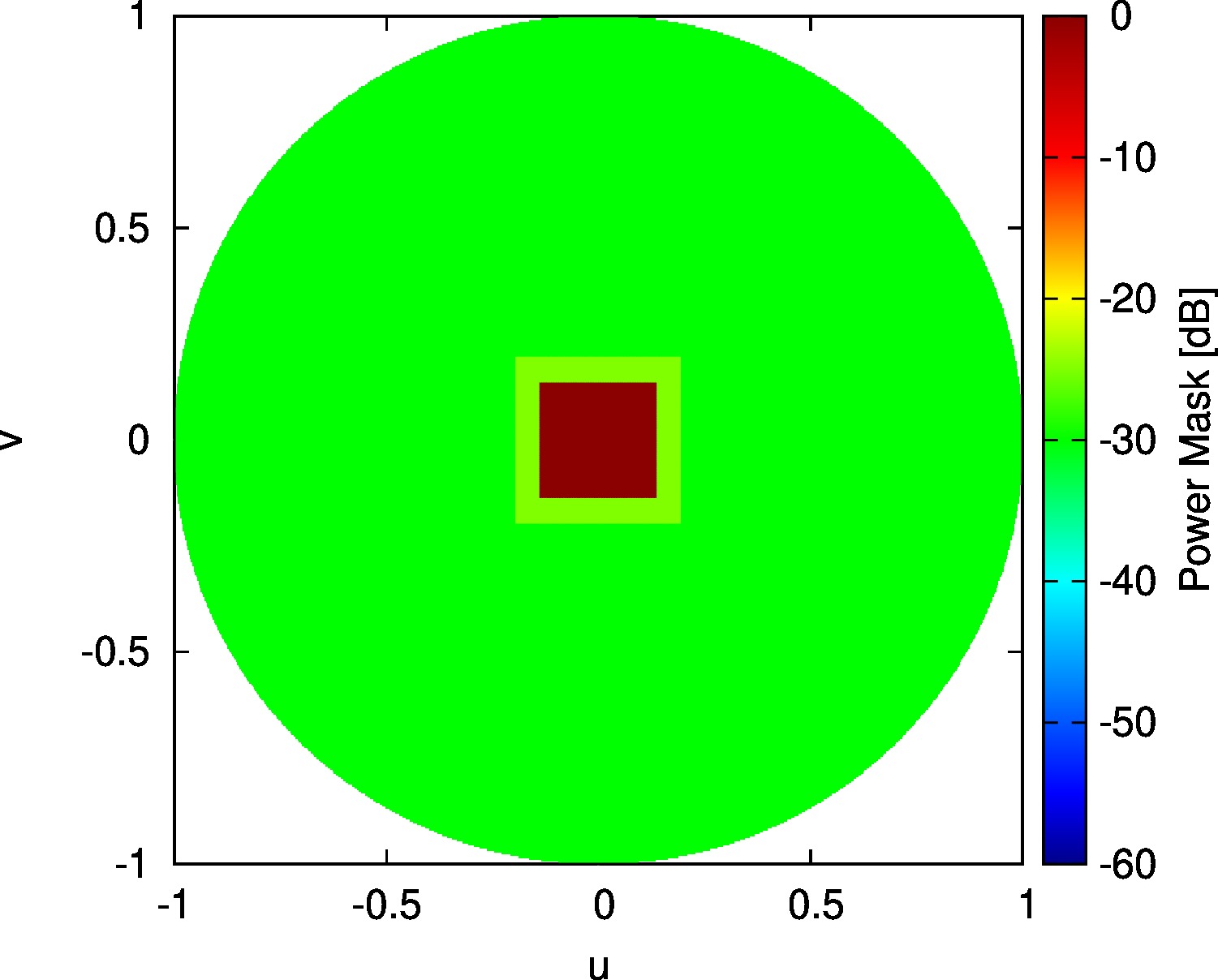}}\tabularnewline
\textcolor{black}{(}\textcolor{black}{\emph{a}}\textcolor{black}{)}\tabularnewline
\textcolor{black}{\includegraphics[%
  width=0.48\columnwidth]{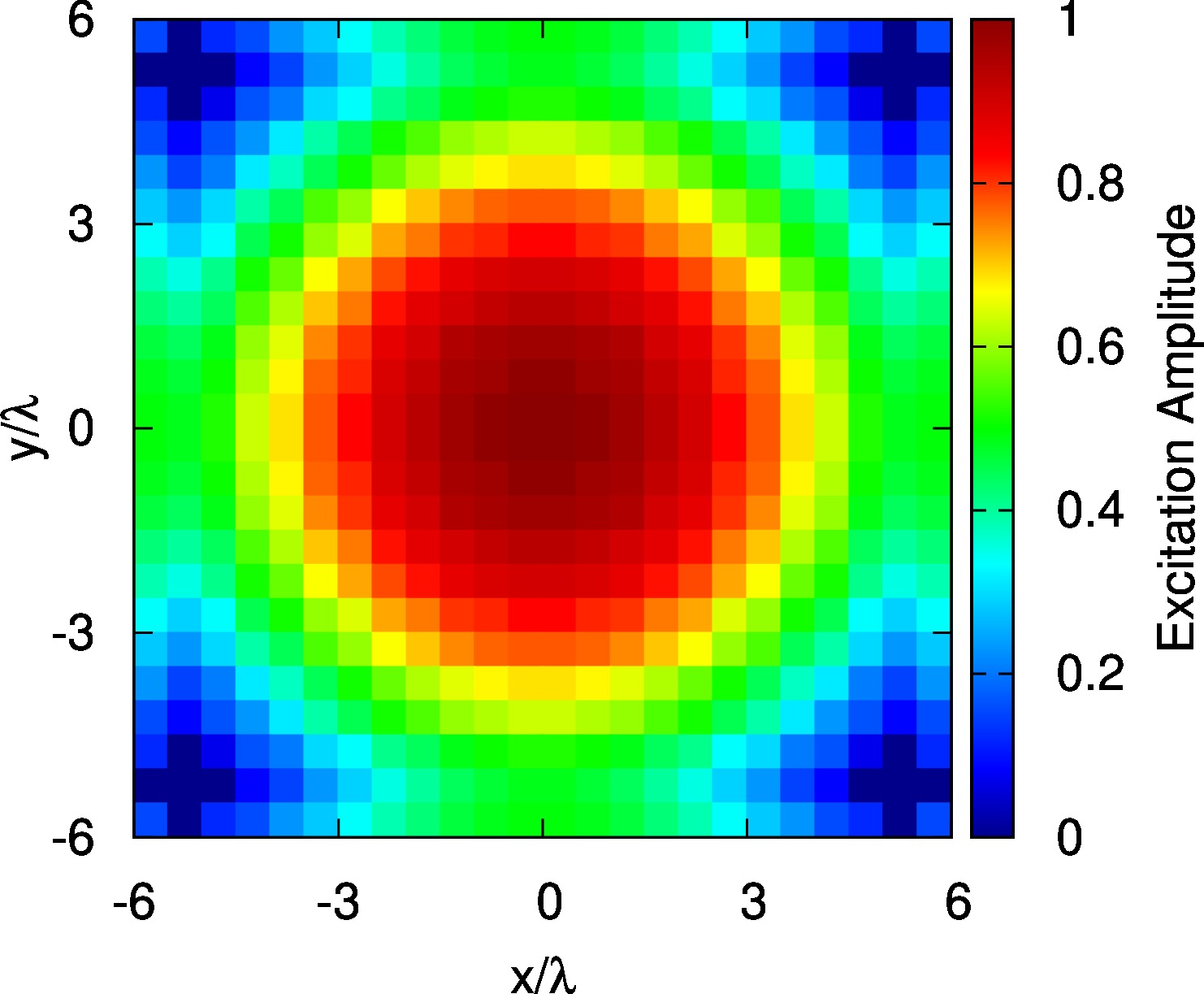}}\tabularnewline
\textcolor{black}{(}\textcolor{black}{\emph{b}}\textcolor{black}{)}\tabularnewline
\end{tabular}\end{center}

\begin{center}\textcolor{black}{~\vfill}\end{center}

\begin{center}\textbf{\textcolor{black}{Fig. 9 - N. Anselmi}} \textbf{\textcolor{black}{\emph{et
al.}}}\textbf{\textcolor{black}{,}} \textbf{\textcolor{black}{\emph{{}``}}}\textcolor{black}{A
Self-Replicating Single-Shape Tiling Technique ...''}\end{center}

\newpage
\begin{center}\textcolor{black}{~\vfill}\end{center}

\begin{center}\textcolor{black}{}\begin{tabular}{c}
\textcolor{black}{\includegraphics[%
  width=0.80\columnwidth]{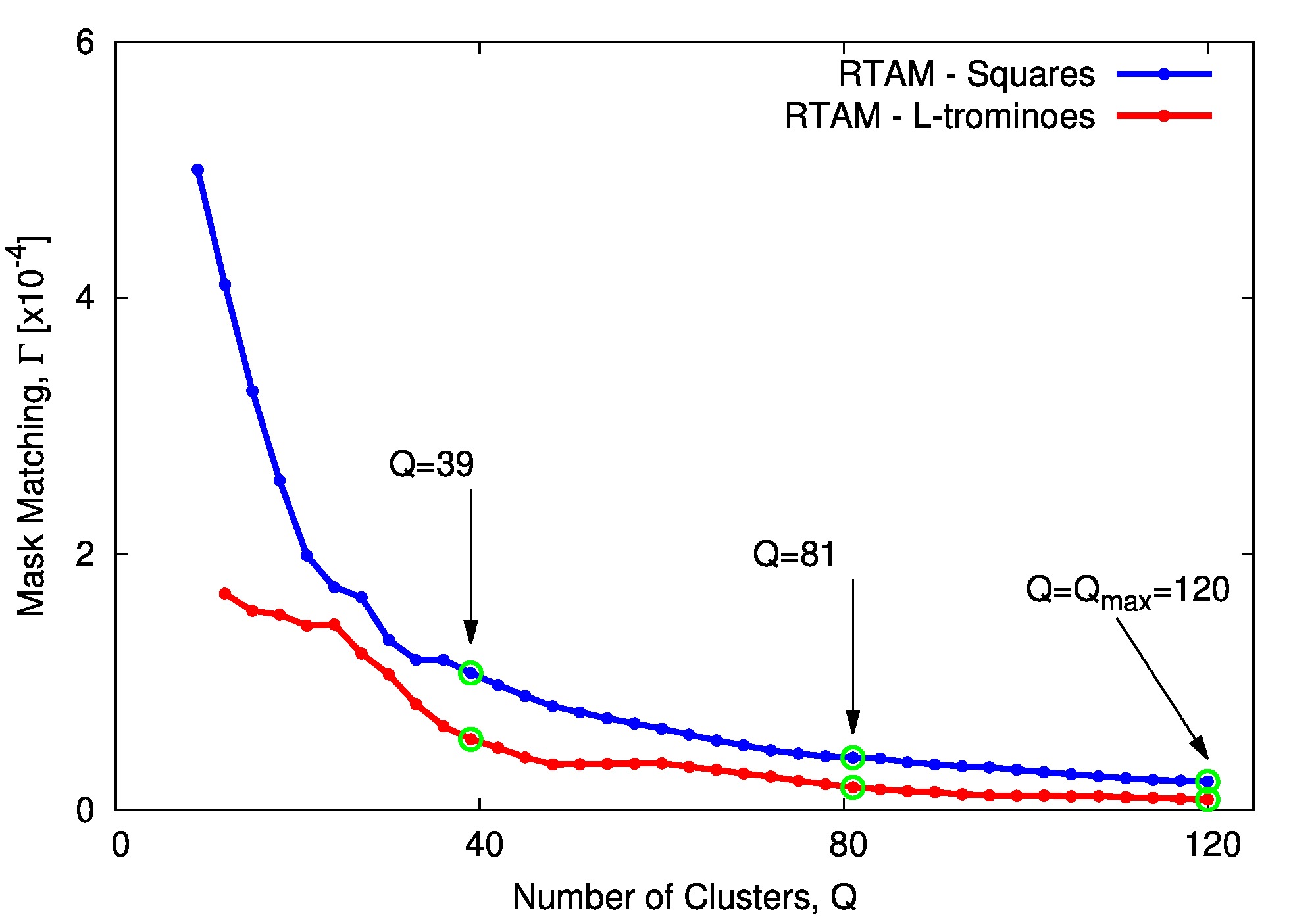}}\tabularnewline
\end{tabular}\end{center}

\begin{center}\textcolor{black}{~\vfill}\end{center}

\begin{center}\textbf{\textcolor{black}{Fig. 10 - N. Anselmi}} \textbf{\textcolor{black}{\emph{et
al.}}}\textbf{\textcolor{black}{,}} \textbf{\textcolor{black}{\emph{{}``}}}\textcolor{black}{A
Self-Replicating Single-Shape Tiling Technique ...''}\end{center}

\newpage
\begin{center}\textcolor{black}{~\vfill}\end{center}

\begin{center}\textcolor{black}{}\begin{tabular}{cc}
\textcolor{black}{\includegraphics[%
  width=0.45\columnwidth]{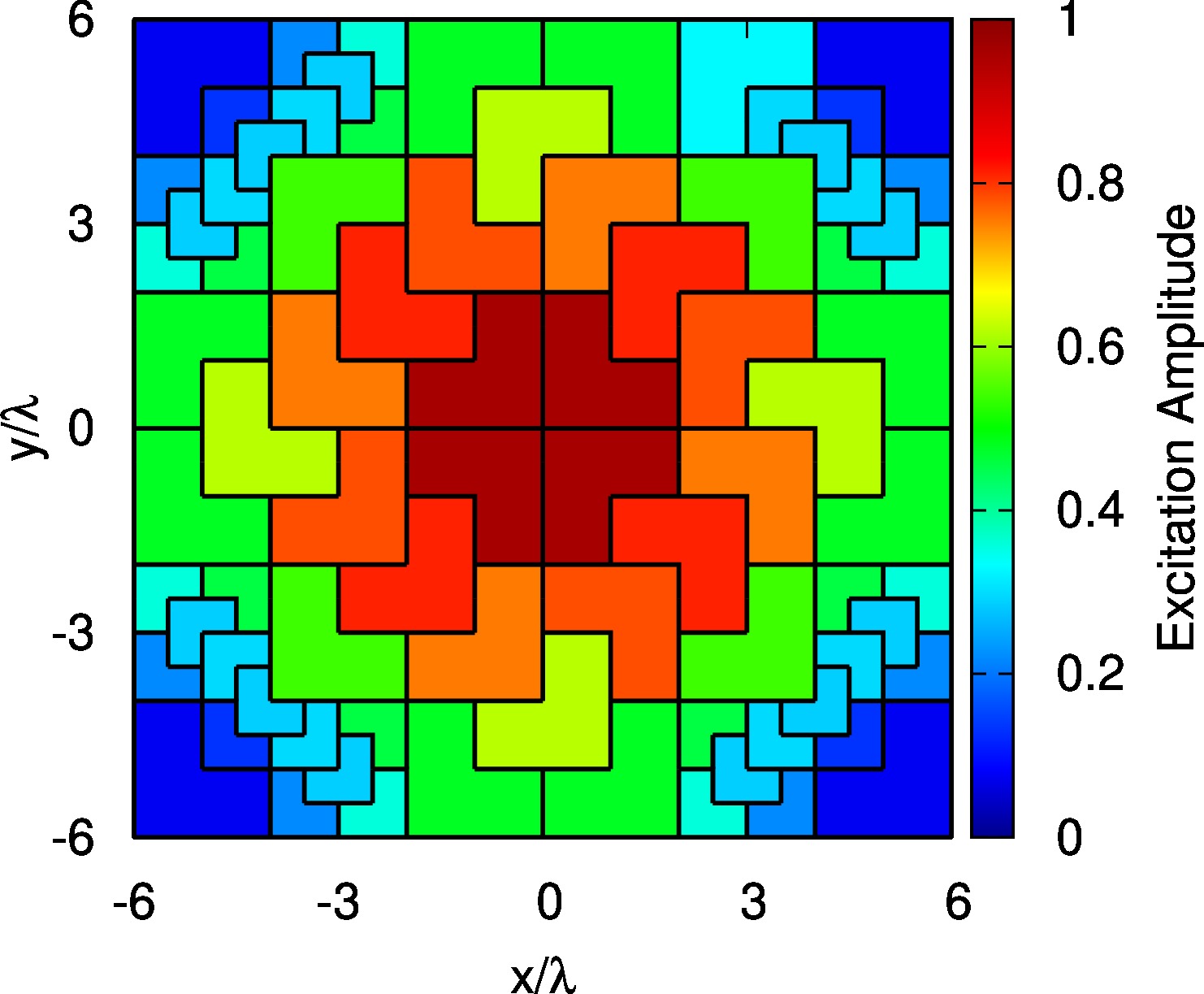}}&
\textcolor{black}{\includegraphics[%
  width=0.45\columnwidth]{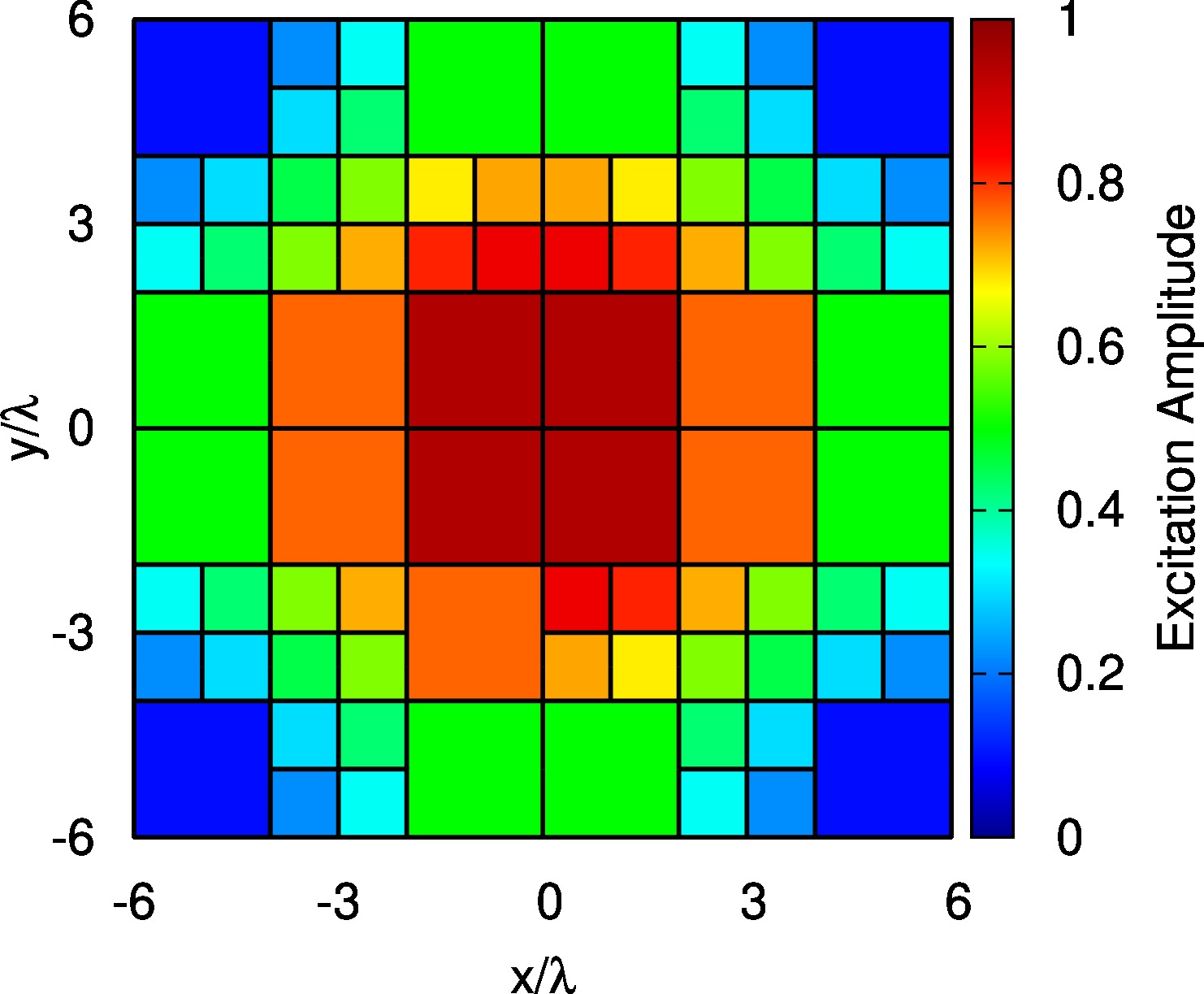}}\tabularnewline
\textcolor{black}{(}\textcolor{black}{\emph{a}}\textcolor{black}{)}&
\textcolor{black}{(}\textcolor{black}{\emph{b}}\textcolor{black}{)}\tabularnewline
\textcolor{black}{\includegraphics[%
  width=0.45\columnwidth]{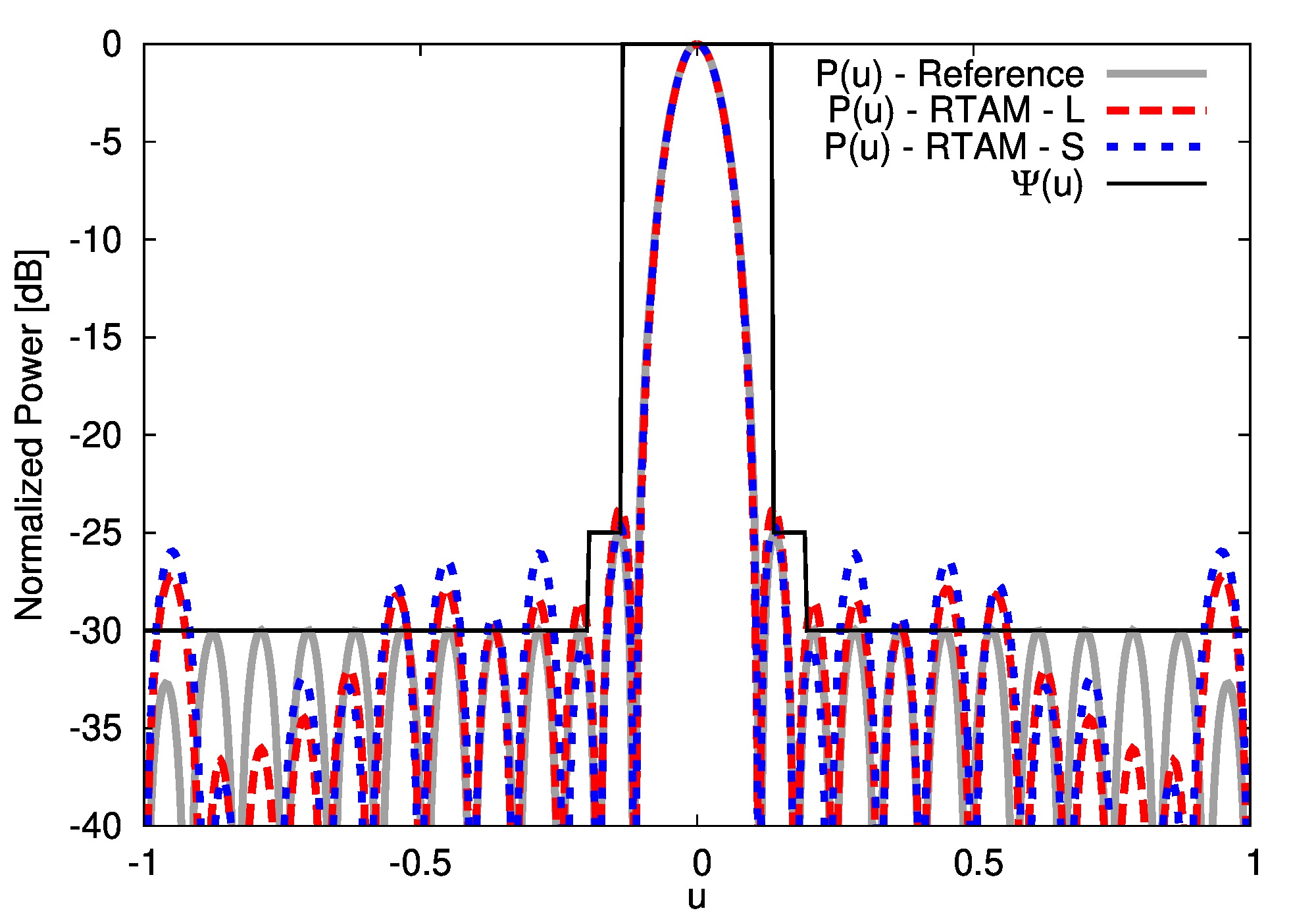}}&
\textcolor{black}{\includegraphics[%
  width=0.45\columnwidth]{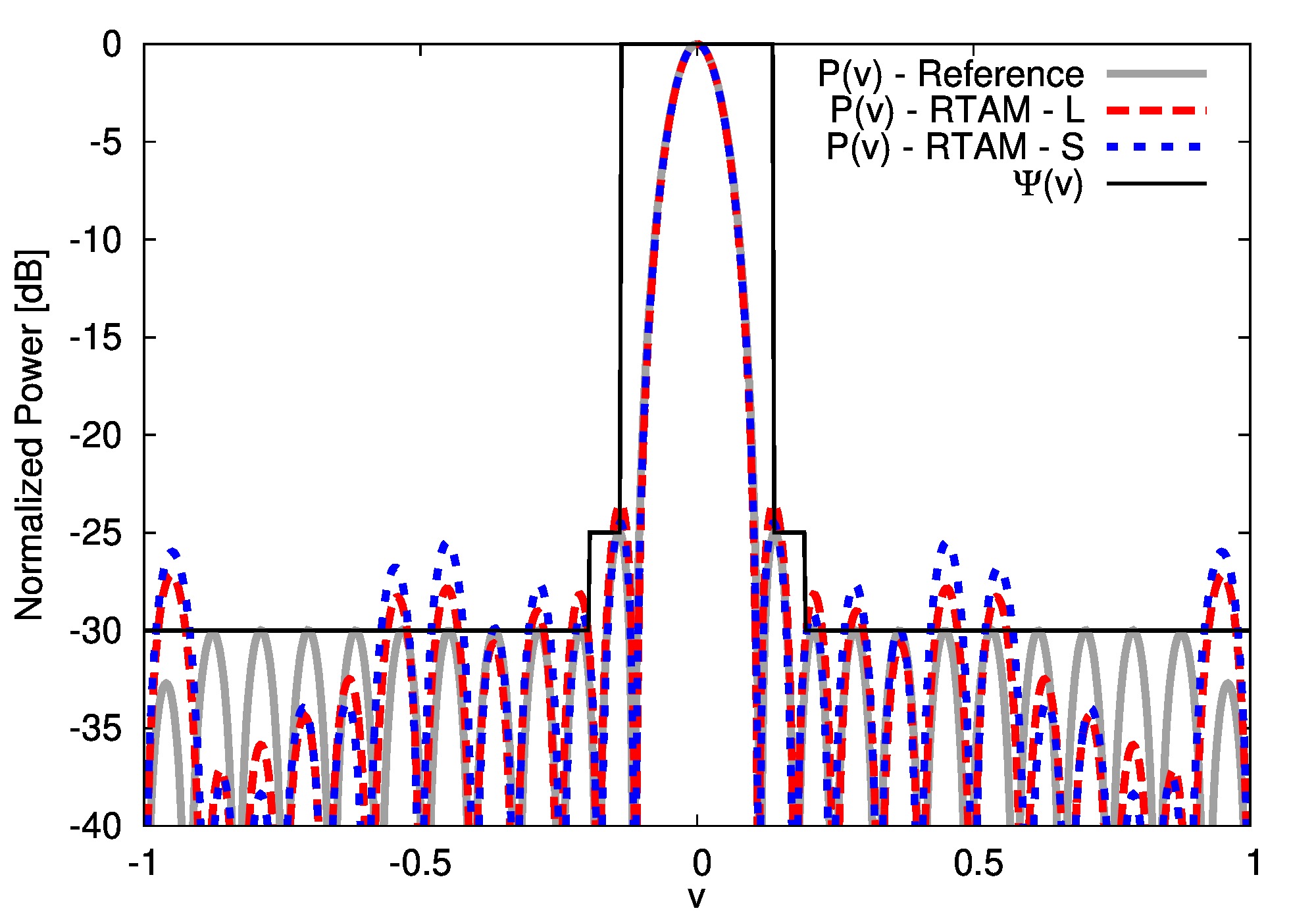}}\tabularnewline
\textcolor{black}{(}\textcolor{black}{\emph{c}}\textcolor{black}{)}&
\textcolor{black}{(}\textcolor{black}{\emph{d}}\textcolor{black}{)}\tabularnewline
\end{tabular}\end{center}

\begin{center}\textcolor{black}{~\vfill}\end{center}

\begin{center}\textbf{\textcolor{black}{Fig. 11 - N. Anselmi}} \textbf{\textcolor{black}{\emph{et
al.}}}\textbf{\textcolor{black}{,}} \textbf{\textcolor{black}{\emph{{}``}}}\textcolor{black}{A
Self-Replicating Single-Shape Tiling Technique ...''}\end{center}
\newpage

\begin{center}\textcolor{black}{~\vfill}\end{center}

\begin{center}\textcolor{black}{}\begin{tabular}{c}
\textcolor{black}{\includegraphics[%
  width=0.48\columnwidth]{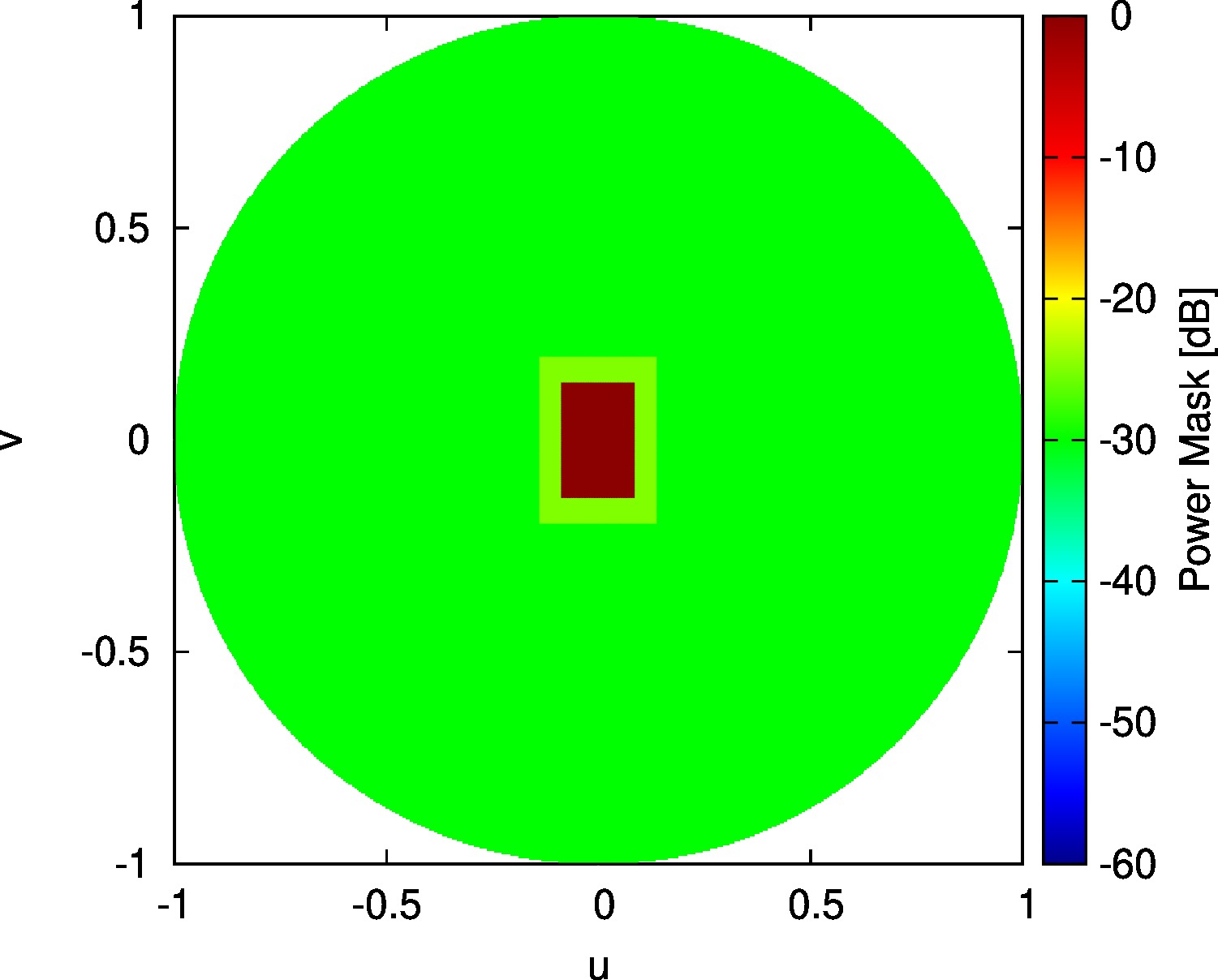}}\tabularnewline
\textcolor{black}{(}\textcolor{black}{\emph{a}}\textcolor{black}{)}\tabularnewline
\textcolor{black}{\includegraphics[%
  width=0.48\columnwidth]{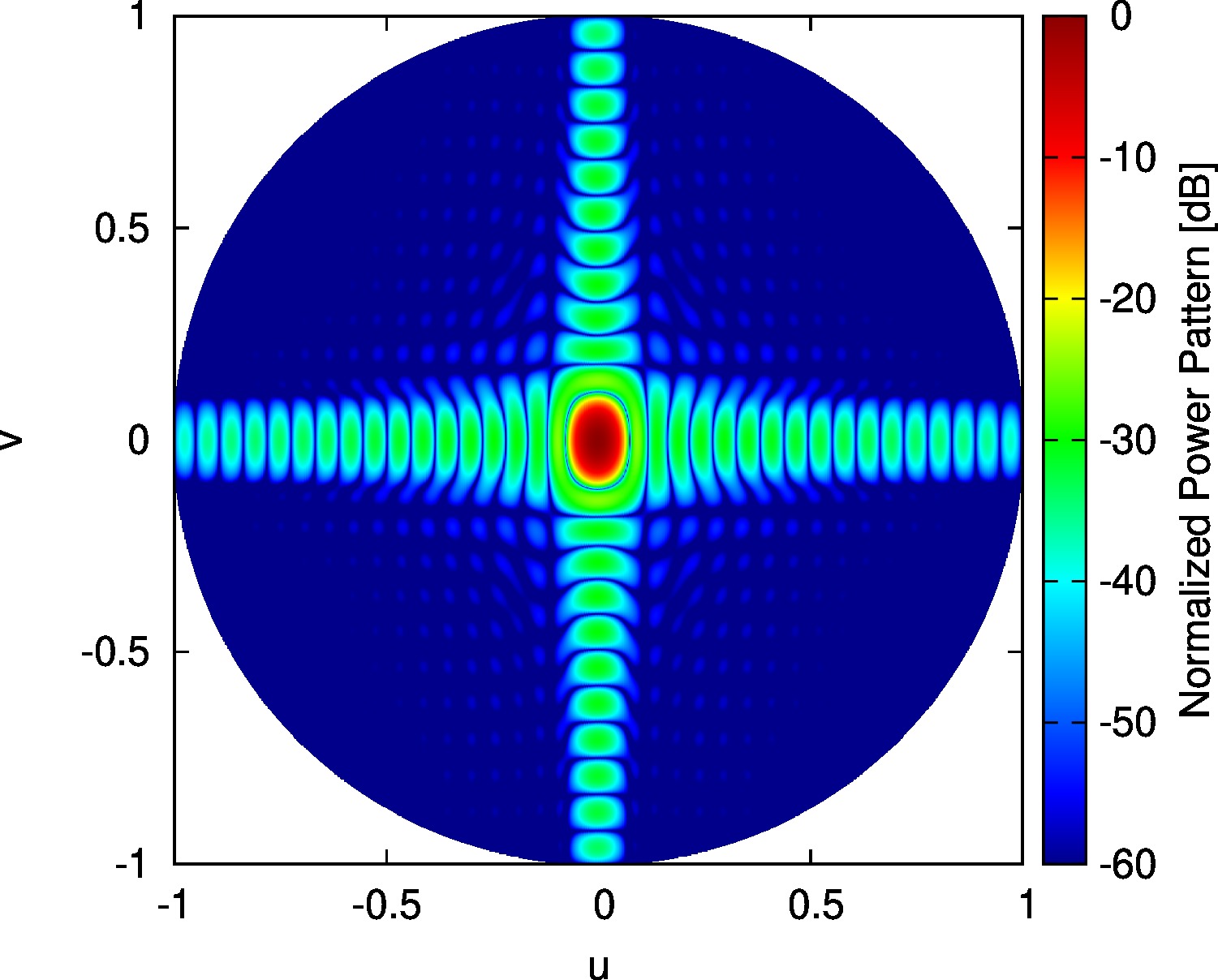}}\tabularnewline
\textcolor{black}{(}\textcolor{black}{\emph{b}}\textcolor{black}{)}\tabularnewline
\textcolor{black}{\includegraphics[%
  width=0.48\columnwidth]{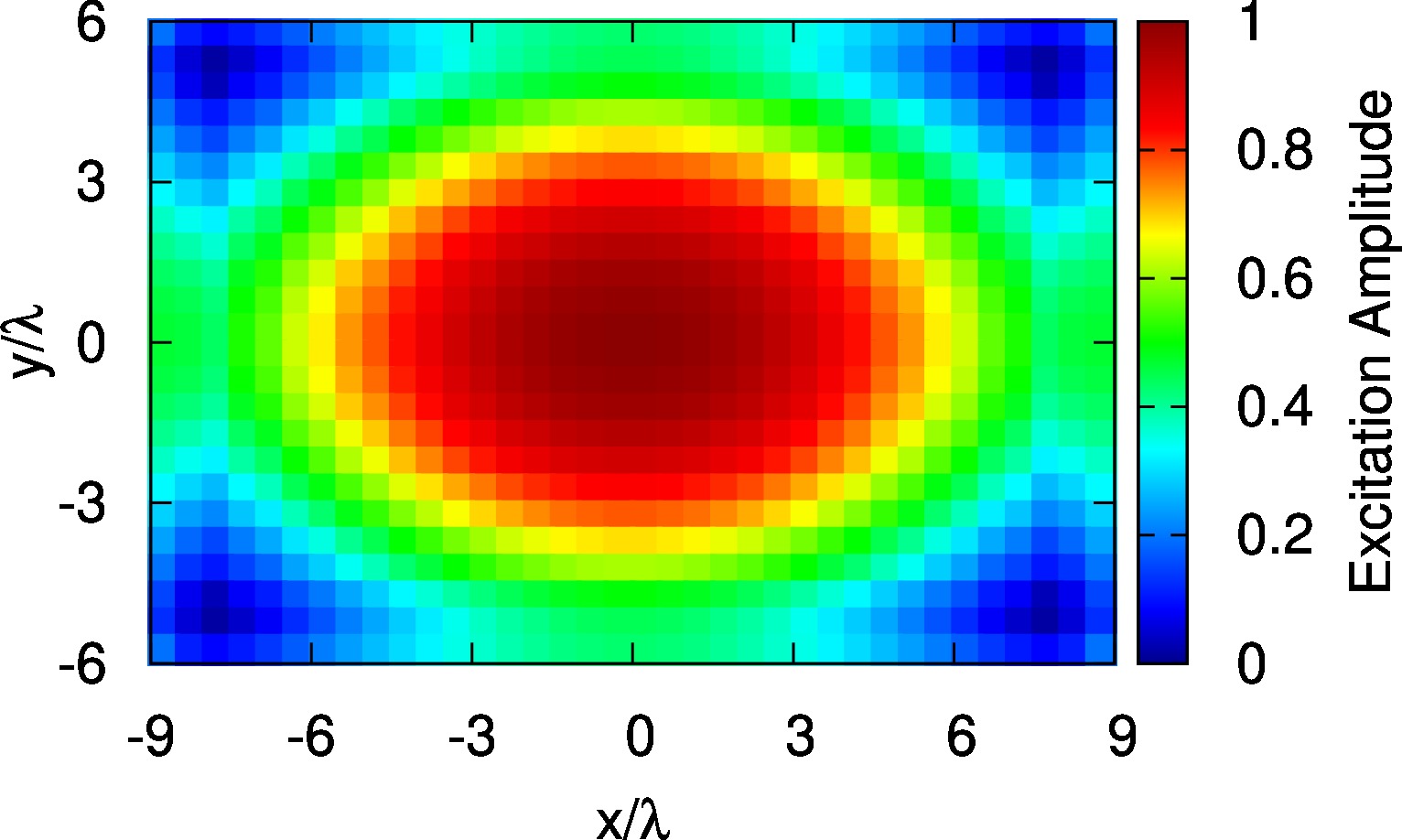}}\tabularnewline
\textcolor{black}{(}\textcolor{black}{\emph{c}}\textcolor{black}{)}\tabularnewline
\end{tabular}\end{center}

\begin{center}\textcolor{black}{~\vfill}\end{center}

\begin{center}\textbf{\textcolor{black}{Fig. 12 - N. Anselmi}} \textbf{\textcolor{black}{\emph{et
al.}}}\textbf{\textcolor{black}{,}} \textbf{\textcolor{black}{\emph{{}``}}}\textcolor{black}{A
Self-Replicating Single-Shape Tiling Technique ...''}\end{center}

\newpage
\begin{center}\textcolor{black}{~\vfill}\end{center}

\begin{center}\textcolor{black}{}\begin{tabular}{c}
\textcolor{black}{\includegraphics[%
  width=0.80\columnwidth]{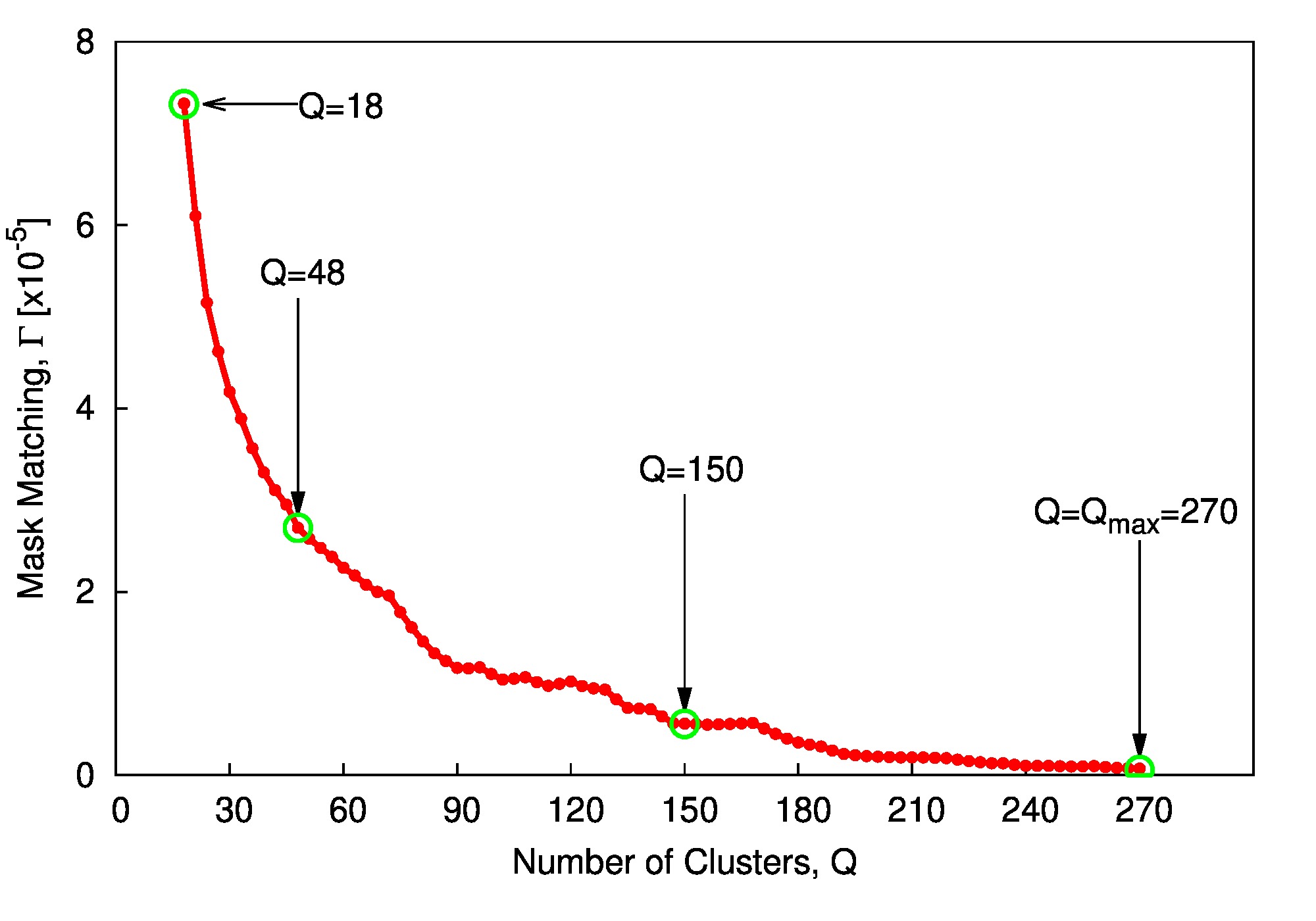}}\tabularnewline
\end{tabular}\end{center}

\begin{center}\textcolor{black}{~\vfill}\end{center}

\begin{center}\textbf{\textcolor{black}{Fig. 13 - N. Anselmi}} \textbf{\textcolor{black}{\emph{et
al.}}}\textbf{\textcolor{black}{,}} \textbf{\textcolor{black}{\emph{{}``}}}\textcolor{black}{A
Self-Replicating Single-Shape Tiling Technique ...''}\end{center}

\newpage
\begin{center}\textcolor{black}{~\vfill}\end{center}

\begin{center}\textcolor{black}{}\begin{sideways}
\begin{tabular}{cccc}
\textcolor{black}{\includegraphics[%
  width=0.30\columnwidth]{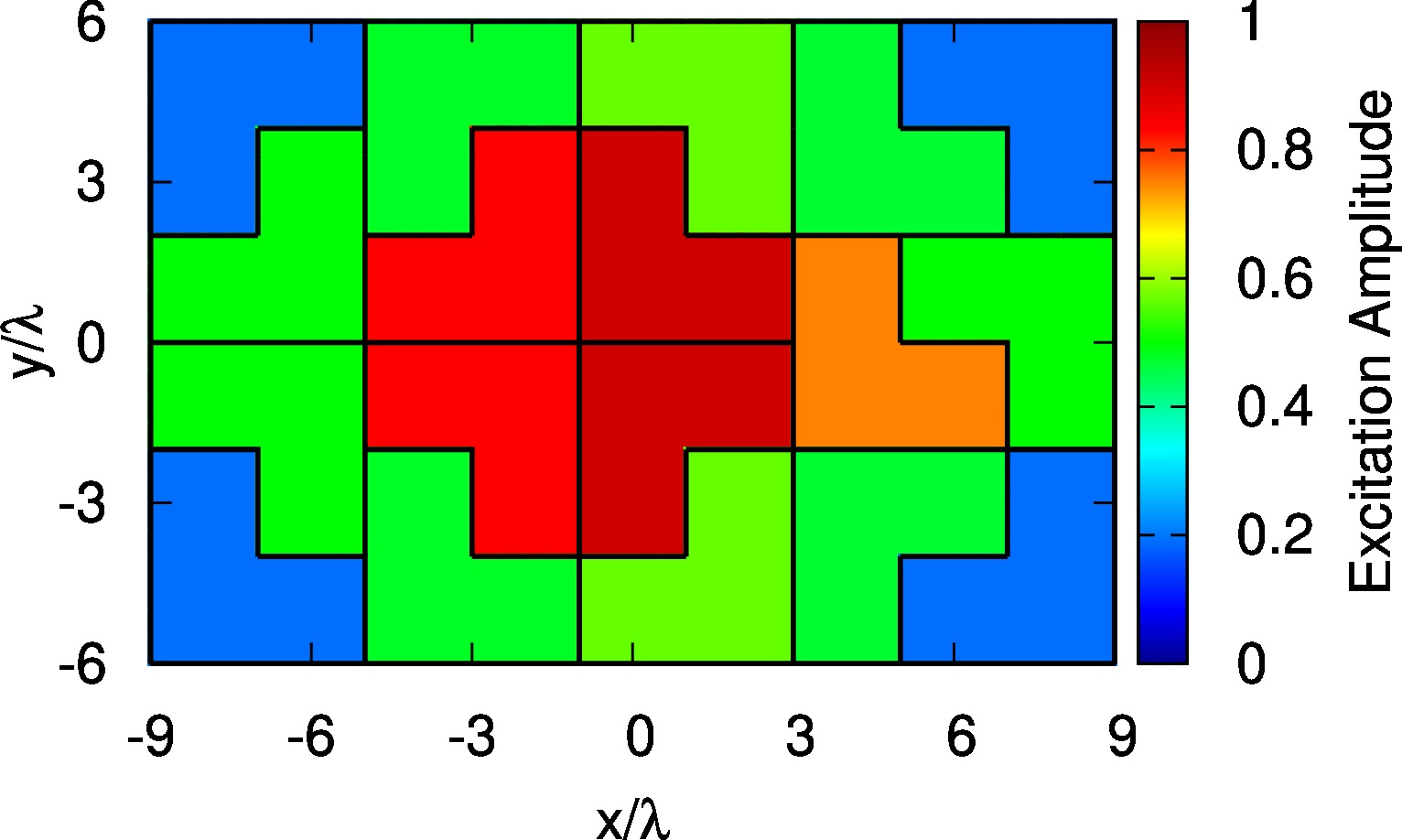}}&
\textcolor{black}{\includegraphics[%
  width=0.30\columnwidth]{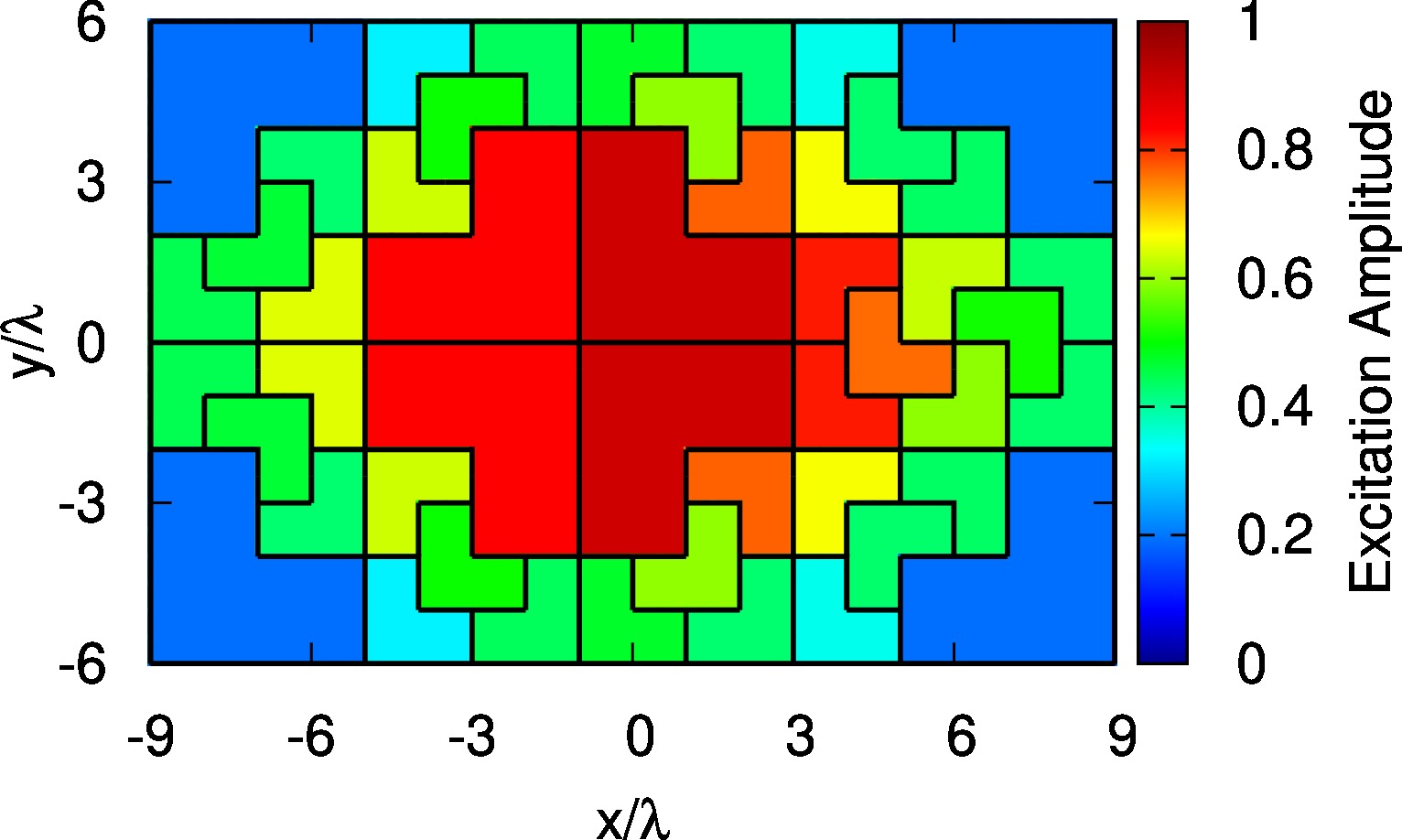}}&
\textcolor{black}{\includegraphics[%
  width=0.30\columnwidth]{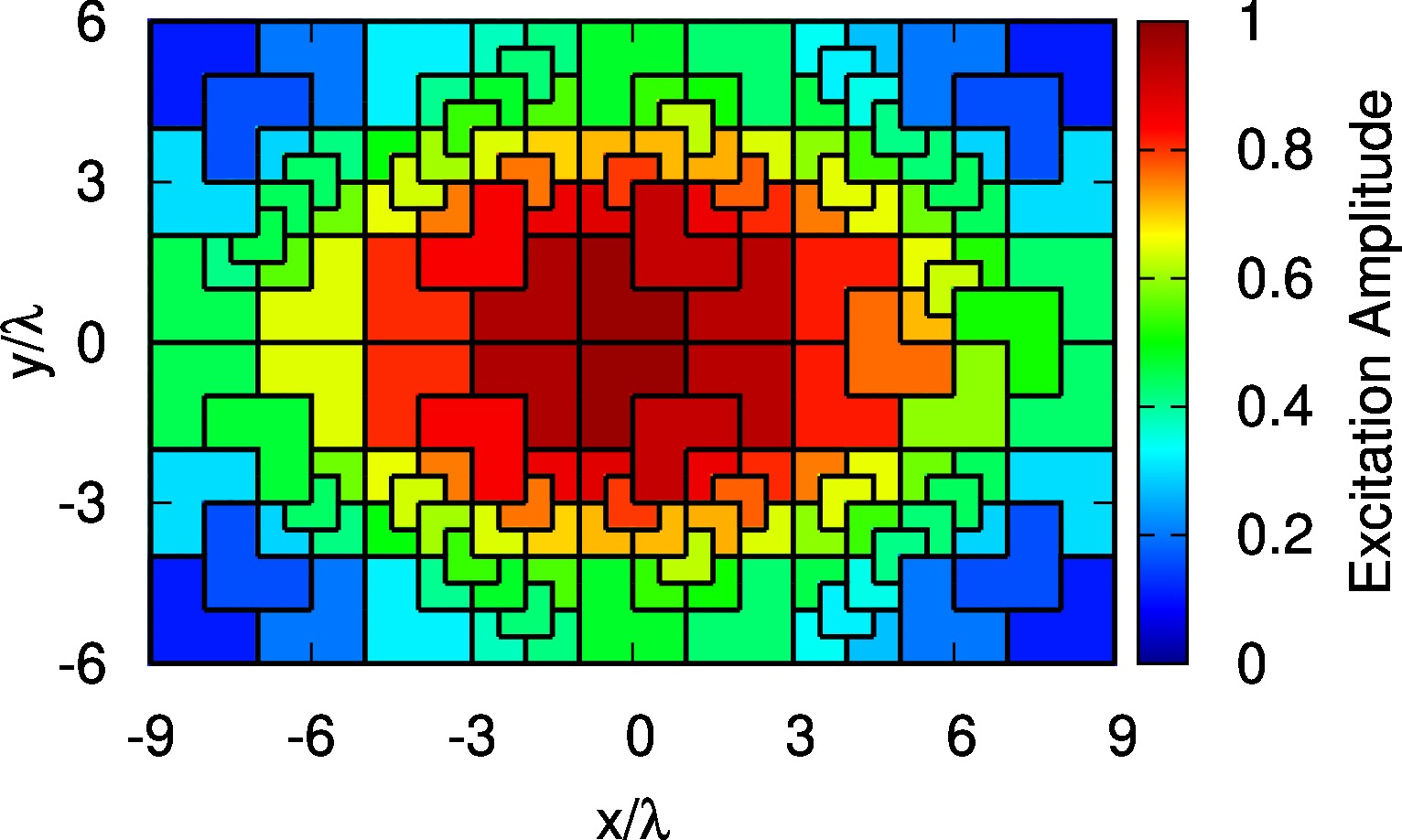}}&
\textcolor{black}{\includegraphics[%
  width=0.30\columnwidth]{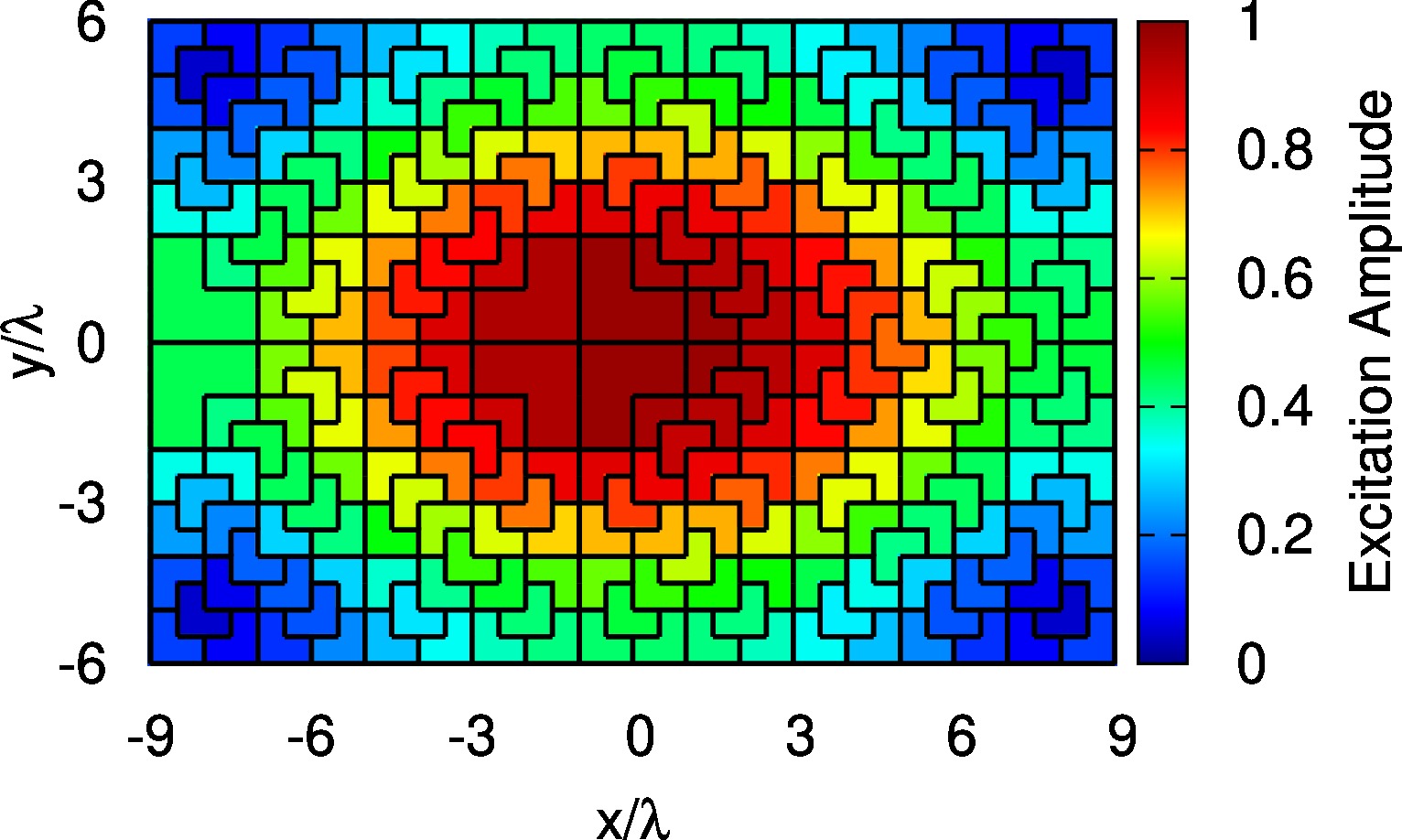}}\tabularnewline
\textcolor{black}{(}\textcolor{black}{\emph{a}}\textcolor{black}{)}&
\textcolor{black}{(}\textcolor{black}{\emph{b}}\textcolor{black}{)}&
\textcolor{black}{(}\textcolor{black}{\emph{c}}\textcolor{black}{)}&
\textcolor{black}{(}\textcolor{black}{\emph{d}}\textcolor{black}{)}\tabularnewline
\textcolor{black}{\includegraphics[%
  width=0.30\columnwidth]{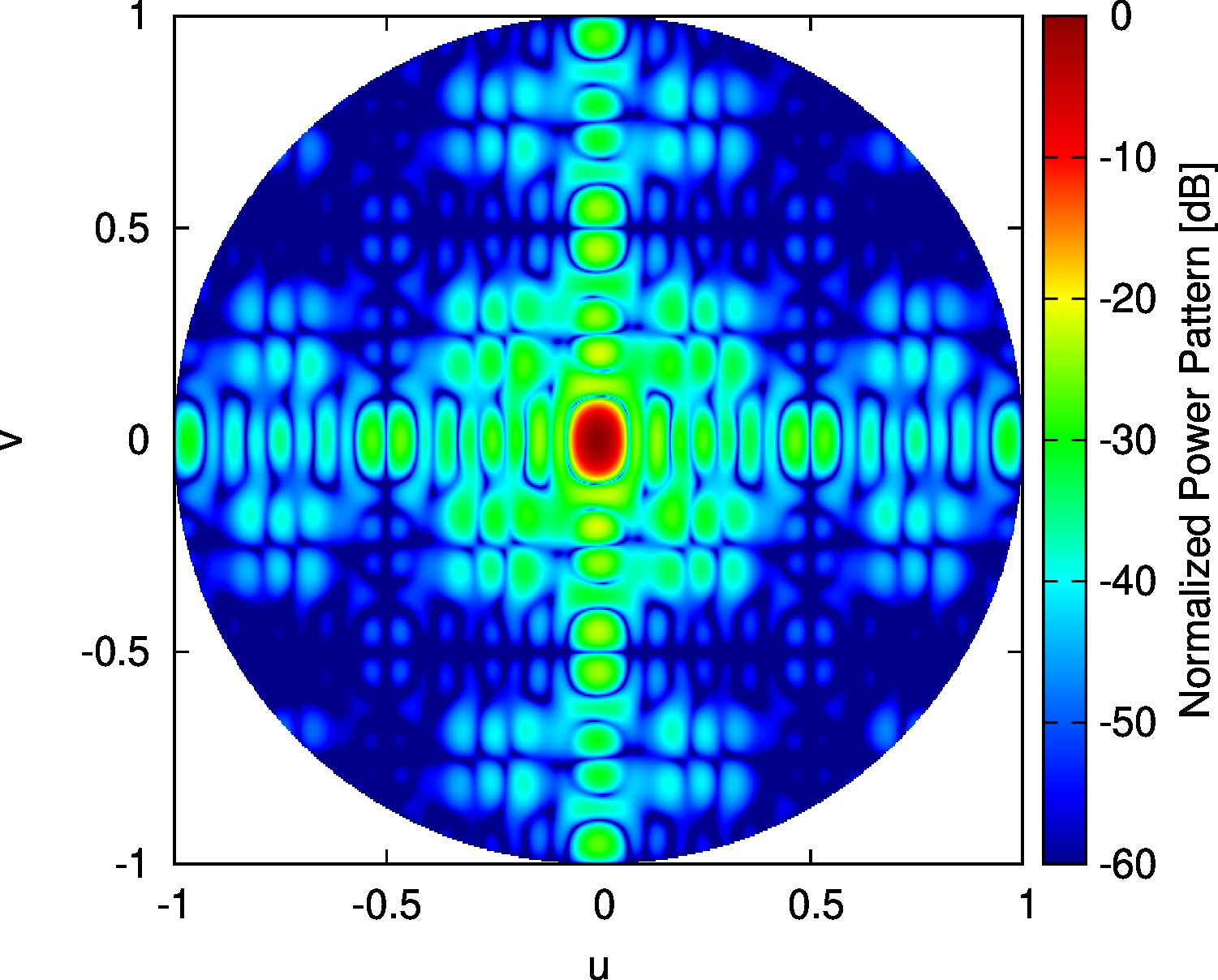}}&
\textcolor{black}{\includegraphics[%
  width=0.30\columnwidth]{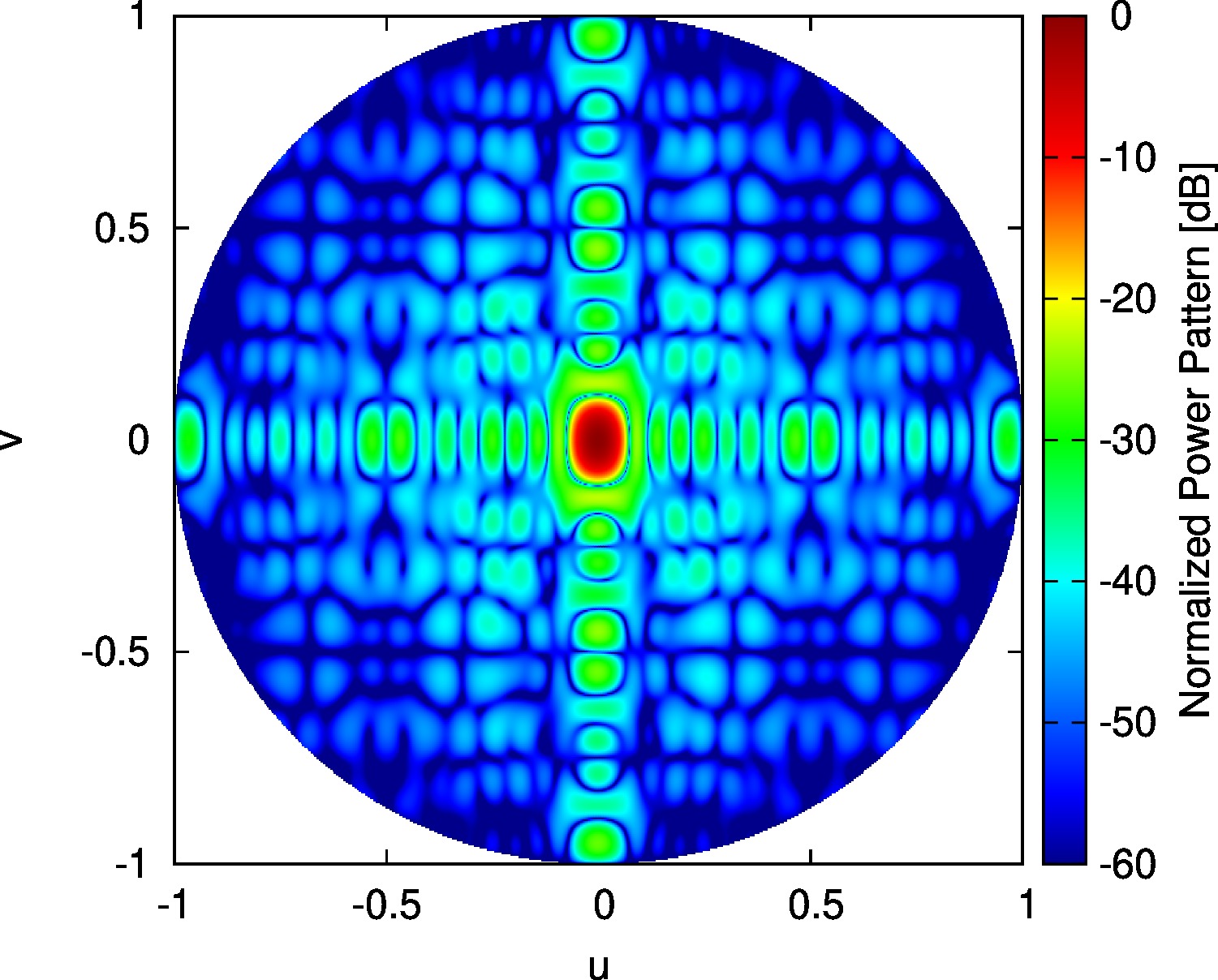}}&
\textcolor{black}{\includegraphics[%
  width=0.30\columnwidth]{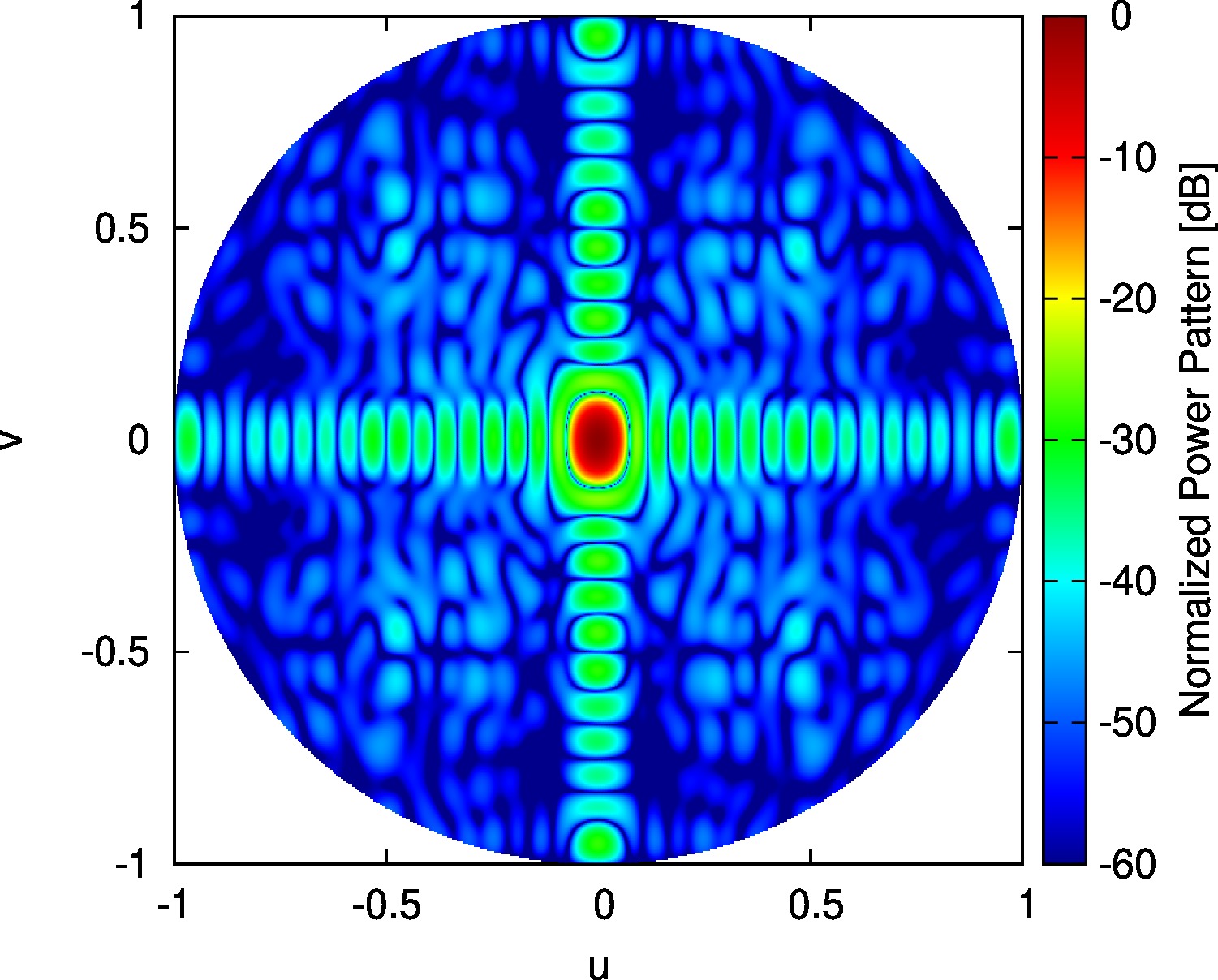}}&
\textcolor{black}{\includegraphics[%
  width=0.30\columnwidth]{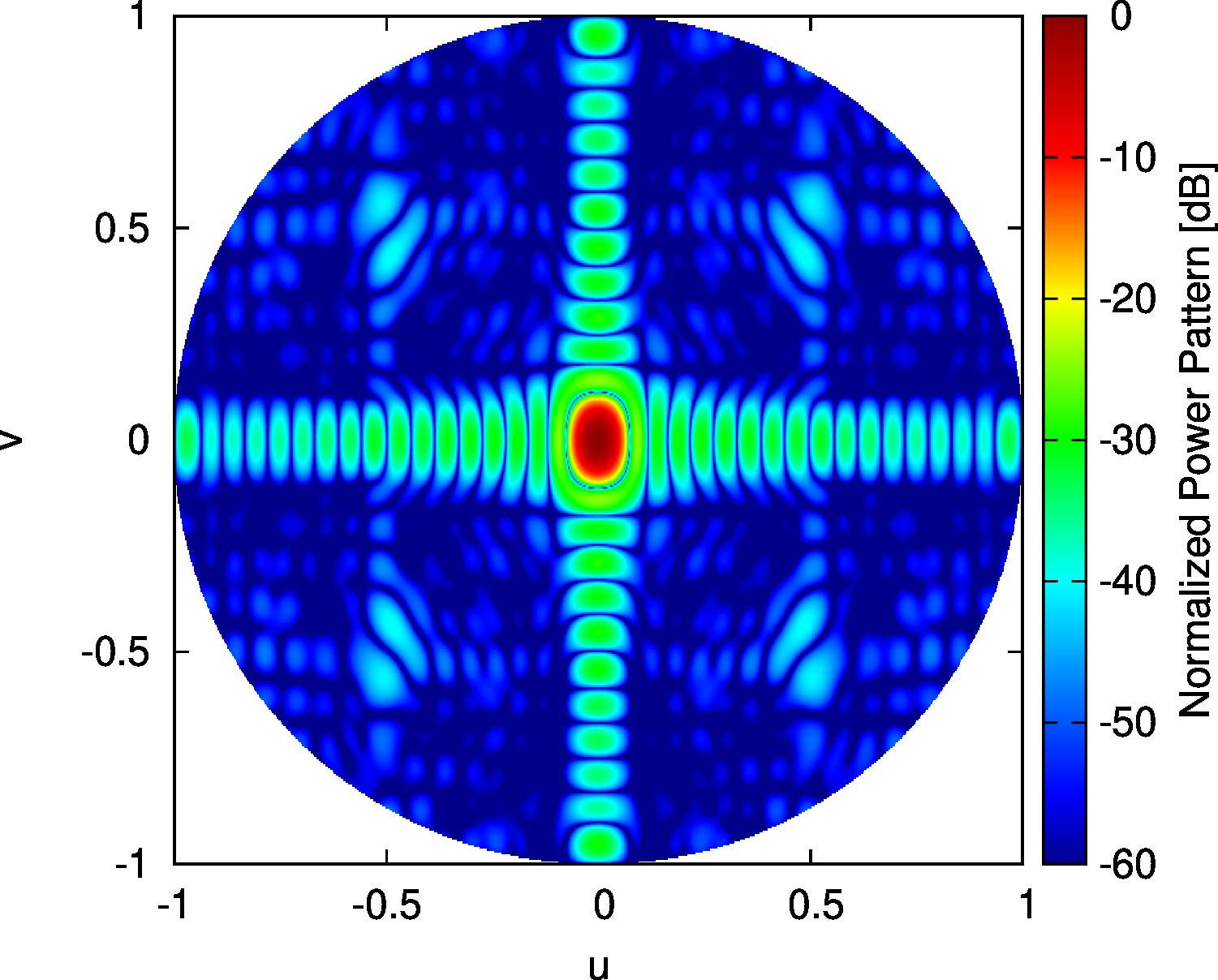}}\tabularnewline
\textcolor{black}{(}\textcolor{black}{\emph{e}}\textcolor{black}{)}&
\textcolor{black}{(}\textcolor{black}{\emph{f}}\textcolor{black}{)}&
\textcolor{black}{(}\textcolor{black}{\emph{g}}\textcolor{black}{)}&
\textcolor{black}{(}\textcolor{black}{\emph{h}}\textcolor{black}{)}\tabularnewline
\end{tabular}
\end{sideways}\end{center}

\begin{center}\textcolor{black}{~\vfill}\end{center}

\begin{center}\textbf{\textcolor{black}{Fig. 14 - N. Anselmi}} \textbf{\textcolor{black}{\emph{et
al.}}}\textbf{\textcolor{black}{,}} \textbf{\textcolor{black}{\emph{{}``}}}\textcolor{black}{A
Self-Replicating Single-Shape Tiling Technique ...''}\end{center}

\newpage
\begin{center}\textcolor{black}{~\vfill}\end{center}

\begin{center}\textcolor{black}{}\begin{tabular}{c}
\textcolor{black}{\includegraphics[%
  width=0.70\columnwidth]{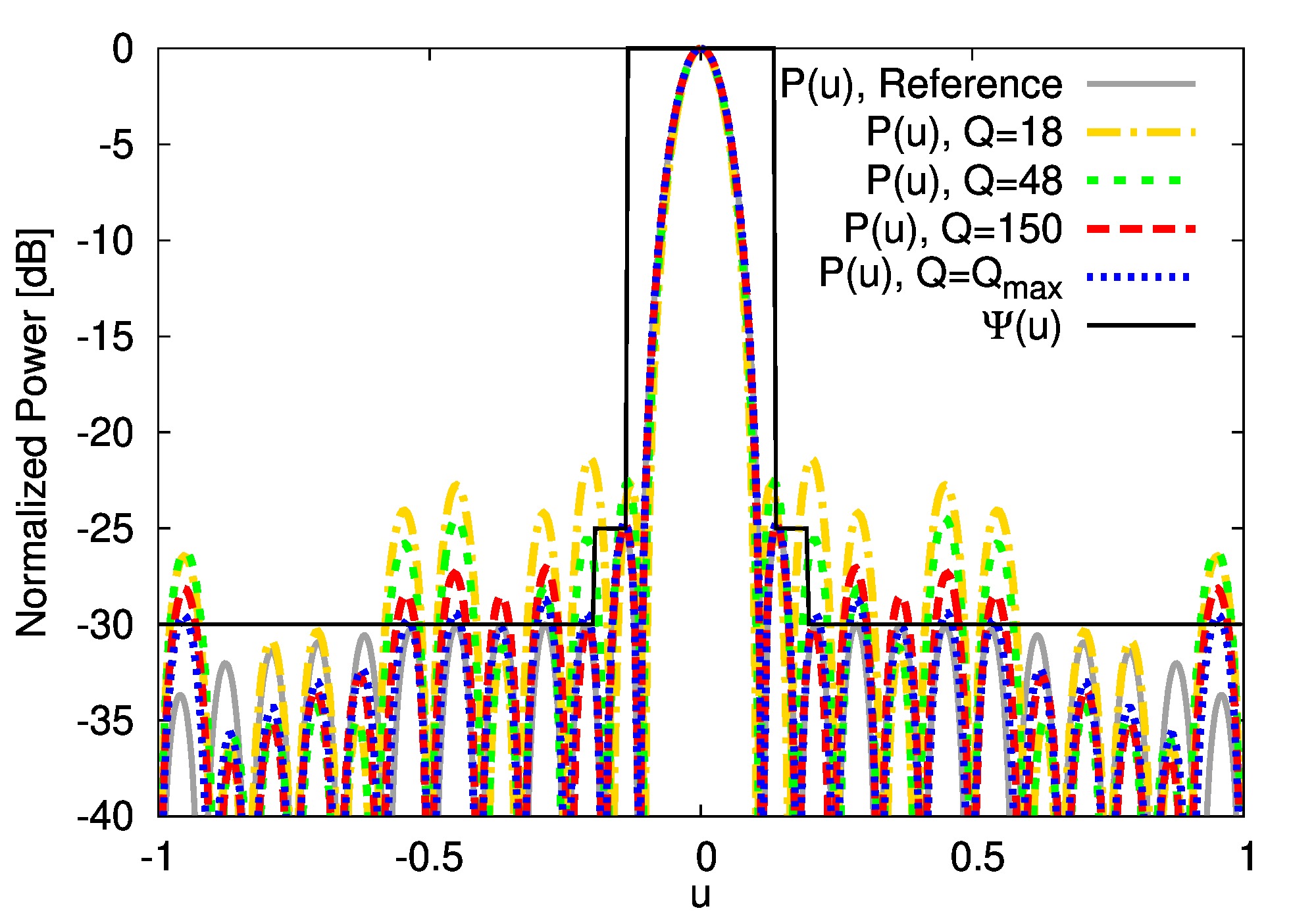}}\tabularnewline
\textcolor{black}{(}\textcolor{black}{\emph{a}}\textcolor{black}{)}\tabularnewline
\textcolor{black}{\includegraphics[%
  width=0.70\columnwidth]{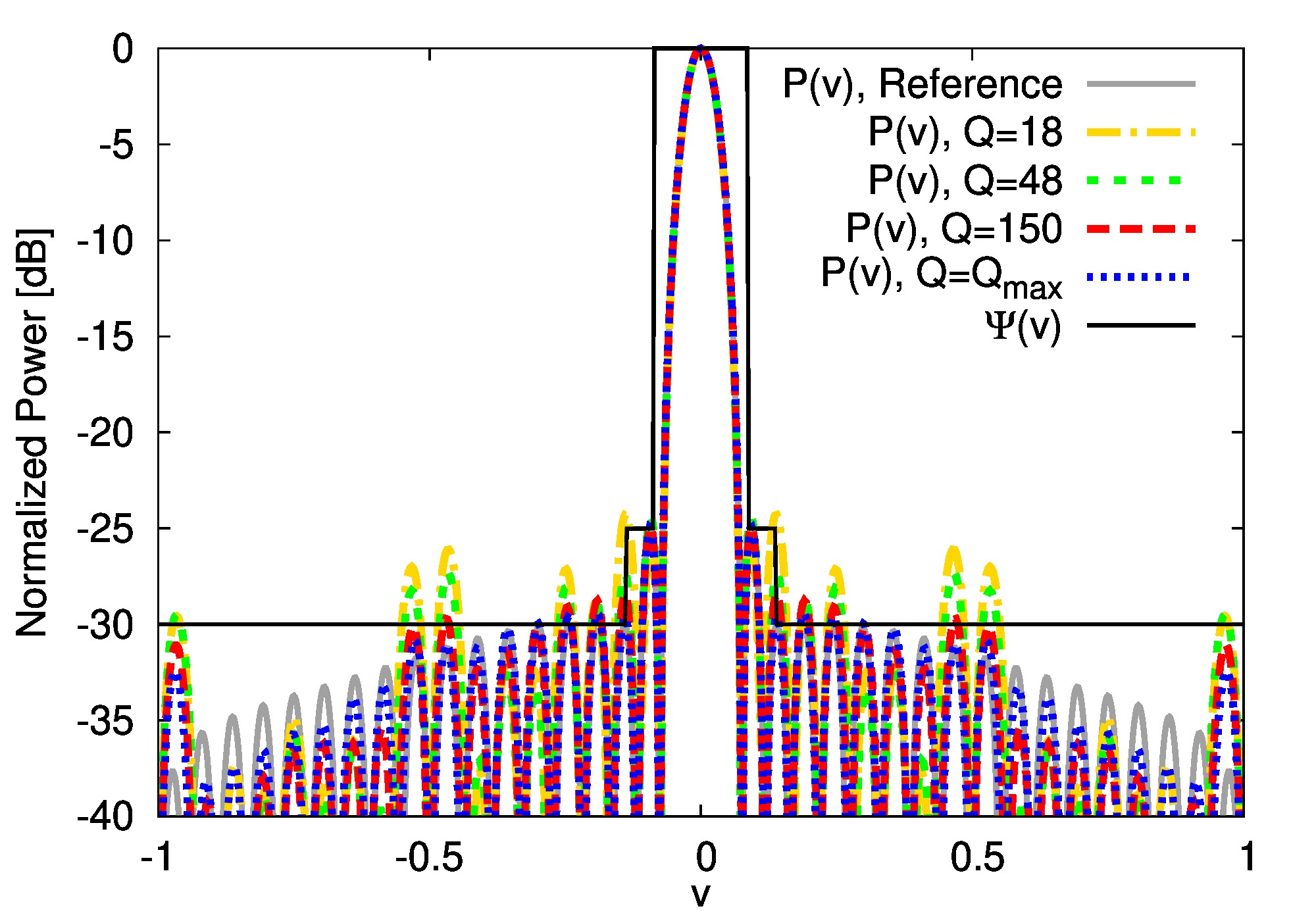}}\tabularnewline
\textcolor{black}{(}\textcolor{black}{\emph{b}}\textcolor{black}{)}\tabularnewline
\end{tabular}\end{center}

\begin{center}\textcolor{black}{~\vfill}\end{center}

\begin{center}\textbf{\textcolor{black}{Fig. 15 - N. Anselmi}} \textbf{\textcolor{black}{\emph{et
al.}}}\textbf{\textcolor{black}{,}} \textbf{\textcolor{black}{\emph{{}``}}}\textcolor{black}{A
Self-Replicating Single-Shape Tiling Technique ...''}\end{center}

\newpage
\begin{center}\textcolor{black}{~\vfill}\end{center}

\begin{center}\textcolor{black}{}\begin{tabular}{cc}
\textcolor{black}{\includegraphics[%
  width=0.48\columnwidth]{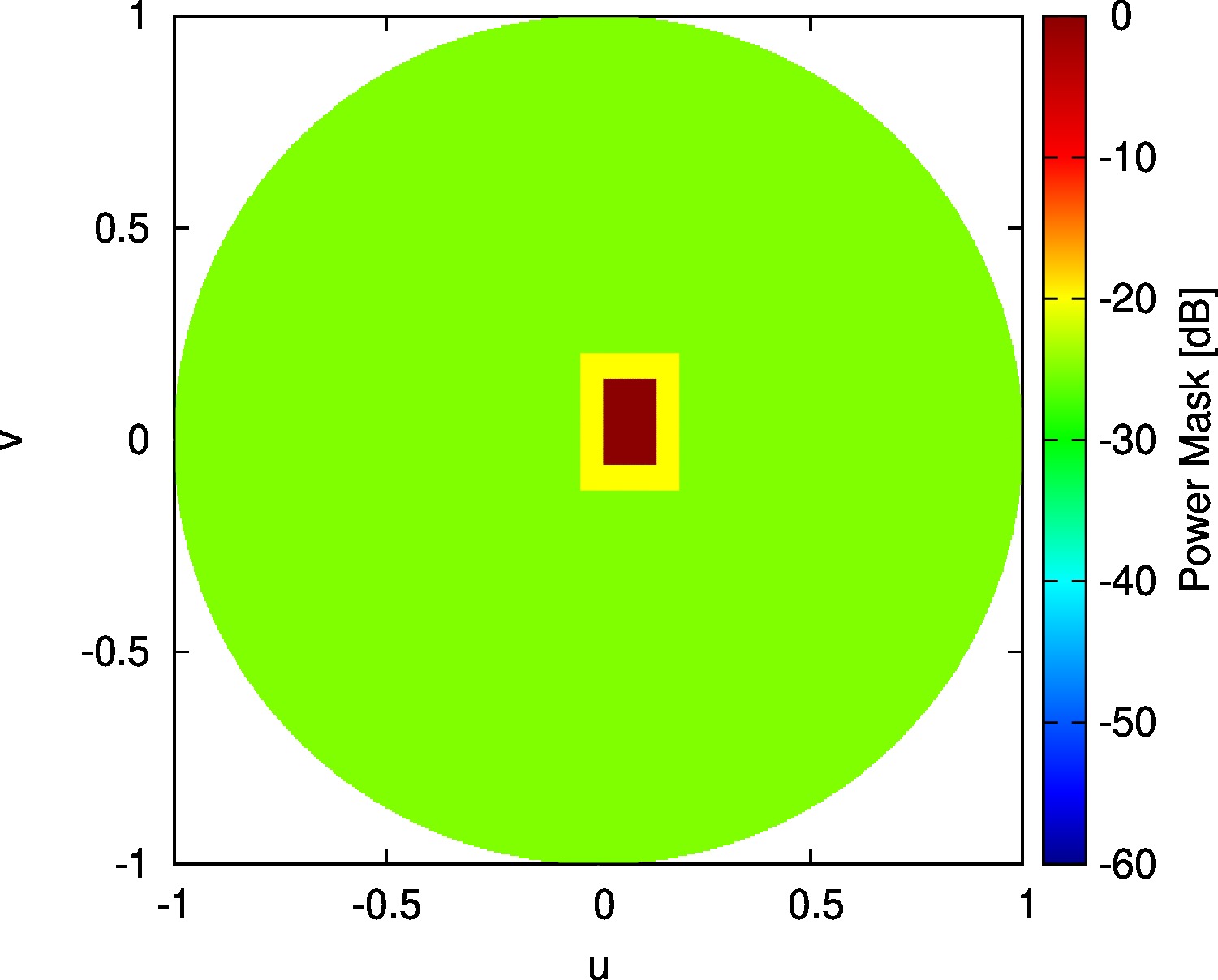}}&
\textcolor{black}{\includegraphics[%
  width=0.49\columnwidth]{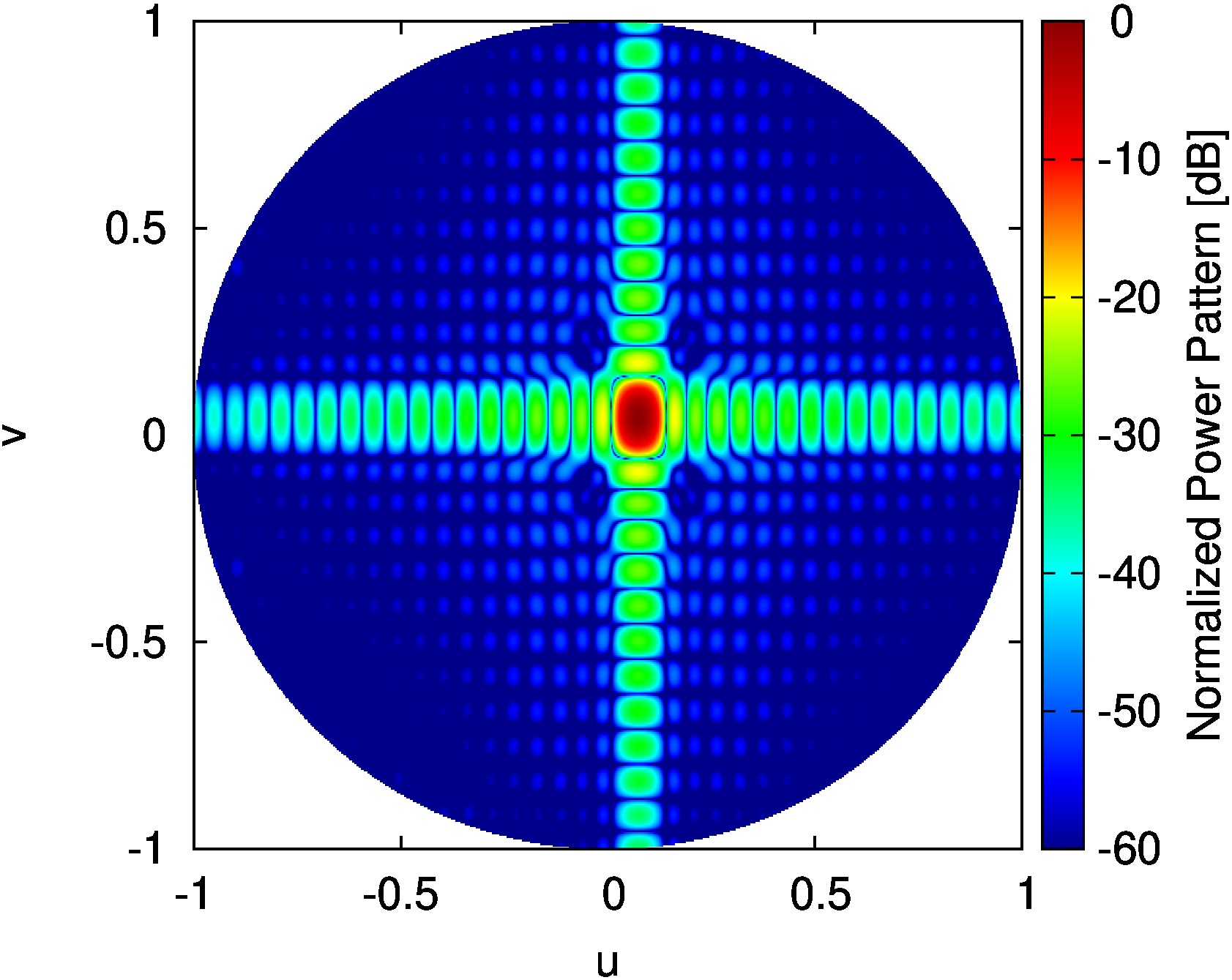}}\tabularnewline
\textcolor{black}{(}\textcolor{black}{\emph{a}}\textcolor{black}{)}&
\textcolor{black}{(}\textcolor{black}{\emph{b}}\textcolor{black}{)}\tabularnewline
\textcolor{black}{\includegraphics[%
  width=0.48\columnwidth]{Fig.Array.24x36.Mask.25dB.Symmetric.Reference.Amplitude.jpg}}&
\textcolor{black}{\includegraphics[%
  width=0.48\columnwidth]{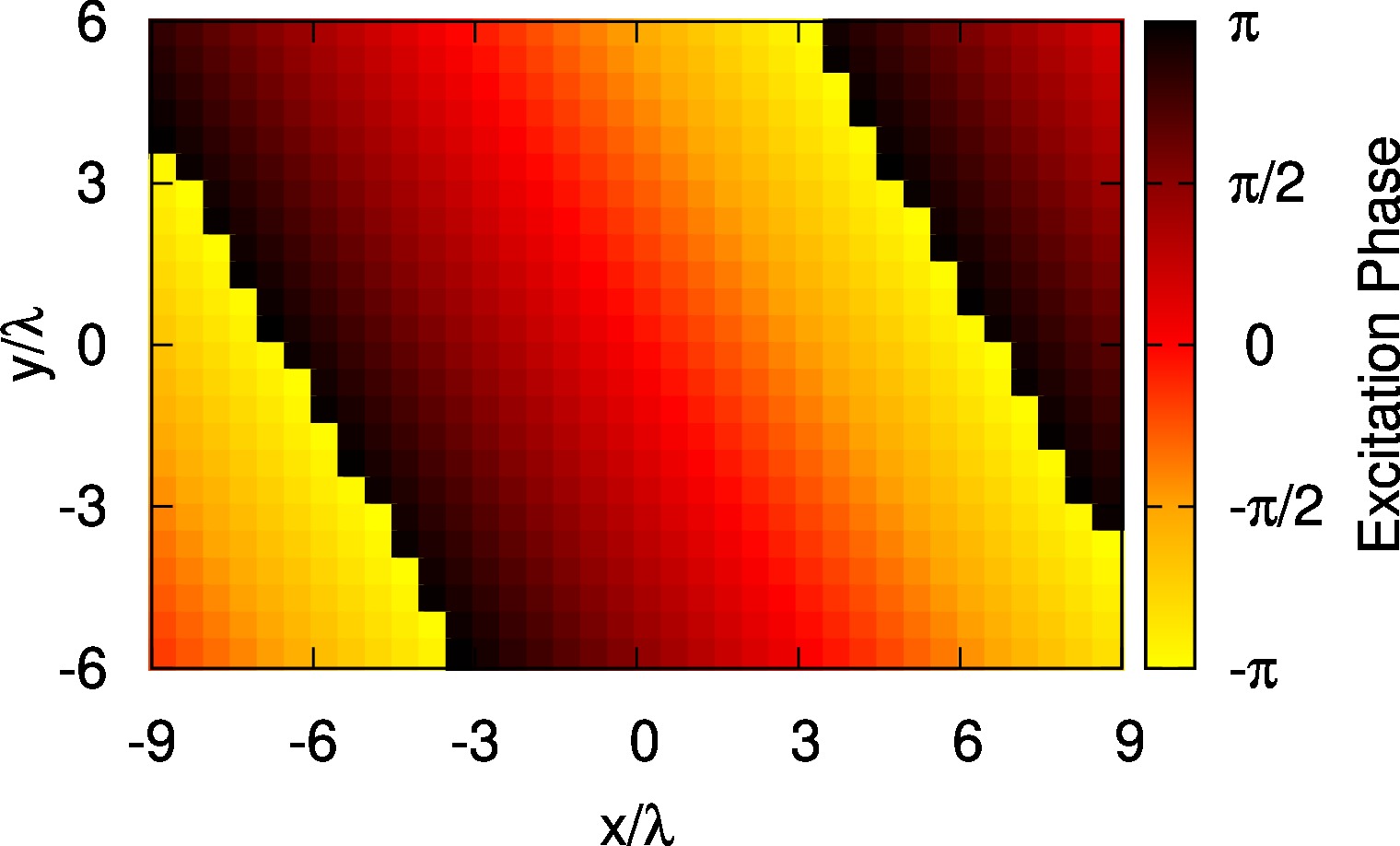}}\tabularnewline
\textcolor{black}{(}\textcolor{black}{\emph{c}}\textcolor{black}{)}&
\textcolor{black}{(}\textcolor{black}{\emph{d}}\textcolor{black}{)}\tabularnewline
\end{tabular}\end{center}

\begin{center}\textcolor{black}{~\vfill}\end{center}

\begin{center}\textbf{\textcolor{black}{Fig. 16 - N. Anselmi}} \textbf{\textcolor{black}{\emph{et
al.}}}\textbf{\textcolor{black}{,}} \textbf{\textcolor{black}{\emph{{}``}}}\textcolor{black}{A
Self-Replicating Single-Shape Tiling Technique ...''}\end{center}

\newpage
\begin{center}\textcolor{black}{~\vfill}\end{center}

\begin{center}\textcolor{black}{}\begin{tabular}{c}
\textcolor{black}{\includegraphics[%
  width=0.80\columnwidth]{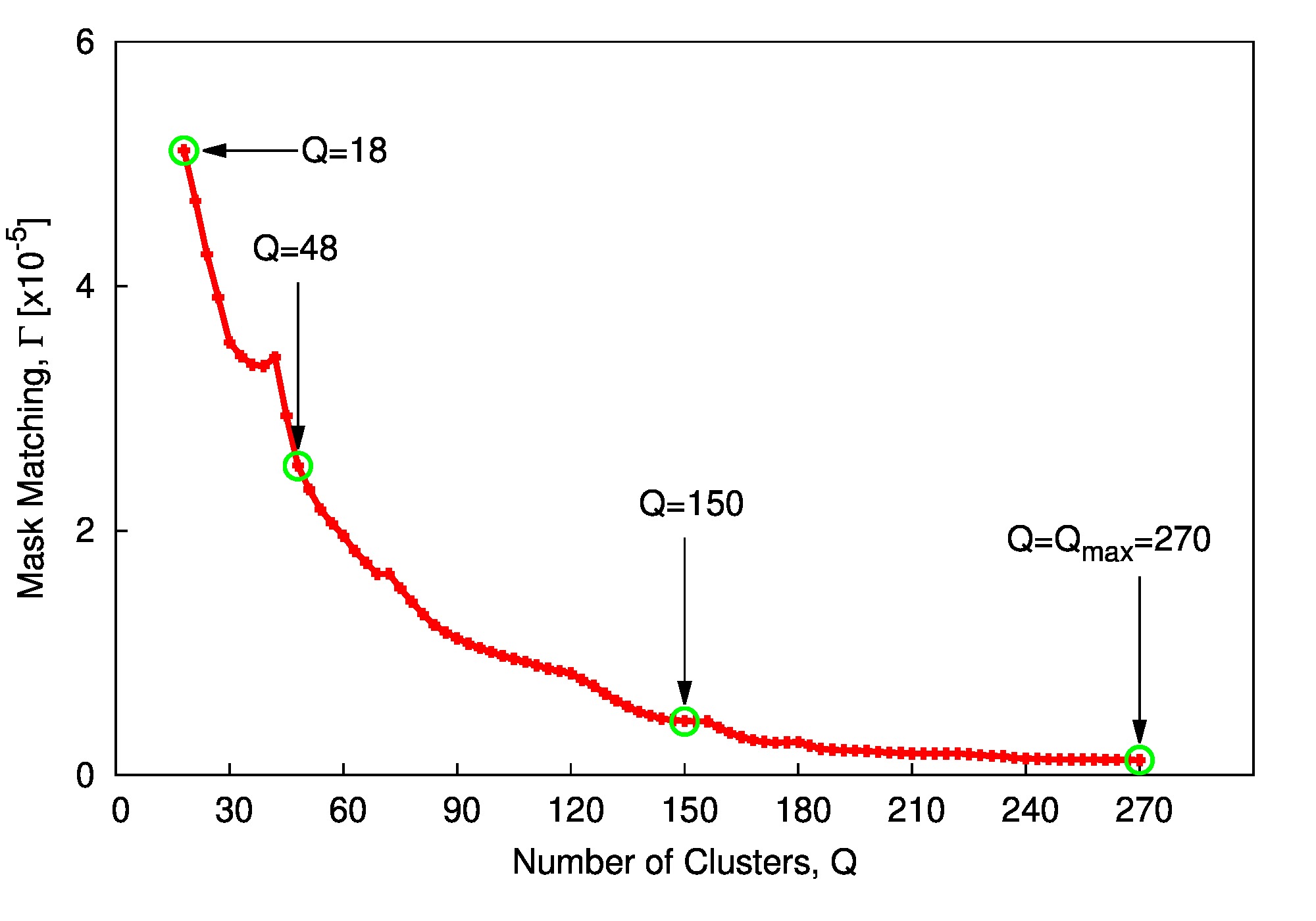}}\tabularnewline
\end{tabular}\end{center}

\begin{center}\textcolor{black}{~\vfill}\end{center}

\begin{center}\textbf{\textcolor{black}{Fig. 17 - N. Anselmi}} \textbf{\textcolor{black}{\emph{et
al.}}}\textbf{\textcolor{black}{,}} \textbf{\textcolor{black}{\emph{{}``}}}\textcolor{black}{A
Self-Replicating Single-Shape Tiling Technique ...''}\end{center}

\newpage
\begin{center}\textcolor{black}{~\vfill}\end{center}

\begin{center}\textcolor{black}{}\begin{tabular}{cc}
\textcolor{black}{\includegraphics[%
  width=0.45\columnwidth]{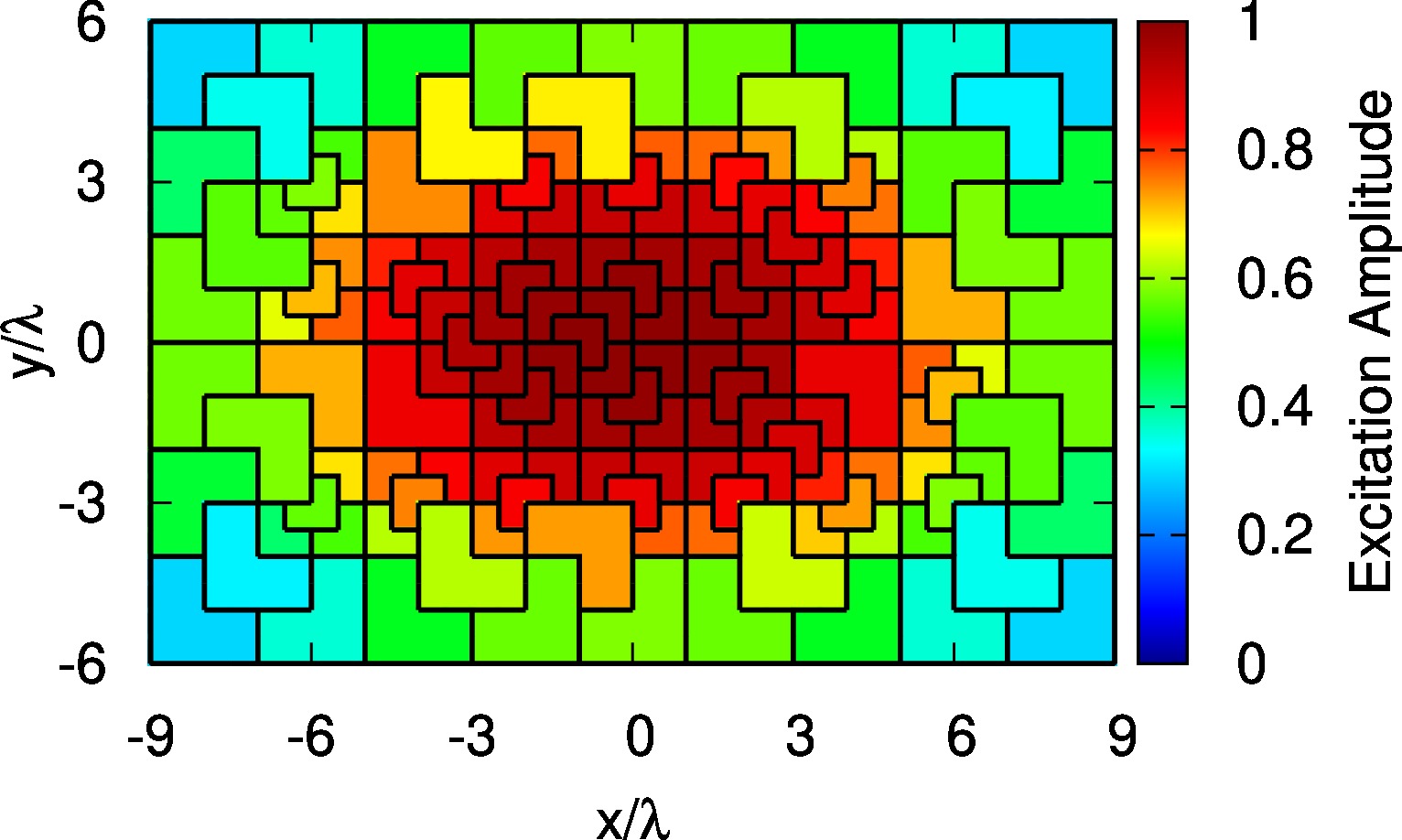}}&
\textcolor{black}{\includegraphics[%
  width=0.45\columnwidth]{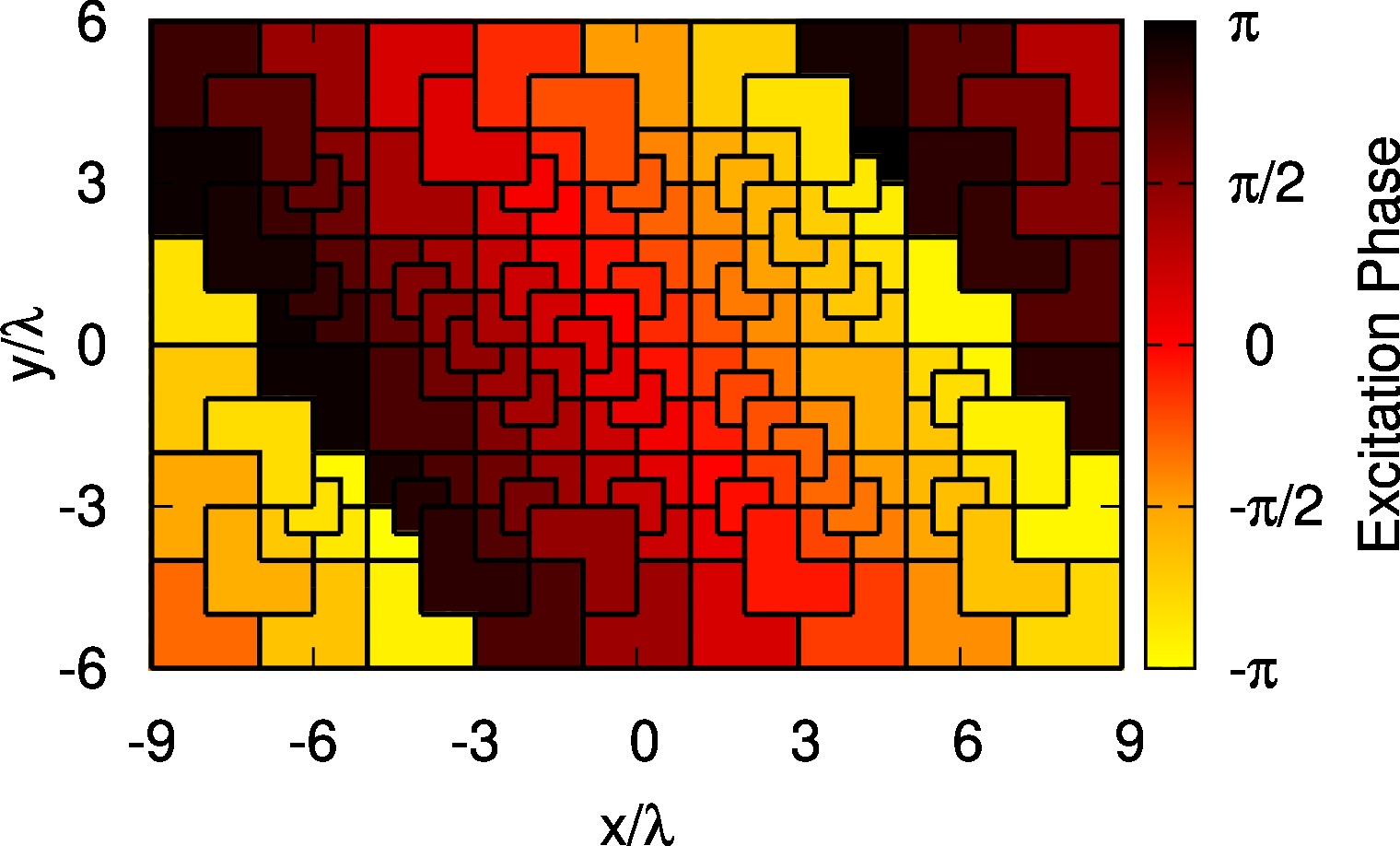}}\tabularnewline
\textcolor{black}{(}\textcolor{black}{\emph{a}}\textcolor{black}{)}&
\textcolor{black}{(}\textcolor{black}{\emph{b}}\textcolor{black}{)}\tabularnewline
\textcolor{black}{\includegraphics[%
  width=0.45\columnwidth]{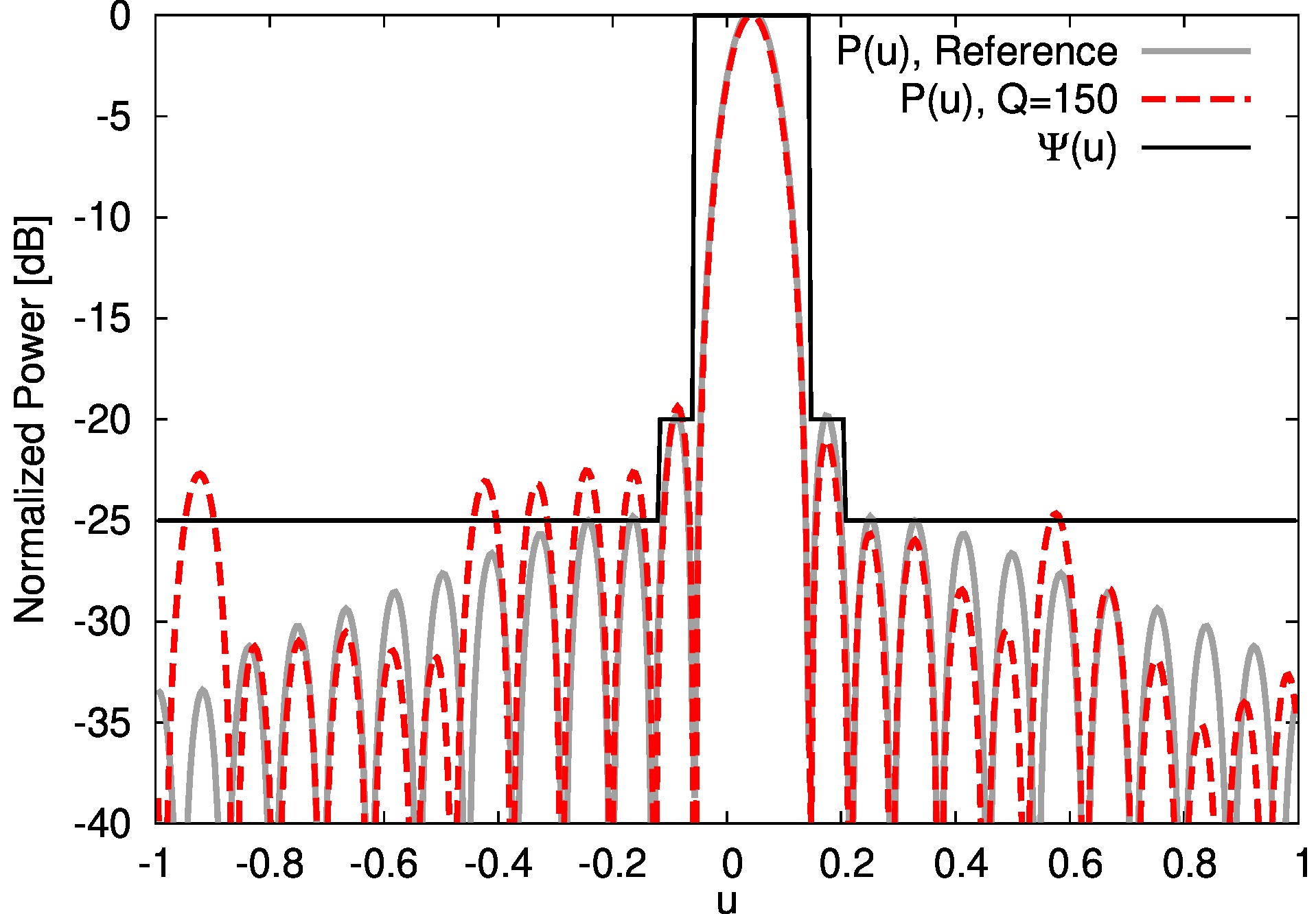}}&
\textcolor{black}{\includegraphics[%
  width=0.45\columnwidth]{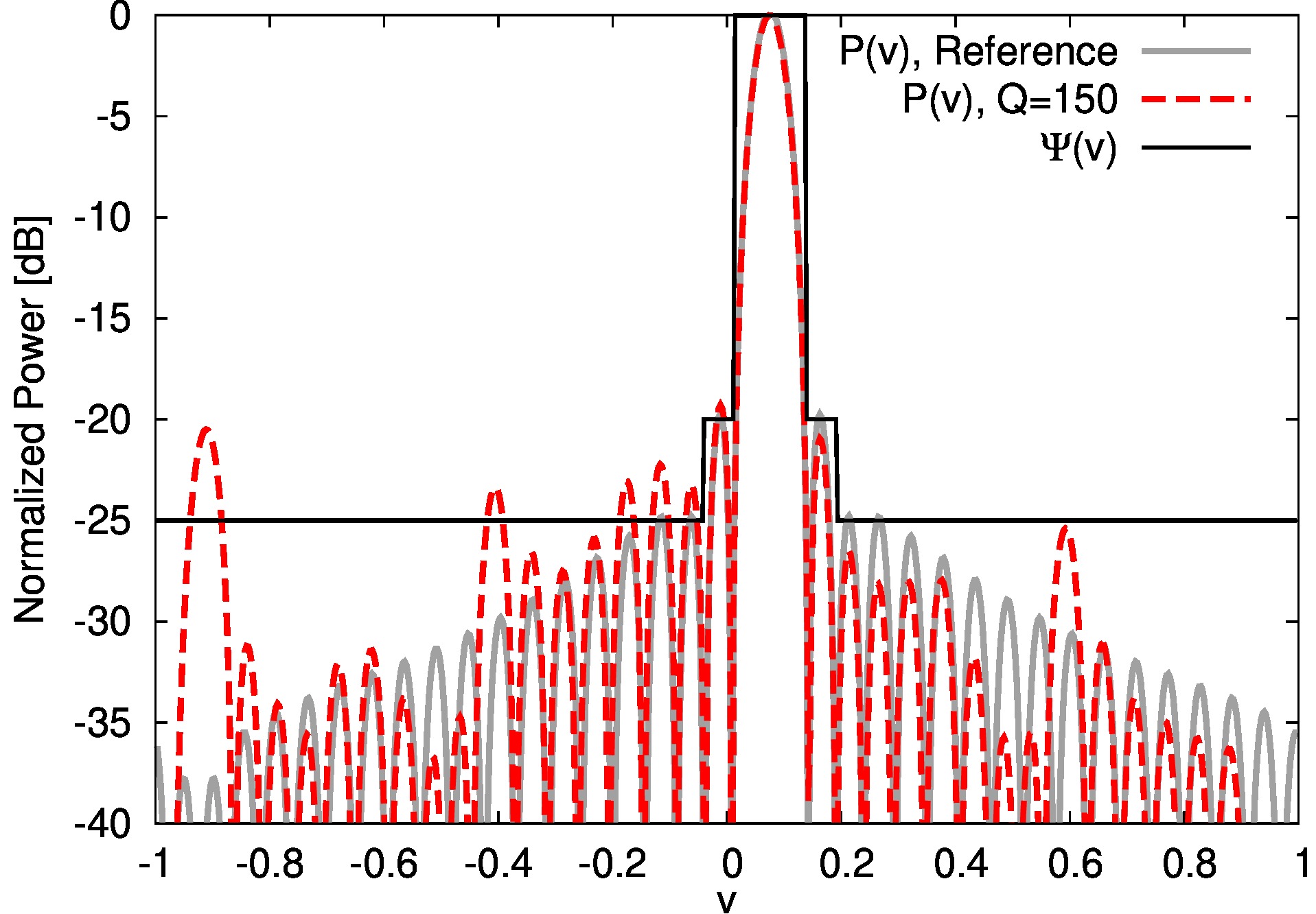}}\tabularnewline
\textcolor{black}{(}\textcolor{black}{\emph{c}}\textcolor{black}{)}&
\textcolor{black}{(}\textcolor{black}{\emph{d}}\textcolor{black}{)}\tabularnewline
\end{tabular}\end{center}

\begin{center}\textcolor{black}{~\vfill}\end{center}

\begin{center}\textbf{\textcolor{black}{Fig. 18 - N. Anselmi}} \textbf{\textcolor{black}{\emph{et
al.}}}\textbf{\textcolor{black}{,}} \textbf{\textcolor{black}{\emph{{}``}}}\textcolor{black}{A
Self-Replicating Single-Shape Tiling Technique ...''}\end{center}

\newpage
\begin{center}\textcolor{black}{~\vfill}\end{center}

\begin{center}\textcolor{black}{}\begin{tabular}{c}
\textcolor{black}{\includegraphics[%
  width=0.70\columnwidth]{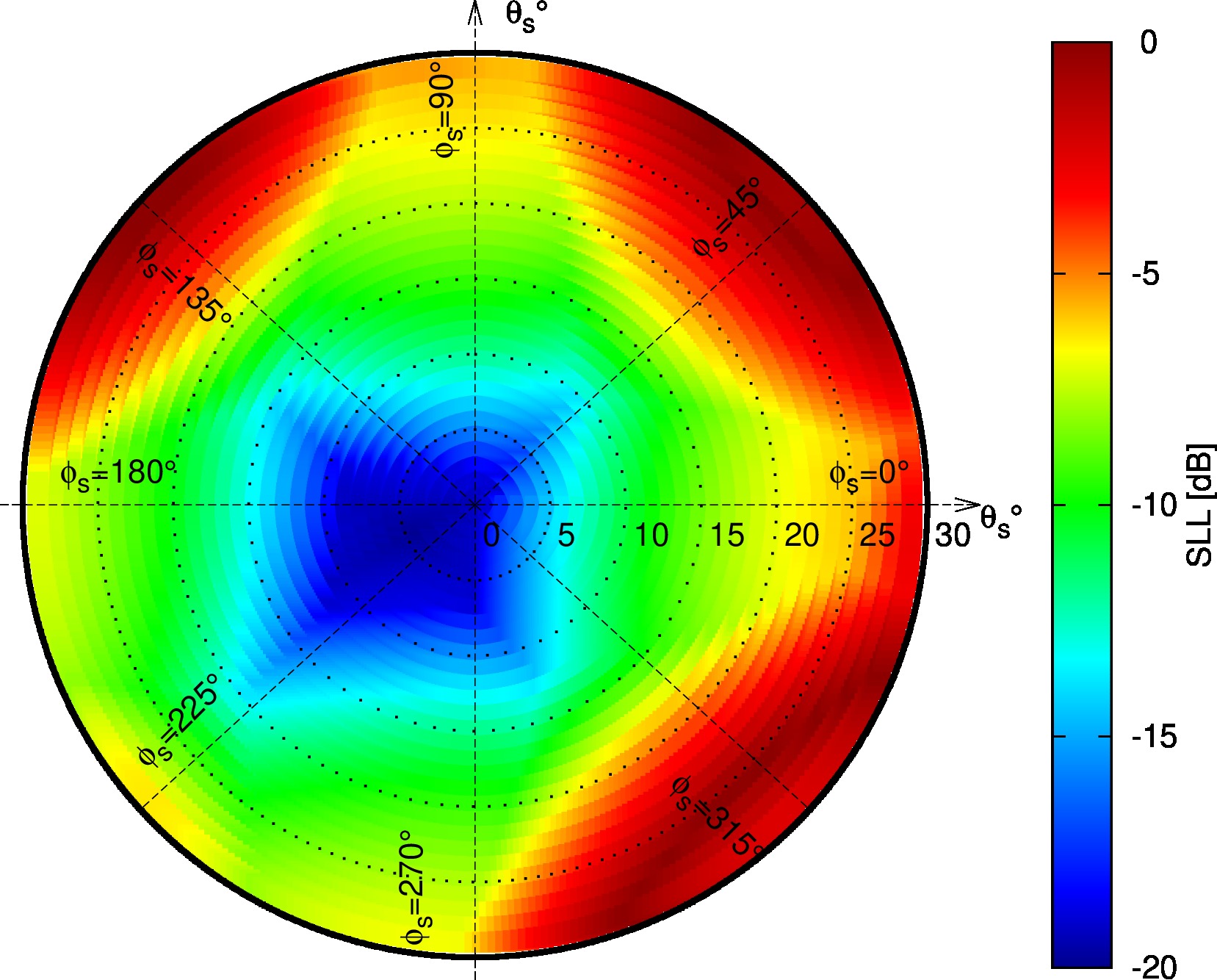}}\tabularnewline
\end{tabular}\end{center}

\begin{center}\textcolor{black}{~\vfill}\end{center}

\begin{center}\textbf{\textcolor{black}{Fig. 19 - N. Anselmi}} \textbf{\textcolor{black}{\emph{et
al.}}}\textbf{\textcolor{black}{,}} \textbf{\textcolor{black}{\emph{{}``}}}\textcolor{black}{A
Self-Replicating Single-Shape Tiling Technique ...''}\end{center}

\newpage
\begin{center}\textcolor{black}{~\vfill}\end{center}

\begin{center}\textcolor{black}{}\begin{tabular}{cc}
\textcolor{black}{\includegraphics[%
  width=0.48\columnwidth]{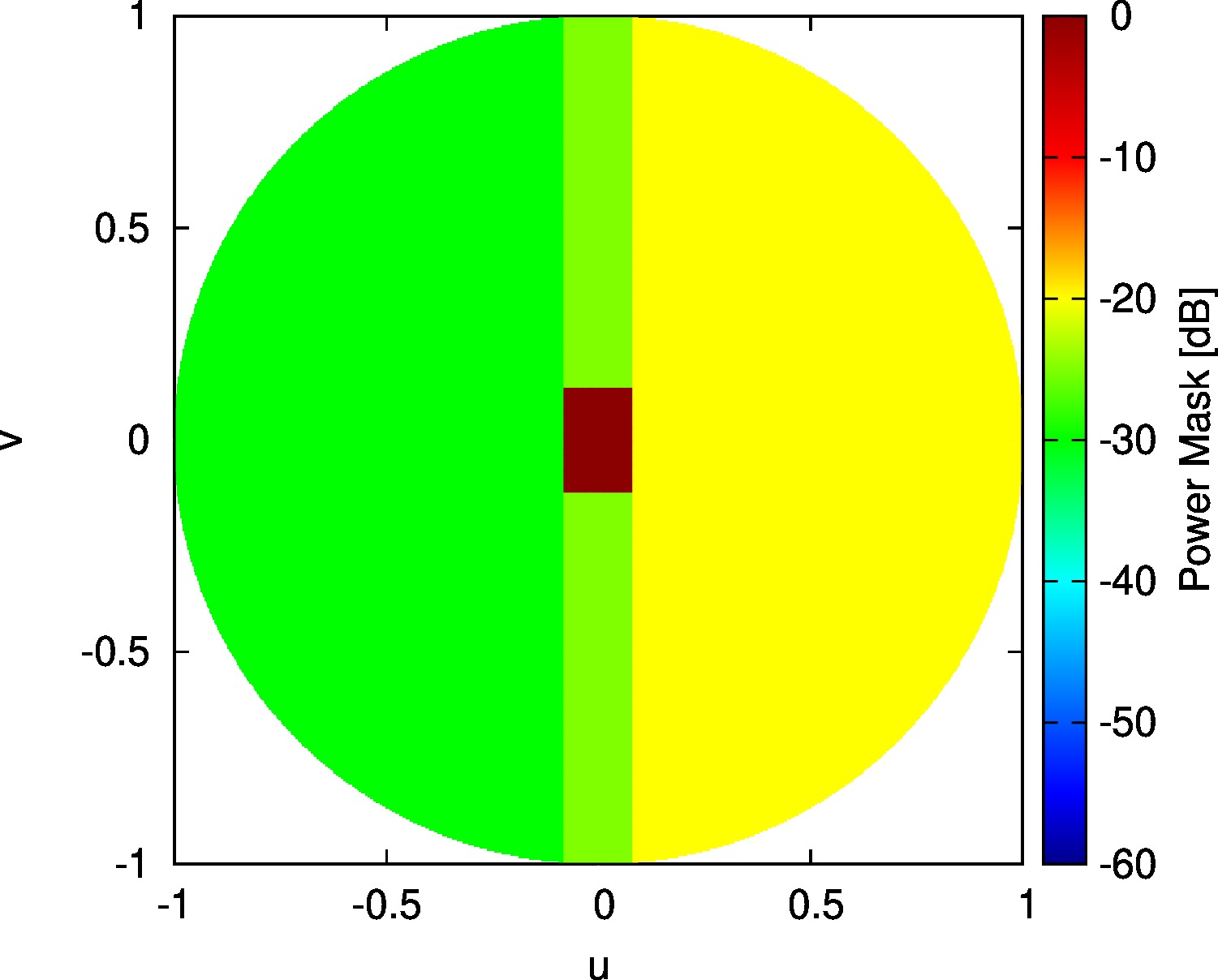}}&
\textcolor{black}{\includegraphics[%
  width=0.48\columnwidth]{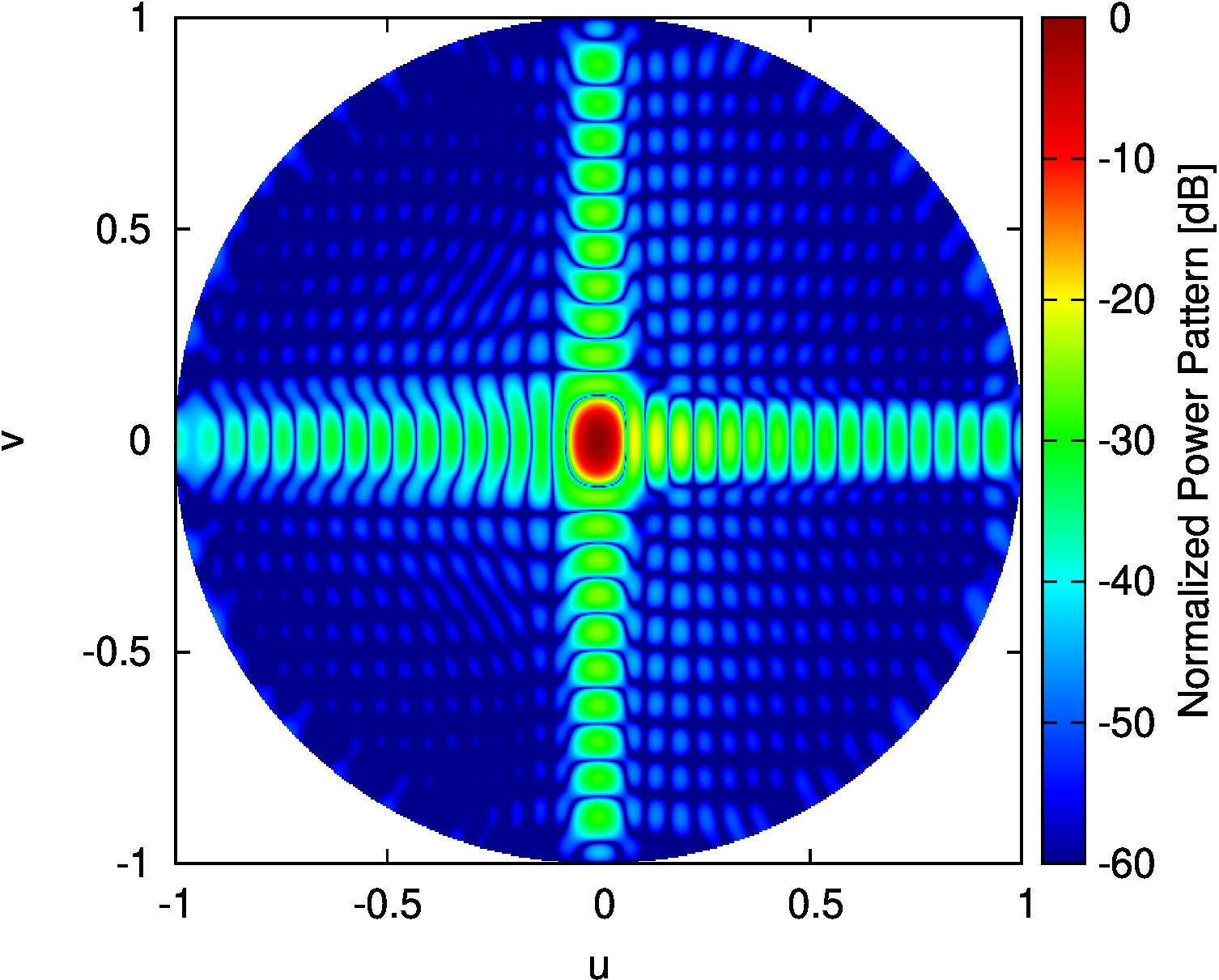}}\tabularnewline
\textcolor{black}{(}\textcolor{black}{\emph{a}}\textcolor{black}{)}&
\textcolor{black}{(}\textcolor{black}{\emph{b}}\textcolor{black}{)}\tabularnewline
\textcolor{black}{\includegraphics[%
  width=0.45\columnwidth]{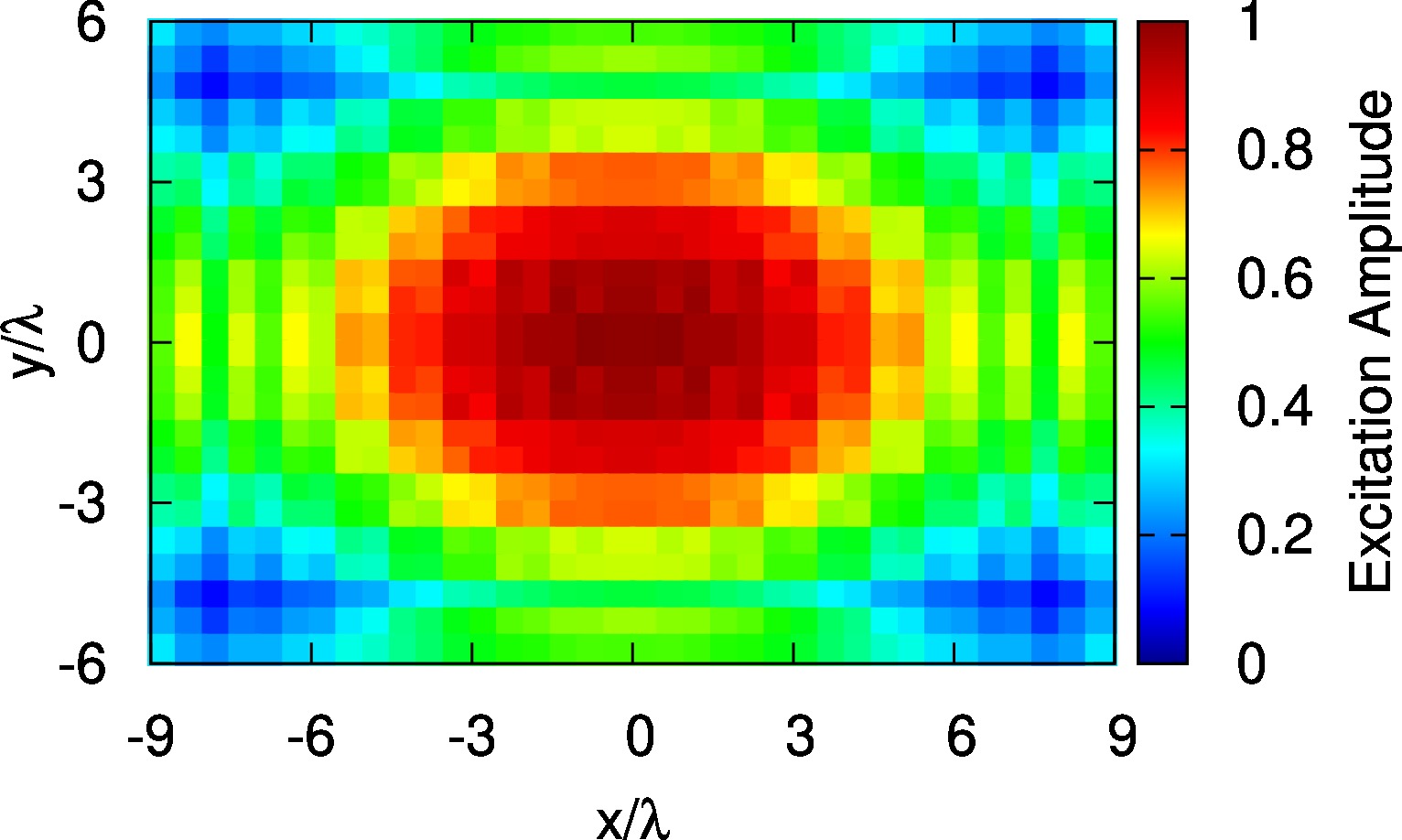}}&
\textcolor{black}{\includegraphics[%
  width=0.45\columnwidth]{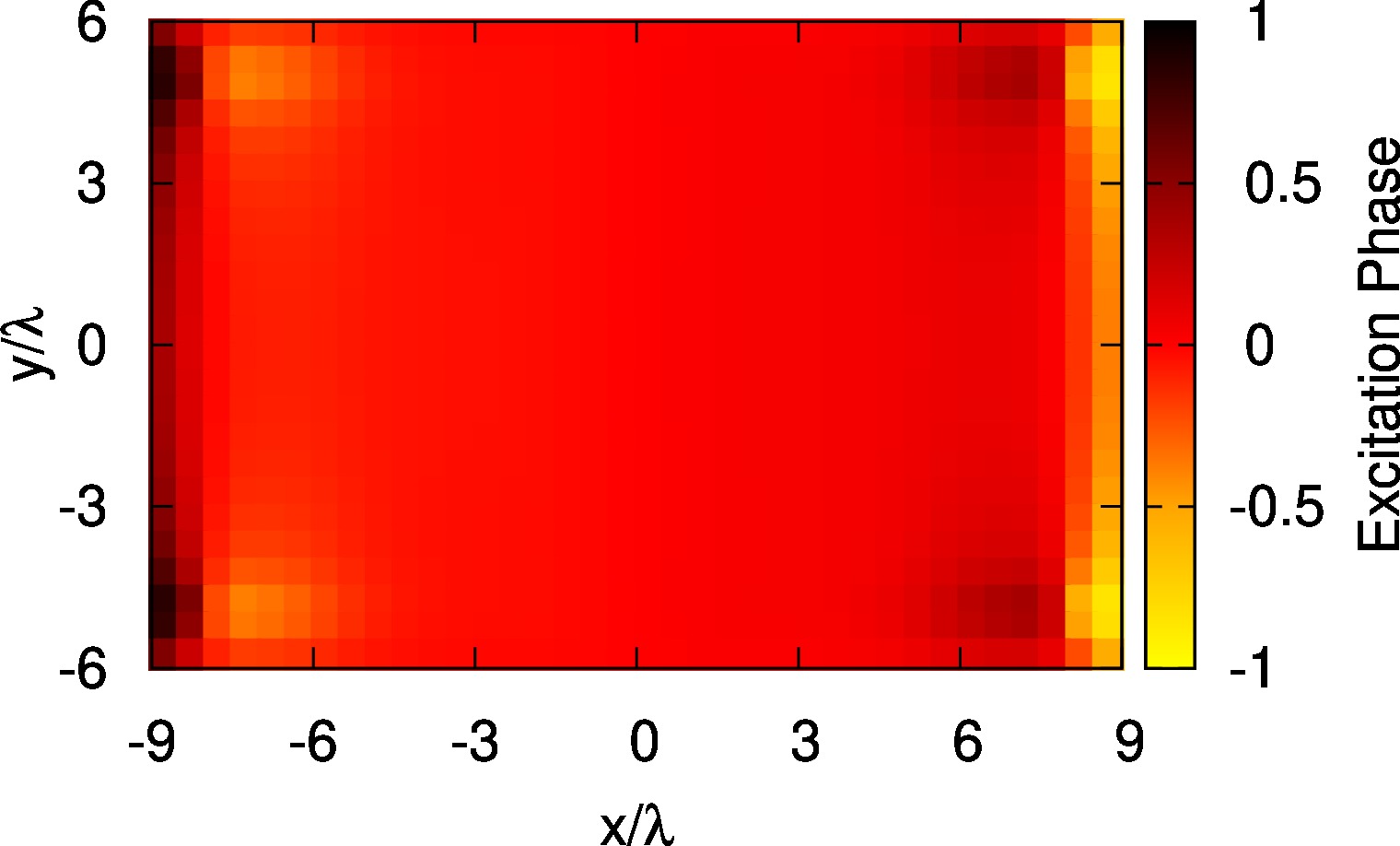}}\tabularnewline
\textcolor{black}{(}\textcolor{black}{\emph{c}}\textcolor{black}{)}&
\textcolor{black}{(}\textcolor{black}{\emph{d}}\textcolor{black}{)}\tabularnewline
\end{tabular}\end{center}

\begin{center}\textcolor{black}{~\vfill}\end{center}

\begin{center}\textbf{\textcolor{black}{Fig. 20 - N. Anselmi}} \textbf{\textcolor{black}{\emph{et
al.}}}\textbf{\textcolor{black}{,}} \textbf{\textcolor{black}{\emph{{}``}}}\textcolor{black}{A
Self-Replicating Single-Shape Tiling Technique ...''}\end{center}

\newpage
\begin{center}\textcolor{black}{~\vfill}\end{center}

\begin{center}\textcolor{black}{}\begin{tabular}{cc}
\textcolor{black}{\includegraphics[%
  width=0.45\columnwidth]{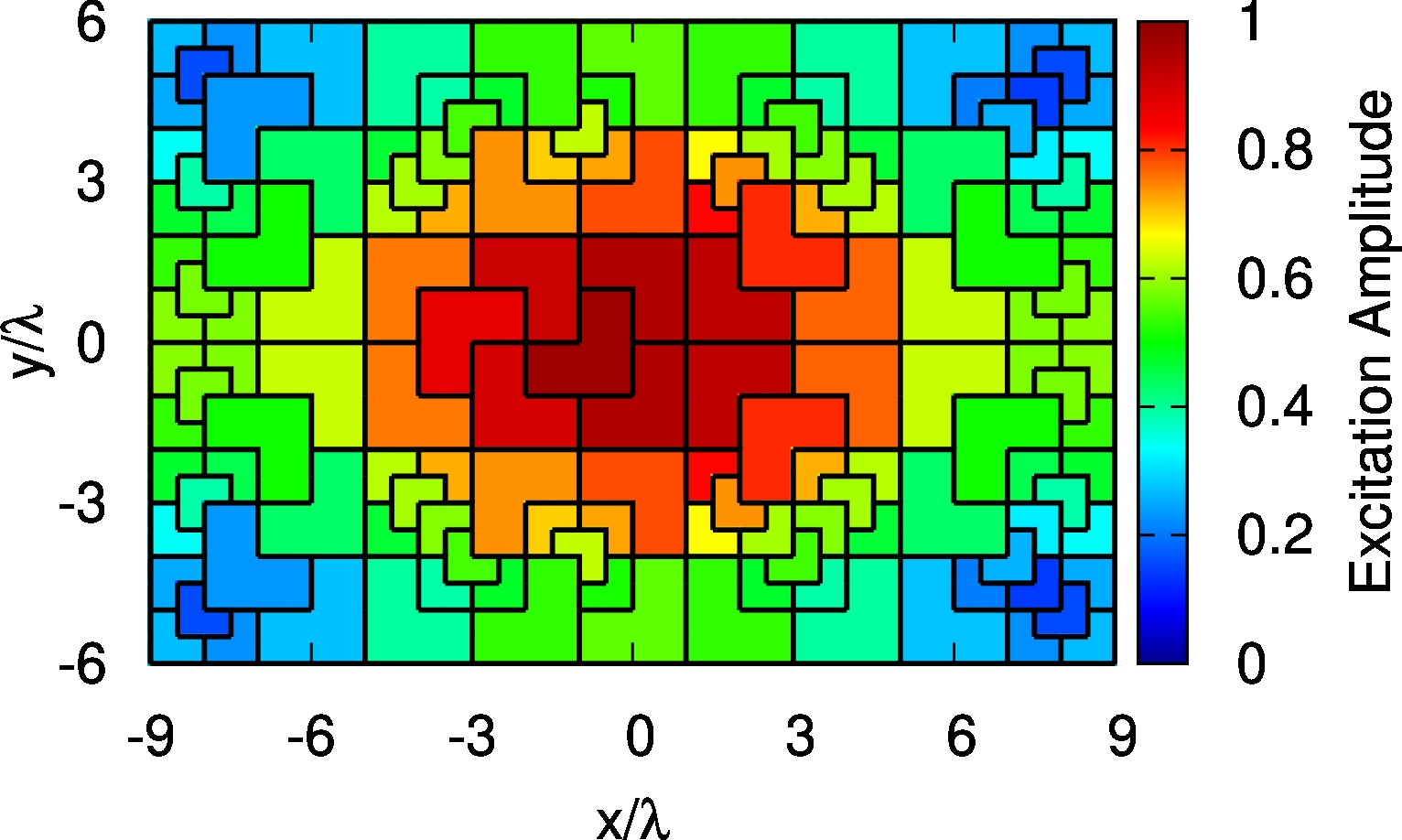}}&
\textcolor{black}{\includegraphics[%
  width=0.45\columnwidth]{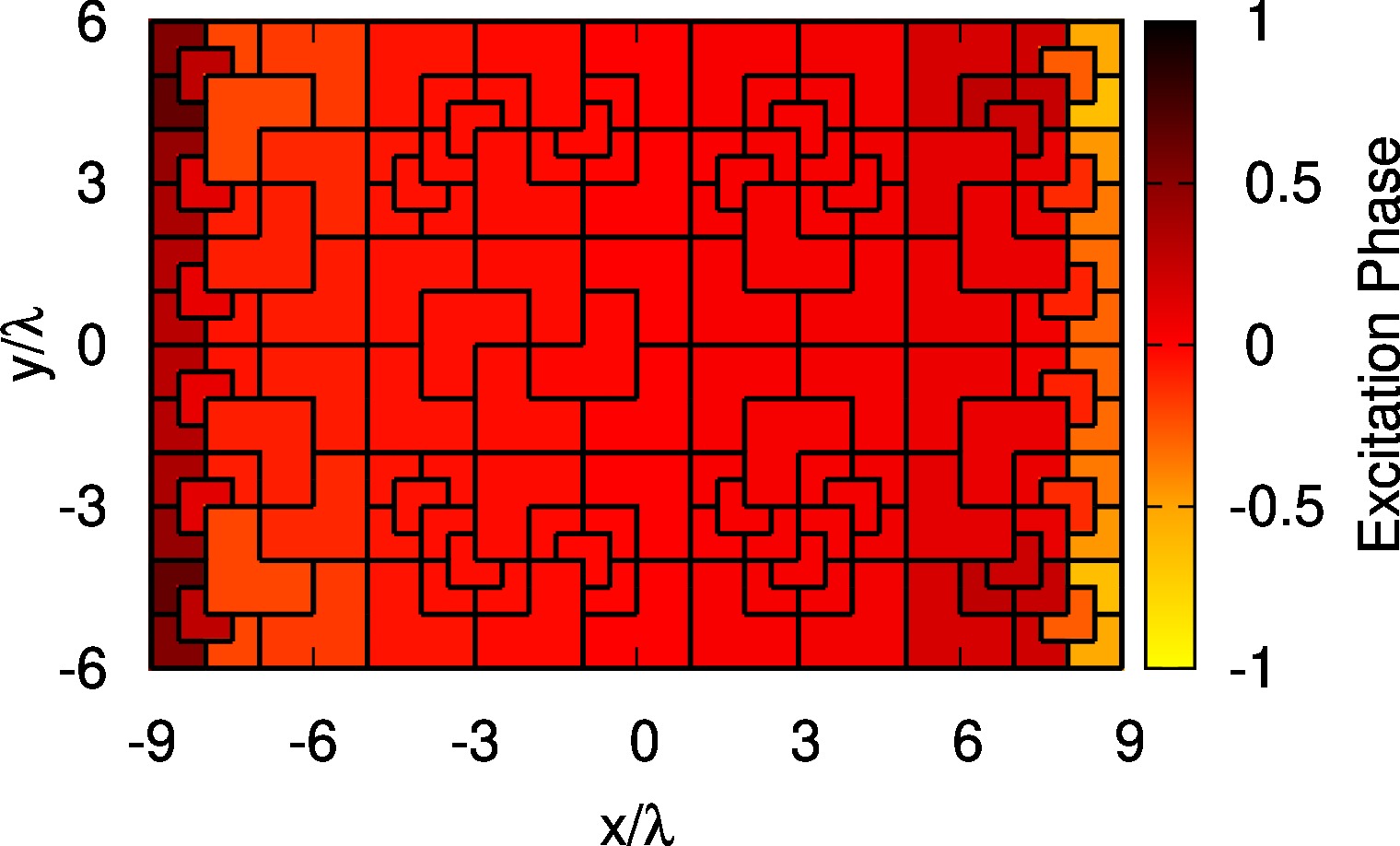}}\tabularnewline
\textcolor{black}{(}\textcolor{black}{\emph{a}}\textcolor{black}{)}&
\textcolor{black}{(}\textcolor{black}{\emph{b}}\textcolor{black}{)}\tabularnewline
\textcolor{black}{\includegraphics[%
  width=0.45\columnwidth]{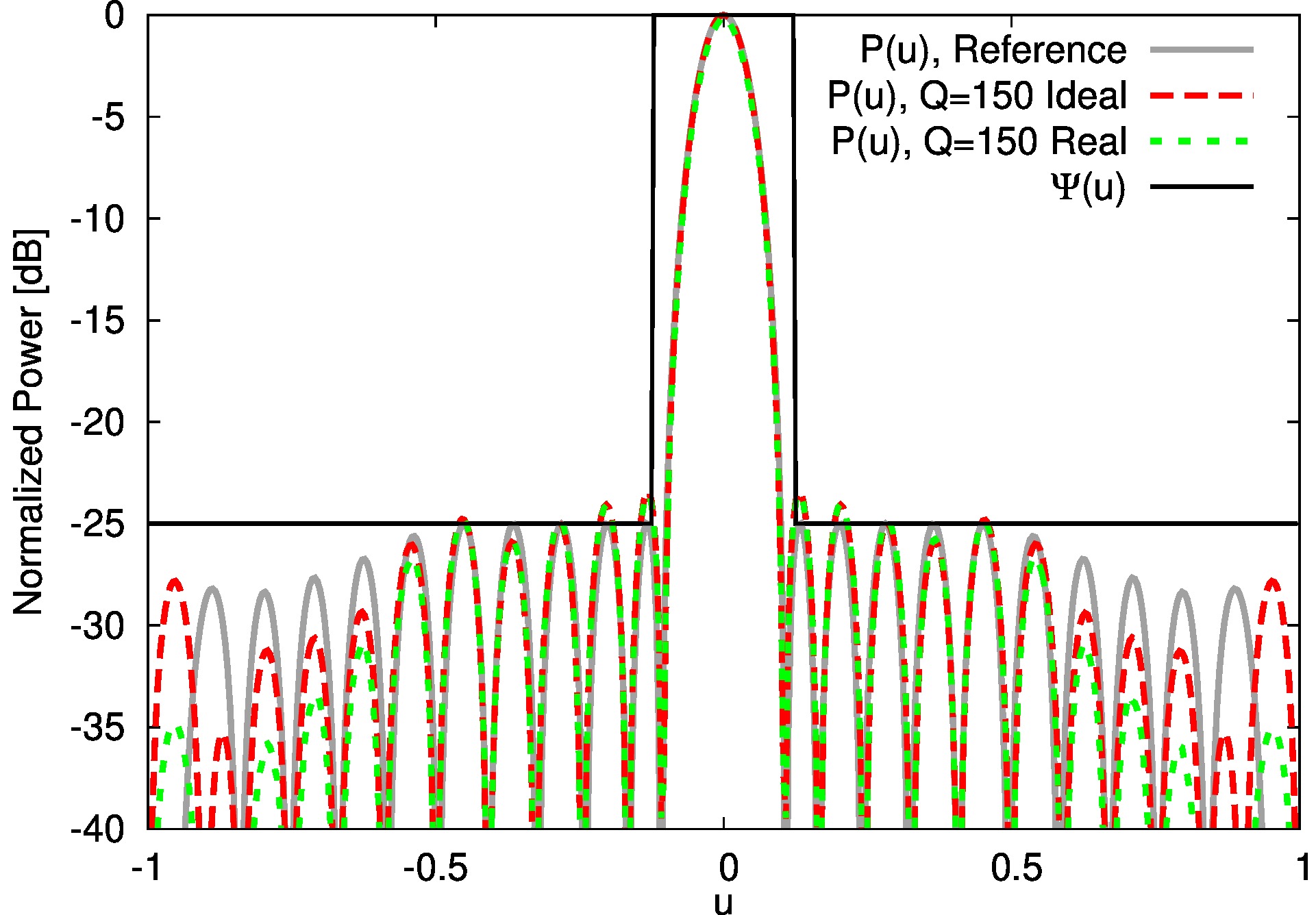}}&
\textcolor{black}{\includegraphics[%
  width=0.45\columnwidth]{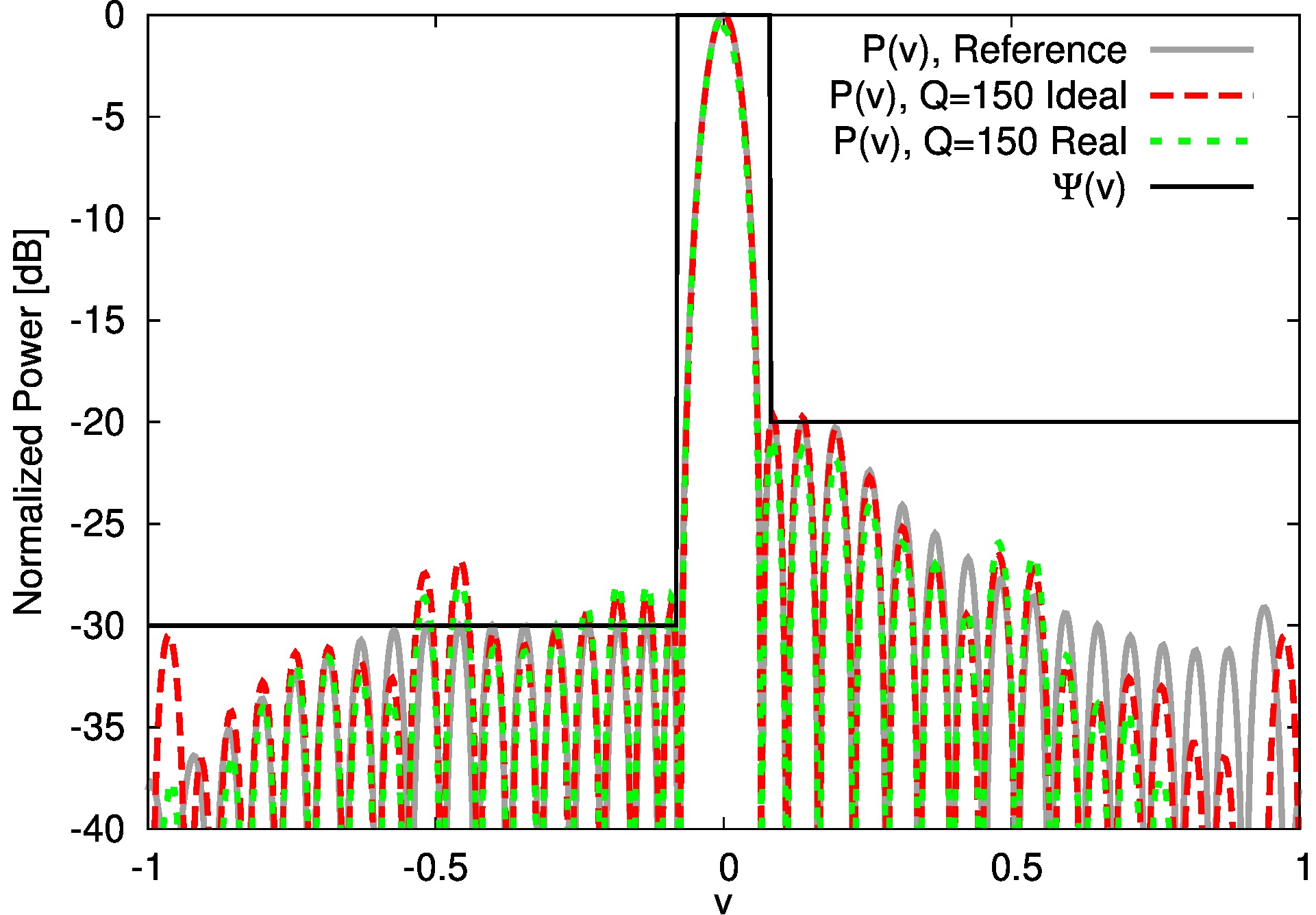}}\tabularnewline
\textcolor{black}{(}\textcolor{black}{\emph{c}}\textcolor{black}{)}&
\textcolor{black}{(}\textcolor{black}{\emph{d}}\textcolor{black}{)}\tabularnewline
\end{tabular}\end{center}

\begin{center}\textcolor{black}{~\vfill}\end{center}

\begin{center}\textbf{\textcolor{black}{Fig. 21 - N. Anselmi}} \textbf{\textcolor{black}{\emph{et
al.}}}\textbf{\textcolor{black}{,}} \textbf{\textcolor{black}{\emph{{}``}}}\textcolor{black}{A
Self-Replicating Single-Shape Tiling Technique ...''}\end{center}

\newpage
\begin{center}\textcolor{black}{~\vfill}\end{center}

\begin{center}\textcolor{black}{}\begin{tabular}{cc}
\textcolor{black}{\includegraphics[%
  width=0.45\columnwidth]{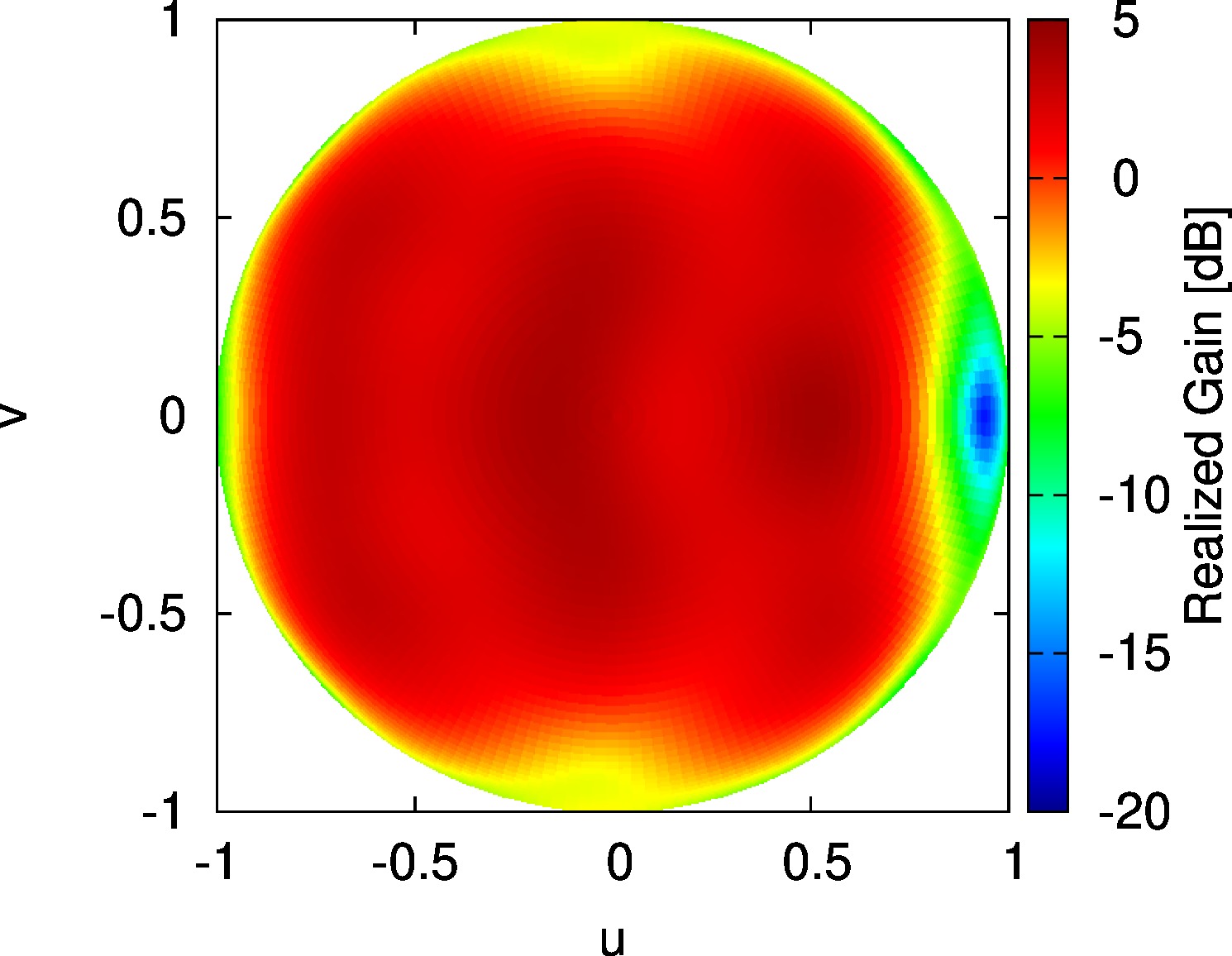}}&
\textcolor{black}{\includegraphics[%
  width=0.45\columnwidth]{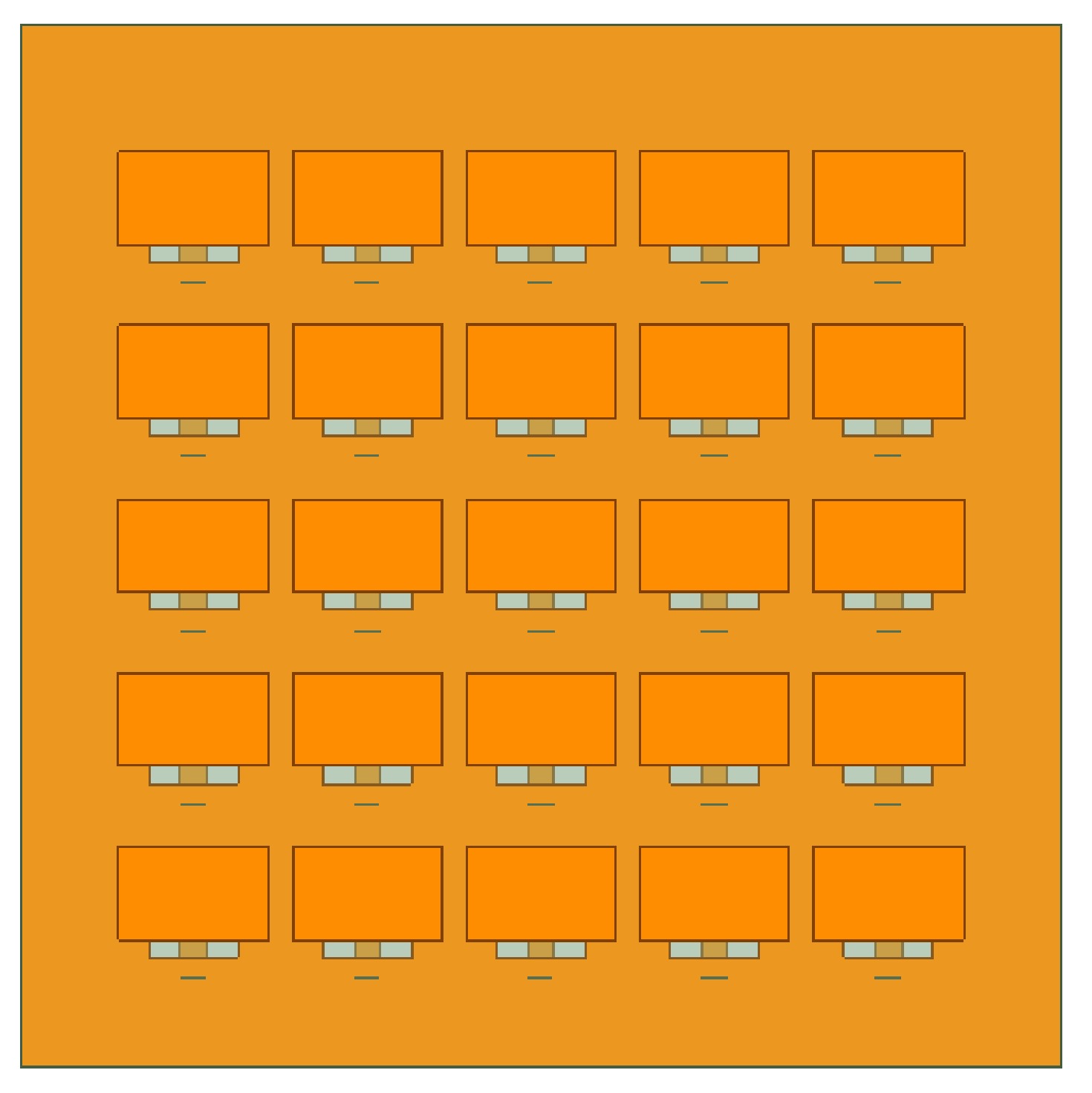}}\tabularnewline
\textcolor{black}{(}\textcolor{black}{\emph{a}}\textcolor{black}{)}&
\textcolor{black}{(}\textcolor{black}{\emph{b}}\textcolor{black}{)}\tabularnewline
\end{tabular}\end{center}

\begin{center}\textcolor{black}{~\vfill}\end{center}

\begin{center}\textbf{\textcolor{black}{Fig. 22 - N. Anselmi}} \textbf{\textcolor{black}{\emph{et
al.}}}\textbf{\textcolor{black}{,}} \textbf{\textcolor{black}{\emph{{}``}}}\textcolor{black}{A
Self-Replicating Single-Shape Tiling Technique ...''}\end{center}

\newpage
\begin{center}\textcolor{black}{~\vfill}\end{center}

\begin{center}\textcolor{black}{}\begin{tabular}{|c|c||c|c|c|c|c|c|c|c|}
\hline 
\multicolumn{1}{|c}{}&
\multicolumn{1}{c||}{}&
\multicolumn{8}{c|}{\textcolor{black}{$\widehat{N}$}}\tabularnewline
\cline{3-10} 
\multicolumn{1}{|c}{}&
\multicolumn{1}{c||}{}&
\textcolor{black}{\emph{2}}&
\textcolor{black}{\emph{3}}&
\textcolor{black}{\emph{4}}&
\textcolor{black}{\emph{5}}&
\textcolor{black}{\emph{6}}&
\textcolor{black}{\emph{7}}&
\textcolor{black}{\emph{8}}&
\textcolor{black}{\emph{9}}\tabularnewline
\hline
\hline 
\multicolumn{1}{|c|}{}&
\textcolor{black}{\emph{2}}&
\textcolor{black}{0}&
\textcolor{black}{2}&
\textcolor{black}{0}&
\textcolor{black}{0}&
\textcolor{black}{4}&
\textcolor{black}{0}&
\textcolor{black}{0}&
\textcolor{black}{8}\tabularnewline
\cline{2-2} \cline{3-3} \cline{4-4} \cline{5-5} \cline{6-6} \cline{7-7} \cline{8-8} \cline{9-9} \cline{10-10} 
\multicolumn{1}{|c|}{}&
\textcolor{black}{\emph{3}}&
\textcolor{black}{2}&
\textcolor{black}{0}&
\textcolor{black}{4}&
\textcolor{black}{0}&
\textcolor{black}{8}&
\textcolor{black}{0}&
\textcolor{black}{16}&
\textcolor{black}{0}\tabularnewline
\cline{2-2} \cline{3-3} \cline{4-4} \cline{5-5} \cline{6-6} \cline{7-7} \cline{8-8} \cline{9-9} \cline{10-10} 
\multicolumn{1}{|c|}{}&
\textcolor{black}{\emph{4}}&
\textcolor{black}{0}&
\textcolor{black}{4}&
\textcolor{black}{0}&
\textcolor{black}{0}&
\textcolor{black}{18}&
\textcolor{black}{0}&
\textcolor{black}{0}&
\textcolor{black}{88}\tabularnewline
\cline{2-2} \cline{3-3} \cline{4-4} \cline{5-5} \cline{6-6} \cline{7-7} \cline{8-8} \cline{9-9} \cline{10-10} 
\multicolumn{1}{|c|}{\textcolor{black}{$\widehat{M}$}}&
\textcolor{black}{\emph{5}}&
\textcolor{black}{0}&
\textcolor{black}{0}&
\textcolor{black}{0}&
\textcolor{black}{0}&
\textcolor{black}{72}&
\textcolor{black}{0}&
\textcolor{black}{0}&
\textcolor{black}{384}\tabularnewline
\cline{2-2} \cline{3-3} \cline{4-4} \cline{5-5} \cline{6-6} \cline{7-7} \cline{8-8} \cline{9-9} \cline{10-10} 
\multicolumn{1}{|c|}{}&
\textcolor{black}{\emph{6}}&
\textcolor{black}{4}&
\textcolor{black}{8}&
\textcolor{black}{18}&
\textcolor{black}{72}&
\textcolor{black}{162}&
\textcolor{black}{520}&
\textcolor{black}{1514}&
\textcolor{black}{4312}\tabularnewline
\cline{2-2} \cline{3-3} \cline{4-4} \cline{5-5} \cline{6-6} \cline{7-7} \cline{8-8} \cline{9-9} \cline{10-10} 
\multicolumn{1}{|c|}{}&
\textcolor{black}{\emph{7}}&
\textcolor{black}{0}&
\textcolor{black}{0}&
\textcolor{black}{0}&
\textcolor{black}{0}&
\textcolor{black}{520}&
\textcolor{black}{0}&
\textcolor{black}{0}&
\textcolor{black}{$2.27\times10^{4}$}\tabularnewline
\cline{2-2} \cline{3-3} \cline{4-4} \cline{5-5} \cline{6-6} \cline{7-7} \cline{8-8} \cline{9-9} \cline{10-10} 
\multicolumn{1}{|c|}{}&
\textcolor{black}{\emph{8}}&
\textcolor{black}{0}&
\textcolor{black}{16}&
\textcolor{black}{0}&
\textcolor{black}{0}&
\textcolor{black}{1514}&
\textcolor{black}{0}&
\textcolor{black}{0}&
\textcolor{black}{$2.04\times10^{5}$}\tabularnewline
\cline{2-2} \cline{3-3} \cline{4-4} \cline{5-5} \cline{6-6} \cline{7-7} \cline{8-8} \cline{9-9} \cline{10-10} 
\multicolumn{1}{|c|}{}&
\textcolor{black}{\emph{9}}&
\textcolor{black}{8}&
\textcolor{black}{0}&
\textcolor{black}{88}&
\textcolor{black}{384}&
\textcolor{black}{4312}&
\textcolor{black}{$2.27\times10^{4}$}&
\textcolor{black}{$2.04\times10^{5}$}&
\textcolor{black}{$1.19\times10^{6}$}\tabularnewline
\hline
\end{tabular}\end{center}

\begin{center}\textcolor{black}{~\vfill}\end{center}

\begin{center}\textbf{\textcolor{black}{Tab. I - N. Anselmi}} \textbf{\textcolor{black}{\emph{et
al.}}}\textbf{\textcolor{black}{,}} \textbf{\textcolor{black}{\emph{{}``}}}\textcolor{black}{A
Self-Replicating Single-Shape Tiling Technique ...''}\end{center}

\newpage
\begin{center}\textcolor{black}{~\vfill}\end{center}

\begin{center}\textcolor{black}{}\begin{tabular}{|c|c|c|c|c|c|c|}
\hline 
&
\textcolor{black}{$Q$}&
\textcolor{black}{$\Gamma$}&
\textcolor{black}{$SLL$}&
\textcolor{black}{$D$}&
\textcolor{black}{$HPBW_{az}$}&
\textcolor{black}{$HPBW_{el}$}\tabularnewline
&
\textcolor{black}{-}&
\textcolor{black}{-}&
\textcolor{black}{{[}dB{]}}&
\textcolor{black}{{[}dB{]}}&
 \textcolor{black}{{[}deg{]}}&
\textcolor{black}{{[}deg{]}}\tabularnewline
\hline
\hline 
\textcolor{black}{Reference}&
\textcolor{black}{576}&
\textcolor{black}{$7.01\times10^{-9}$}&
\textcolor{black}{-24.99}&
\textcolor{black}{31.58}&
\textcolor{black}{5.12}&
\textcolor{black}{5.11}\tabularnewline
\hline
\hline 
\textcolor{black}{\emph{L}}\textcolor{black}{-trominoes}&
\textcolor{black}{39}&
\textcolor{black}{$5.53\times10^{-5}$}&
\textcolor{black}{-19.31}&
\textcolor{black}{31.65}&
\textcolor{black}{4.97}&
\textcolor{black}{4.97}\tabularnewline
\hline 
\textcolor{black}{Squares}&
\textcolor{black}{39}&
\textcolor{black}{$1.07\times10^{-4}$}&
\textcolor{black}{-22.55}&
\textcolor{black}{31.61}&
\textcolor{black}{5.02}&
\textcolor{black}{5.02}\tabularnewline
\hline
\hline 
\textcolor{black}{\emph{L}}\textcolor{black}{-trominoes}&
\textcolor{black}{81}&
\textcolor{black}{$1.77\times10^{-5}$}&
\textcolor{black}{-23.54}&
\textcolor{black}{31.60}&
\textcolor{black}{5.05}&
\textcolor{black}{5.05}\tabularnewline
\hline 
\textcolor{black}{Squares}&
\textcolor{black}{81}&
\textcolor{black}{$4.08\times10^{-5}$}&
\textcolor{black}{-24.49}&
\textcolor{black}{31.59}&
\textcolor{black}{5.06}&
\textcolor{black}{5.07}\tabularnewline
\hline
\hline 
\textcolor{black}{\emph{L}}\textcolor{black}{-trominoes}&
\textcolor{black}{120}&
\textcolor{black}{$8.11\times10^{-6}$}&
\textcolor{black}{-24.17}&
\textcolor{black}{31.60}&
\textcolor{black}{5.08}&
\textcolor{black}{5.08}\tabularnewline
\hline 
\textcolor{black}{Squares}&
\textcolor{black}{120}&
\textcolor{black}{$2.23\times10^{-5}$}&
\textcolor{black}{-24.10}&
\textcolor{black}{31.58}&
\textcolor{black}{5.09}&
\textcolor{black}{5.08}\tabularnewline
\hline
\end{tabular}\end{center}

\begin{center}\textcolor{black}{~\vfill}\end{center}

\begin{center}\textbf{\textcolor{black}{Tab. II - N. Anselmi}} \textbf{\textcolor{black}{\emph{et
al.}}}\textbf{\textcolor{black}{,}} \textbf{\textcolor{black}{\emph{{}``}}}\textcolor{black}{A
Self-Replicating Single-Shape Tiling Technique ...''}\end{center}

\newpage
\begin{center}\textcolor{black}{~\vfill}\end{center}

\begin{center}\textcolor{black}{}\begin{tabular}{|c|c|c|c|c|c|c|}
\hline 
&
\textcolor{black}{$Q$}&
 \textcolor{black}{$\Gamma$}&
\textcolor{black}{$SLL$}&
\textcolor{black}{$D$}&
\textcolor{black}{$HPBW_{az}$}&
\textcolor{black}{$HPBW_{el}$}\tabularnewline
&
\textcolor{black}{-}&
\textcolor{black}{-}&
 \textcolor{black}{{[}dB{]}}&
\textcolor{black}{{[}dB{]}}&
 \textcolor{black}{{[}deg{]}}&
\textcolor{black}{{[}deg{]}}\tabularnewline
\hline
\hline 
\textcolor{black}{Reference}&
\textcolor{black}{864}&
\textcolor{black}{$3.39\times10^{-10}$}&
\textcolor{black}{-25.00}&
\textcolor{black}{33.34}&
\textcolor{black}{3.39}&
\textcolor{black}{5.13}\tabularnewline
\hline
\hline 
\textcolor{black}{\emph{RTA}}&
\textcolor{black}{18}&
\textcolor{black}{$7.33\times10^{-5}$}&
\textcolor{black}{-21.33}&
\textcolor{black}{33.46}&
\textcolor{black}{3.34}&
\textcolor{black}{4.78}\tabularnewline
\hline 
\textcolor{black}{\emph{RTA}}&
\textcolor{black}{48}&
\textcolor{black}{$2.58\times10^{-5}$}&
\textcolor{black}{-22.63}&
\textcolor{black}{33.39}&
\textcolor{black}{3.37}&
\textcolor{black}{4.95}\tabularnewline
\hline 
\textcolor{black}{\emph{RTA}}&
\textcolor{black}{150}&
\textcolor{black}{$5.63\times10^{-5}$}&
\textcolor{black}{-24.78}&
\textcolor{black}{33.35}&
\textcolor{black}{3.38}&
\textcolor{black}{5.08}\tabularnewline
\hline 
\textcolor{black}{\emph{RTA}}&
\textcolor{black}{270}&
\textcolor{black}{$7.17\times10^{-6}$}&
\textcolor{black}{-24.70}&
\textcolor{black}{33.35}&
\textcolor{black}{3.39}&
\textcolor{black}{5.11}\tabularnewline
\hline
\end{tabular}\end{center}

\begin{center}\textcolor{black}{~\vfill}\end{center}

\begin{center}\textbf{\textcolor{black}{Tab. III - N. Anselmi}} \textbf{\textcolor{black}{\emph{et
al.}}}\textbf{\textcolor{black}{,}} \textbf{\textcolor{black}{\emph{{}``}}}\textcolor{black}{A
Self-Replicating Single-Shape Tiling Technique ...''}\end{center}

\newpage
\begin{center}\textcolor{black}{~\vfill}\end{center}

\begin{center}\textcolor{black}{}\begin{tabular}{|c|c|c|c|c|c|c|}
\hline 
&
\textcolor{black}{$Q$}&
\textcolor{black}{$\Gamma$}&
\textcolor{black}{$SLL$}&
\textcolor{black}{$D$}&
\textcolor{black}{$HPBW_{az}$}&
\textcolor{black}{$HPBW_{el}$}\tabularnewline
&
\textcolor{black}{-}&
 \textcolor{black}{-}&
\textcolor{black}{{[}dB{]}}&
\textcolor{black}{{[}dB{]}}&
\textcolor{black}{{[}deg{]}}&
\textcolor{black}{{[}deg{]}}\tabularnewline
\hline
\hline 
\textcolor{black}{Reference}&
\textcolor{black}{864}&
\textcolor{black}{$3.99\times10^{-9}$}&
\textcolor{black}{-20.00}&
\textcolor{black}{33.81}&
\textcolor{black}{3.18}&
\textcolor{black}{4.78}\tabularnewline
\hline
\hline 
\textcolor{black}{\emph{RTA}}&
\textcolor{black}{18}&
\textcolor{black}{$7.73\times10^{-4}$}&
\textcolor{black}{-12.57}&
\textcolor{black}{32.70}&
\textcolor{black}{3.11}&
\textcolor{black}{4.53}\tabularnewline
\hline 
\textcolor{black}{\emph{RTA}}&
\textcolor{black}{48}&
\textcolor{black}{$2.85\times10^{-4}$}&
\textcolor{black}{-15.91}&
\textcolor{black}{33.10}&
\textcolor{black}{3.18}&
\textcolor{black}{4.65}\tabularnewline
\hline 
\textcolor{black}{\emph{RTA}}&
\textcolor{black}{150}&
\textcolor{black}{$2.33\times10^{-5}$}&
\textcolor{black}{-19.30}&
\textcolor{black}{33.53}&
\textcolor{black}{3.18}&
\textcolor{black}{4.76}\tabularnewline
\hline 
\textcolor{black}{\emph{RTA}}&
\textcolor{black}{270}&
\textcolor{black}{$1.52\times10^{-6}$}&
\textcolor{black}{-19.71}&
\textcolor{black}{33.71}&
\textcolor{black}{3.18}&
\textcolor{black}{4.77}\tabularnewline
\hline
\end{tabular}\end{center}

\begin{center}\textcolor{black}{~\vfill}\end{center}

\begin{center}\textbf{\textcolor{black}{Tab. IV - N. Anselmi}} \textbf{\textcolor{black}{\emph{et
al.}}}\textbf{\textcolor{black}{,}} \textbf{\textcolor{black}{\emph{{}``}}}\textcolor{black}{A
Self-Replicating Single-Shape Tiling Technique ...''}\end{center}

\newpage
\begin{center}\textcolor{black}{~\vfill}\end{center}

\begin{center}\textcolor{black}{}\begin{tabular}{|c|c|c|c|c|c|c|}
\hline 
&
\textcolor{black}{$Q$}&
 \textcolor{black}{$\Gamma$}&
\textcolor{black}{$SLL$}&
\textcolor{black}{$D$}&
\textcolor{black}{$HPBW_{az}$}&
\textcolor{black}{$HPBW_{el}$}\tabularnewline
&
\textcolor{black}{-}&
\textcolor{black}{-}&
\textcolor{black}{{[}dB{]}}&
\textcolor{black}{{[}dB{]}}&
\textcolor{black}{{[}deg{]}}&
\textcolor{black}{{[}deg{]}}\tabularnewline
\hline
\hline 
\textcolor{black}{Reference - Ideal}&
\textcolor{black}{864}&
\textcolor{black}{$1.70\times10^{-9}$}&
\textcolor{black}{-20.00}&
\textcolor{black}{33.57}&
\textcolor{black}{3.25}&
\textcolor{black}{4.90}\tabularnewline
\hline
\textcolor{black}{Reference - Real}&
\textcolor{black}{864}&
\textcolor{black}{$1.92\times10^{-7}$}&
\textcolor{black}{-21.10}&
\textcolor{black}{33.85}&
\textcolor{black}{3.07}&
\textcolor{black}{4.65}\tabularnewline
\hline
\hline 
\textcolor{black}{\emph{RTA}} \textcolor{black}{- Ideal}&
\textcolor{black}{18}&
\textcolor{black}{$5.12\times10^{-5}$}&
\textcolor{black}{-20.00}&
\textcolor{black}{33.57}&
\textcolor{black}{3.20}&
\textcolor{black}{4.65}\tabularnewline
\hline 
\textcolor{black}{\emph{RTA}} \textcolor{black}{- Real}&
\textcolor{black}{18}&
\textcolor{black}{$5.24\times10^{-5}$}&
\textcolor{black}{-19.97}&
\textcolor{black}{34.04}&
\textcolor{black}{3.02}&
\textcolor{black}{4.43}\tabularnewline
\hline
\hline 
\textcolor{black}{\emph{RTA}} \textcolor{black}{- Ideal}&
\textcolor{black}{48}&
\textcolor{black}{$2.53\times10^{-5}$}&
\textcolor{black}{-19.12}&
\textcolor{black}{33.67}&
\textcolor{black}{3.20}&
\textcolor{black}{4.76}\tabularnewline
\hline 
\textcolor{black}{\emph{RTA}} \textcolor{black}{- Real}&
\textcolor{black}{48}&
\textcolor{black}{$2.46\times10^{-5}$}&
\textcolor{black}{-20.16}&
\textcolor{black}{33.97}&
\textcolor{black}{3.03}&
\textcolor{black}{4.53}\tabularnewline
\hline
\hline 
\textcolor{black}{\emph{RTA}} \textcolor{black}{- Ideal}&
\textcolor{black}{150}&
\textcolor{black}{$4.44\times10^{-6}$}&
\textcolor{black}{-19.68}&
\textcolor{black}{33.61}&
\textcolor{black}{3.22}&
\textcolor{black}{4.86}\tabularnewline
\hline 
\textcolor{black}{\emph{RTA}} \textcolor{black}{- Real}&
\textcolor{black}{150}&
\textcolor{black}{$3.77\times10^{-6}$}&
\textcolor{black}{-20.72}&
\textcolor{black}{33.90}&
\textcolor{black}{3.05}&
\textcolor{black}{4.61}\tabularnewline
\hline
\hline 
\textcolor{black}{\emph{RTA}} \textcolor{black}{- Ideal}&
\textcolor{black}{270}&
\textcolor{black}{$1.25\times10^{-6}$}&
\textcolor{black}{-19.94}&
\textcolor{black}{33.60}&
\textcolor{black}{3.24}&
\textcolor{black}{4.88}\tabularnewline
\hline
\textcolor{black}{\emph{RTA}} \textcolor{black}{- Real}&
\textcolor{black}{270}&
\textcolor{black}{$1.74\times10^{-6}$}&
\textcolor{black}{-20.98}&
\textcolor{black}{33.86}&
\textcolor{black}{3.06}&
\textcolor{black}{4.63}\tabularnewline
\hline
\end{tabular}\end{center}

\begin{center}\textcolor{black}{~\vfill}\end{center}

\begin{center}\textbf{\textcolor{black}{Tab. V - N. Anselmi}} \textbf{\textcolor{black}{\emph{et
al.}}}\textbf{\textcolor{black}{,}} \textbf{\textcolor{black}{\emph{{}``}}}\textcolor{black}{A
Self-Replicating Single-Shape Tiling Technique ...''}\end{center}

\begin{thebibliography}{10}
\bibitem{Hong.2021}\textcolor{black}{W. Hong, Z. H. Jiang, C. Yu, D. Hou, H. Wang, C.
Guo, Y. Hu, L. Kuai, Y. Yu, Z. Jiang, Z. Chen, J. Chen, Z. Yu, J.
Zhai, N. Zhang, L. Tian, F. Wu, G. Yang, Z.-C. Hao, and J. Y. Zhou,
{}``The role of millimeter-wave technologies in 5G/6G wireless communications,''}
\textcolor{black}{\emph{IEEE J. Microw.}}\textcolor{black}{, vol.
1, no. 1, pp. 101-122, Jan. 2021.}
\bibitem{Oliveri.2019}\textcolor{black}{G. Oliveri, G. Gottardi, and A. Massa, {}``A new
meta-paradigm for the synthesis of antenna arrays for future wireless
communications,''} \textcolor{black}{\emph{IEEE Trans. Antennas Propag.}}\textcolor{black}{,
vol. 67, no. 6, pp. 3774-3788, Jun. 2019.}
\bibitem{Yu.2020}\textcolor{black}{Y. Yu, Z. H. Jiang, H. Zhang, Z. Zhang, and W. Hong,
{}``A low-profile beamforming patch array with a cosecant fourth
power pattern for millimeter-wave synthetic aperture radar applications,''}
\textcolor{black}{\emph{IEEE Trans. Antennas Propag}}\textcolor{black}{.,
vol. 68, no. 9, pp. 6486-6496, Sep. 2020.}
\bibitem{Waldshmidt.2021}\textcolor{black}{C. Waldschmidt, J. Hasch, and W. Menzel, \char`\"{}Automotive
radar - From first efforts to future systems,''} \textcolor{black}{\emph{IEEE
J. Microw.}}\textcolor{black}{, vol. 1, no. 1, pp. 135-148, Jan. 2021.}
\bibitem{Peng.2021}\textcolor{black}{J.-J. Peng, S.-W. Qu, M. Xia, and S. Yang, {}``Conformal
phased array antenna for unmanned aerial vehicle with +/-70 {[}deg{]}
scanning range,''} \textcolor{black}{\emph{IEEE Trans. Antennas Propag.}}\textcolor{black}{,
vol. 69, no. 8, pp. 4580-4587, Aug. 2021.}
\bibitem{Yang.2020}\textcolor{black}{Y. Yang, Z. Fan, T. Hong, M. Chen, X. Tang, J. He,
X. Chen, C. Liu, H. Zhu, and K. Huang, {}``Design of microwave directional
heating system based on phased-array antenna,''} \textcolor{black}{\emph{IEEE
Trans. Microw. Theory Tech.}}\textcolor{black}{, vol. 68, no. 11,
pp. 4896-4904, Nov. 2020.}
\bibitem{Li.2021}\textcolor{black}{S. Li, S. Wang, Q. An, G. Zhao, and H. Sun, {}``Cylindrical
MIMO array-based near-field microwave imaging,''} \textcolor{black}{\emph{IEEE
Trans. Antennas Propag}}\textcolor{black}{., vol. 69, no. 1, pp. 612-617,
Jan. 2021.}
\bibitem{Gu.2015}\textcolor{black}{X. Gu, A. Valdes-Garcia, A. Natarajan, B. Sadhu,
D. Liu, and S. K. Reynolds, \char`\"{}W-band scalable phased arrays
for imaging and communications,\char`\"{}} \textcolor{black}{\emph{IEEE
Comm. Mag.}}\textcolor{black}{, vol. 53, no. 4, pp. 196-204, Apr.
2015.}
\bibitem{Herd.2016}\textcolor{black}{J. S. Herd and M. D. Conwey, {}``The evolution
to modern phased array architectures,''} \textit{\textcolor{black}{Proc.
IEEE,}} \textcolor{black}{vol. 104, no. 3, pp. 519-529, Mar. 2016.}
\bibitem{Abdellatif.2021}\textcolor{black}{A. S. Abdellatif, W. Zhai, H. K. Pothula, and M.
Repeta, {}``Array of arrays: optimizing phased array tiles,''} \textcolor{black}{\emph{IEEE
Antennas Wireless Propag. Lett.}}\textcolor{black}{, vol. 20, no.
5, pp. 718-722, May 2021.}
\bibitem{Rocca.2016}\textcolor{black}{P. Rocca, G. Oliveri, R. J. Mailloux, and A. Massa,
{}``Unconventional phased array architectures and design methodologies
- A review,''} \textcolor{black}{\emph{Proc. IEEE}}\textcolor{black}{,
vol. 104, no. 3, pp. 544-560, Mar. 2016.}
\bibitem{Oliveri.2020}\textcolor{black}{G. Oliveri, G. Gottardi, M. A. Hannan, N. Anselmi,
and L. Poli, {}``Autocorrelation-driven synthesis of antenna arrays
- The case of DS-based planar isophoric thinned arrays,''} \textcolor{black}{\emph{IEEE
Trans. Antennas Propag.}}\textcolor{black}{, vol. 68, no. 4, pp. 2895-2910,
Apr. 2020.}
\bibitem{Poli.2013}\textcolor{black}{L. Poli, P. Rocca, M. Salucci, and A. Massa, {}``Reconfigurable
thinning for the adaptive control of linear arrays,''} \textcolor{black}{\emph{IEEE
Trans. Antennas Propag.}}\textcolor{black}{, vol. 61, no. 10, pp.
5068-5077, Oct. 2013.}
\bibitem{Carlin.2015}\textcolor{black}{M. Carlin, G. Oliveri, and A. Massa, {}``Hybrid
BCS-deterministic approach for sparse concentric ring isophoric arrays,''}
\textcolor{black}{\emph{IEEE Trans. Antennas Propag.}}\textcolor{black}{,
vol. 63, no. 1, pp. 378-383, Jan. 2015.}
\bibitem{Viswanathan.2021}\textcolor{black}{N. Viswanathan, S. Venkatesh, and D. Schurig, {}``Optimization
of a sparse aperture configuration for millimeter-wave computational
imaging,''} \textcolor{black}{\emph{IEEE Trans. Antennas Propag}}\textcolor{black}{,
vol. 69, no. 2, pp. 1107-1117, Feb. 2021.}
\bibitem{Manica.2008}\textcolor{black}{L. Manica, P. Rocca, A. Martini, and A. Massa, \char`\"{}An
innovative approach based on a tree-searching algorithm for the optimal
matching of independently optimum sum and difference excitations,\char`\"{}}
\textcolor{black}{\emph{IEEE Trans. Antennas Propag}}\textcolor{black}{.,
vol. 56, no. 1, pp. 58-66, Jan. 2008.}
\bibitem{Rocca.2019}\textcolor{black}{P. Rocca, M. H. Hannan, L. Poli, N. Anselmi, and
A. Massa, {}``Optimal phase-matching strategy for beam scanning of
sub-arrayed phased arrays,''} \textcolor{black}{\emph{IEEE Trans.
Antennas and Propag.}}\textcolor{black}{, vol. 67, no. 2, pp. 951-959,
Feb. 2019.}
\bibitem{Anselmi.2018.a}\textcolor{black}{N. Anselmi, P. Rocca, M. Salucci, and A. Massa,
{}``Contiguous phase-clustering in multibeam-on-receive scanning
arrays,''} \textcolor{black}{\emph{IEEE Trans. Antennas Propag.}}\textcolor{black}{,
vol. 66, no. 11, pp. 5879-5891, Nov. 2018.}
\bibitem{Anselmi.2019.a}\textcolor{black}{N. Anselmi, G. Gottardi, G. Oliveri, and A. Massa,
{}``A total-variation sparseness-promoting method for the synthesis
of contiguously clustered linear architectures''} \textcolor{black}{\emph{IEEE
Trans. Antennas Propag}}\textcolor{black}{., vol. 67, no. 7, pp. 4589-4601,
Jul. 2019.}
\bibitem{Yang.2015}\textcolor{black}{X. Yang, W. Xi, Y. Sun, T. Zeng, T. Long, and T.
K. Sarkar, {}``Optimization of subarray partition for large planar
phased array radar based on weighted K-means clustering method,''}
\textcolor{black}{\emph{IEEE J. Sel. Topics Signal Process}}\textcolor{black}{.,
vol. 9, no. 8, pp. 1460-1468, Dec. 2015.}
\bibitem{Rocca.2014}\textcolor{black}{P. Rocca, M. D'Urso, and L. Poli, {}``Advanced
sub-arraying strategy for large antenna array design with sub-array-only
control,''} \textcolor{black}{\emph{IEEE Antennas Wireless Propag.
Lett.}}\textcolor{black}{, vol. 13, pp. 91-94, 2014.}
\bibitem{Rocca.2015}\textcolor{black}{P. Rocca, R. J. Mailloux, and G. Toso, {}``GA-based
optimization of irregular sub-array layouts for wideband phased arrays
design,''} \textcolor{black}{\emph{IEEE Antennas Wireless Propag.
Lett.}}\textcolor{black}{, vol. 14, pp. 131-134, 2015.}
\bibitem{D'Urso.2012}\textcolor{black}{M. D'Urso, M. G. Labate, A. Buonanno, and P. Vinetti,
{}``Effective beam forming networks for large arbitrary array of
antennas,''} \textcolor{black}{\emph{IEEE Trans. Antennas Propag.}}\textcolor{black}{,
vol. 60, no. 11, pp. 5129-5135, Nov. 2012.}
\bibitem{Xiong.2013}\textcolor{black}{Z-Y. Xiong, Z-H. Xu, S-W. Chen, and S-P. Xiao, {}``Subarray
partition in array antenna based on the algorithm X,''} \textcolor{black}{\emph{IEEE
Antennas Wireless Propag. Lett.}}\textcolor{black}{, vol. 12, pp.
906-909, 2013.}
\bibitem{Anselmi.2017}\textcolor{black}{N. Anselmi, P. Rocca, M. Salucci, and A. Massa,
{}``Irregular phased array tiling by means of analytic schemata-driven
optimization,''} \textcolor{black}{\emph{IEEE Trans. Antennas Propag.}}\textcolor{black}{,
vol. 65, no. 9, pp. 4495-4510, Sep. 2017.}
\bibitem{Zhang.2019}\textcolor{black}{R. Zhang, H. Zhang, W. Xu, and X. You, {}``Subarray-based
simultaneous beam training for multiuser mmwave massive MIMO systems,''}
\textcolor{black}{\emph{IEEE Wireless Commun. Lett}}\textcolor{black}{.,
vol. 8, no. 4, pp. 976-979, Aug. 2019.}
\bibitem{Ma.2019}\textcolor{black}{Y. Ma, S. Yang, Y. Chen, S. Qu, and J. Hu, {}``Pattern
synthesis of 4-D irregular antenna arrays based on maximum-entropy
model,''} \textcolor{black}{\emph{IEEE Trans. Antennas Propag.}}\textcolor{black}{,
vol. 67, no. 5, pp. 3048-3057, May 2019.}
\bibitem{Rocca.2020.a}\textcolor{black}{P. Rocca, N. Anselmi, A. Polo, and A. Massa, {}``Modular
design of hexagonal phased arrays through diamond tiles,''} \textcolor{black}{\emph{IEEE
Trans. Antennas Propag.}}\textcolor{black}{, vol. 68, no. 5, pp. 3598-3612,
May 2020.}
\bibitem{Bekker.2020}\textcolor{black}{E. Bekker, J. W. Odendaal, and J. Joubert, {}``Optimized
polarization for rotationally tiled, wideband, dual-polarized Vivaldi
arrays,''} \textcolor{black}{\emph{Proc. 14th European Conf. Antennas
Propag.}} \textcolor{black}{(EuCAP), Copenhagen, Denmark, 15-20 Mar.
2020.}
\bibitem{Dong.2020}\textcolor{black}{W. Dong, Z. Xu, X. Liu, L. Wang, and S. Xiao, {}``Irregular
subarray tiling via heuristic iterative convex relaxation programming,''}
\textcolor{black}{\emph{IEEE Trans. Antennas Propag.}}\textcolor{black}{,
vol. 68, no. 4, pp. 2842-2852, Apr. 2020.}
\bibitem{Oliveri.2016}\textcolor{black}{G. Oliveri, M. Salucci, and A. Massa, {}``Synthesis
of modular contiguously clustered linear arrays through a sparseness-regularized
solver,''} \textcolor{black}{\emph{IEEE Trans. Antennas Propag}}\textcolor{black}{.,
vol. 64, no. 10, pp. 4277-4287, Oct. 2016.}
\bibitem{Oliveri.2017}\textcolor{black}{G. Oliveri, G. Gottardi, F. Robol, A. Polo, L. Poli,
M. Salucci, M. Chuan, C. Massagrande, P. Vinetti, M. Mattivi, R. Lombardi,
and A Massa, {}``Codesign of unconventional array architectures and
antenna elements for 5G base stations,''} \textcolor{black}{\emph{IEEE
Trans. Antennas Propag.}}\textcolor{black}{, vol. 65, no. 12, pp.
6752-6767, Dec. 2017.}
\bibitem{Bekers.2019}\textcolor{black}{D. J. Bekers, S. Jacobs, S. Monni, R. J. Bolt, D.
Fortini. P. Capece, and G. Toso, {}``A Ka-band spaceborne synthetic
aperture radar instrument: a modular sparse array antenna design,''}
\textcolor{black}{\emph{IEEE Antennas Propag. Mag}}\textcolor{black}{.,
vol. 61, no. 5, pp. 97-104, Oct. 2019.}
\bibitem{Rocca.2020.c}\textcolor{black}{P. Rocca, N. Anselmi, A. Polo, and A. Massa, {}``An
irregular two-sizes square tiling method for the design of isophoric
phased arrays,''} \textcolor{black}{\emph{IEEE Trans. Antennas Propag.}}\textcolor{black}{,
vol. 68, no. 6, pp. 4437-4449, Jun. 2020.}
\bibitem{Rapakula.2020}\textcolor{black}{B. Rupakula, A. H. Aljuhani, and G. M. Rebeiz, {}``Limited
scan-angle phased-arrays using randomly-grouped subarrays and reduced
number of phase-shifters,''} \textcolor{black}{\emph{IEEE Trans.
Antennas Propag.,}} \textcolor{black}{vol. 68, no. 1, Jan. 2020.}
\bibitem{Mailloux.2009}\textcolor{black}{R. J. Mailloux, S. G. Santarelli, T. M. Roberts,
and D. Luu, {}``Irregular polyomino-shaped subarrays for space-based
active arrays'',} \textcolor{black}{\emph{Int. J. Antennas Propag.}}\textcolor{black}{,
vol. 2009, no. 956524, pp. 1-9, 2009.}
\bibitem{Mailloux.2005}\textcolor{black}{R. J. Mailloux,} \textcolor{black}{\emph{Phased
Array Antenna Handbook}}\textcolor{black}{, 2nd ed. Norwood, MA: Artech
House, 2005.}
\bibitem{Golomb.1964}\textcolor{black}{S. W. Golomb, \char`\"{}Replicating figures in the
plane,\char`\"{}} \textcolor{black}{\emph{Math. Gaz}}\textcolor{black}{.,
vol. 48, no. 366, pp. 403-412, Dec. 1964.}
\bibitem{Gardner.2001}\textcolor{black}{M. Gardner, \char`\"{}Rep-Tiles,''} \textcolor{black}{\emph{The
Colossal Book of Mathematics: Classic Puzzles, Paradoxes, and Problems}}\textcolor{black}{,
W. W. Norton, pp. 46-58, 2001.}
\bibitem{Robinson.1999}\textcolor{black}{E. A. Robinson, {}``On the table and the chair,''}
\textcolor{black}{\emph{Indag. Mathem.}}\textcolor{black}{, vol. 10,
no. 4, pp. 581-599, Dec. 1999.}
\bibitem{Julien.2016}\textcolor{black}{A. Julien and J. Savinien, \char`\"{}K-theory of
the chair tiling via AF-algebras.\char`\"{}} \textcolor{black}{\emph{J.
Geom. Phys.,}} \textcolor{black}{vol. 106, pp. 314-326, May 2016.}
\bibitem{Fathauer.2021}\textcolor{black}{R. Fathauer, {}``Matching rules, aperiodic Tiles,
and substitution tilings,''} \textcolor{black}{\emph{Tessellations:
Mathematics, Art, and Recreation}}\textcolor{black}{, CRC Press, NY,
2021.}
\bibitem{Knuth.2000}\textcolor{black}{D. E. Knuth, {}``Dancing links,''} \textcolor{black}{\emph{Millenial
Perspectives in Computer Science}}\textcolor{black}{, pp. 187-214,
2000.}
\bibitem{Anselmi.2022}\textcolor{black}{N. Anselmi, P. Rocca, S. Feuchtinger, B. Biscontini,
A. Murillo-Barrera, and A. Massa, {}``Optimal capacity-driven design
of aperiodic clustered phased arrays for multi-user MIMO communication
systems,''} \textcolor{black}{\emph{IEEE Trans. Antennas Propag.}}\textcolor{black}{,
vol. 70, no. 7, pp. 5491-5505, Jul. 2022.}
\bibitem{Chu.1958}\textcolor{black}{I. Chu and R. Johnsonbaugh, {}``Tiling boards with
trominoes,''} \textcolor{black}{\emph{J. Recreat. Math}}\textcolor{black}{.,
vol. 18, no. 3, pp. 188-193, 1958.}
\bibitem{Ueno.2008}\textcolor{black}{C. Ueno, {}``Matrices and tilings with right trominoes,
''} \textcolor{black}{\emph{Math. Mag.,}} \textcolor{black}{vol.
81, no. 5, pp. 319-331, 2008.}
\bibitem{Isernia.2004}\textcolor{black}{T. Isernia, F. A. Pena, O. M. Bucci, M. D'Urso,
J. F. Gomez, and J. A. Rodriguez, {}``A hybrid approach for the optimal
synthesis of pencil beams through array antennas,''} \textcolor{black}{\emph{IEEE
Trans. Antennas Propag}}\textcolor{black}{., vol. 52, no. 11, pp.
2912-2918, Nov. 2004.}
\bibitem{ANSYS 2019}\textcolor{black}{ANSYS Electromagnetics Suite-HFSS, ANSYS, Canonsburg,
PA, USA, 2019. }\newpage

\end{thebibliography}
\end{document}